\documentclass[11pt]{article}
\usepackage{amsmath}
\usepackage{comment}
\usepackage{amsfonts}
\usepackage{amsmath, amsfonts}
\usepackage{dsfont}
\usepackage{mathrsfs} 
\usepackage[active]{srcltx}
\usepackage{hyperref}
\usepackage[all]{xy}
\usepackage{bm}
\usepackage{dcolumn}
\usepackage[dvipsnames]{xcolor}
\usepackage{graphicx}
\usepackage{siunitx}
\usepackage{yfonts}
\usepackage{textcomp}
\usepackage{textgreek}
\usepackage{physics}
\usepackage[cal=boondox]{mathalfa}
\usepackage[bottom]{footmisc}
\usepackage{amssymb}
\usepackage{mathtools}
\usepackage{tikz}
\usepackage{bigints}

\usepackage[utf8]{inputenc}
\usepackage{array}
\usepackage{multirow}
\usepackage{tabu}

\makeatletter
\newcommand*{\rom}[1]{\expandafter\@slowromancap\romannumeral #1@}
\makeatother

\usepackage{fancyhdr} 
\usepackage[utf8]{inputenc}
\usepackage[english]{babel}

\DeclareMathAlphabet{\mathpzc}{OT1}{pzc}{m}{it}
\graphicspath{ {./images/} }

\hypersetup{
    colorlinks=true,   
    linkcolor=black,
    urlcolor=blue,
    citecolor=black
}
\usepackage{color}
\definecolor{DarkBlue}{rgb}{0.1,0.1,0.5}
\definecolor{Red}{rgb}{0.9,0.1,0.1}
\definecolor{Green}{rgb}{0.3,0.7,0.0}
\definecolor{green2}{rgb}{0.1,0.7,0.2}
\definecolor{blue2}{rgb}{0.0,0.6,0.7}
\definecolor{pink}{rgb}{1,0.0,1}
\definecolor{orange}{rgb}{0.9,0.0,0.1}
 
\usepackage[bottom]{footmisc}

\newtheorem{lemma}{Lemma}
\newtheorem{prop}{Proposition}

\renewcommand{\d}{\mathrm{d}}
\newcommand{\derpar}[2]{\displaystyle\frac{\partial{#1}}{\partial{#2}}}

\newcommand{\Lag}{\mathscr{L}}

\newcommand{\vf}{\mathfrak{X}}
\newcommand{\df}{\Omega}

\newcommand{\Tan}{\mathrm{T}}
\newcommand{\inn}{{\mathop{i}\nolimits}}

\def\proof{( {\sl Proof} )\quad}
\def\qed{\ifvmode\removelastskip\fi
{\unskip\nobreak\hfil\penalty50\hbox{}\nobreak\hfil \hbox{\vrule
height1.2ex width1.2ex}\parfillskip=0pt \finalhyphendemerits=0
\par\smallskip}}

\newcommand{\bal}{\begin{align*}}
\newcommand{\eal}{\end{align*}}
\def\beq{\begin{equation}}
\def\eeq{\end{equation}}
\def\bea{\begin{eqnarray}}
\def\eea{\end{eqnarray}}
\def\beann{\begin{eqnarray*}}
\def\eeann{\end{eqnarray*}}
\def\ben{\begin{enumerate}}
\def\een{\end{enumerate}}
\def\bit{\begin{itemize}}
\def\eit{\end{itemize}}

\def\vf{\mathfrak X}
\def\df{{\mit\Omega}}

\def\d{{\rm d}}

\def\Real{\mathbb{R}}

\def\Tan{{\rm T}}

\def\inn{\mathop{i}\nolimits}
\def\Cinfty{{\rm C}^\infty}

\title{\vskip -10mm
\sc Multisymplectic Constraint Analysis of Scalar Field Theories, 
Chern-Simons Gravity, and Bosonic String Theory}
\author{\sffamily 
\sc 
$^a$ Joaquim Gomis
\thanks{joaquim.gomis@ub.edu\,({\it ORCID}:\,0000-0002-8706-2989).} ,
$^b$ Arnoldo Guerra IV
\thanks{aguerrgu54@alumnes.ub.edu\,({\it ORCID}:\,0000-0001-8738-274X).} ,
$^c$ Narciso Rom\'an-Roy
\thanks{narciso.roman@upc.edu\,({\it ORCID}:\,0000-0003-3663-9861).} .
\\[1ex]
\normalsize\itshape\sffamily 
$^a$Departament de F\'isica Qu\`antica i Astrof\'isica and Institut de Ci\`encies del Cosmos (ICCUB),
\\  \normalsize\itshape\sffamily 
Universitat de Barcelona, 
Barcelona, Spain.
\\[1ex]
\normalsize\itshape\sffamily 
$^b$Departament de F\'isica Qu\`antica i Astrof\'isica,
Universitat de Barcelona, 
Barcelona, Spain.
\\[1ex]
\normalsize\itshape\sffamily 
$^d$Department of Mathematics,
Universitat Polit\`ecnica de Catalunya,
Barcelona, Spain.
}

\begin{document}

\maketitle

\leavevmode
\vadjust{\kern -15mm}

\begin{abstract}
\noindent
The (pre)multisymplectic geometry of the De Donder--Weyl formalism for field theories is further developed for a variety of field theories including a scalar field theory from the canonical Klein-Gordon action, the electric and magnetic Carrollian scalar field theories, bosonic string theory from the Nambu-Goto action, and $2+1$ gravity as a Chern-Simons theory. 
The Lagrangians for the scalar field theories and for $2+1$ Chern-Simons gravity are found to be singular in the De Donder--Weyl sense while the Nambu-Goto Lagrangian is found to be regular.
Furthermore, the constraint structure of the premultisymplectic phase spaces of singular field theories is explained and applied to these theories.
Finally, it is studied how symmetries are developed on the (pre)multisymplectic phase spaces in the presence of constraints. 
\end{abstract}


\bigskip

\vbox{AMS (2020): 
{\it Primary\/}: 53D42, 70S05, 70S10.
{\it Secondary\/}: 35Q75, 53C15, 53Z05, 81T30, 83D05.
\indent PACS (2022): 
03.50.-z, 11.10.Ef, 02.40.-k, 04.50.Kd, 04.20.Fy, 11.25.-w, 11.30.-j, 02.30.Jr.}
\null 

\markright{{\rm J. Gomis, A. Guerra IV,  N. Rom\'an--Roy},
    {\sl Multisymplectic Constraint Analysis \ldots}}

\newpage
\setcounter{tocdepth}{2}
{
\def\baselinestretch{1}
\small
\def\addvspace#1{\vskip 1pt}
\parskip 0pt plus 0.1mm
\tableofcontents
}


\section{Introduction}

Covariant phase space formalisms for field theories have been relevant in theoretical physics since the development of the variational calculus for classical field theories. 
When a field theory under investigation is invariant under Lorentz transformations, it is possible to develop a manifestly Lorentz covariant construction of the variational principle and the symmetries of the Lagrangian. 
This is evident in the Lagrangian formulation of field theories in which the fields are, in general, tensors on spacetime and the spacetime coordinates are treated as indistinguishable independent variables.
In the case of string theory, the fields are scalar fields on the string worldsheet and covariance is similarly maintained. 
One appropriate geometrical setting for constructing covariant phase spaces for field theories in the Lagrangian formalism is given by {\sl jet bundles} which are fiber bundle over spacetime whose fiber coordinates consist of the fields and the spacetime derivatives of the fields. 
The covariant phase space manifold in the Lagrangian setting for first-order field theories (``first-order" in the sense that only first derivatives appear in the Lagrangian) is called the {\sl first-order jet bundle} which will be referred to colloquially here as the {\sl multivelocity phase space},
in analogy with the velocity phase space where the Lagrangian formulation of mechanical systems takes place.
Thus, the coordinates on the first-order jet bundle, whose fibers over spacetime are the spacetime first derivatives of the fields, are called the {\sl multivelocities}. In this paper only first-order field theories will be investigated.

Lorentz covariance in Hamiltonian formulations of Lorentz invariant field theories is much more subtle. The first attempt to develop a covariant Hamiltonian formalism was made in the 1930s by Th\'eophile De Donder and  Hermann Weyl \cite{dD-1930,We-1935} who generalized the variational calculus for the Hamiltonian formalism in mechanics to the multivariable context in field theory, giving rise to what is today referred to as the {\sl De Donder--Weyl approach} in which the variational principle leads to the so-called {\sl (Hamilton)--De Donder--Weyl equations}. 
The phase space manifold on which the De Donder--Weyl formalism for first-order field theories is constructed is called the 
{\sl dual first-order jet bundle} and is referred to as the {\sl multimomentum phase space}.
The coordinates on this phase space are called {\sl multimomenta} and are obtained from the Legendre map given as the partial derivatives of the Lagrangian with respect to all multivelocities.
The De Donder--Weyl formulation of field theories has become better understood by learning about the underlying {\sl (pre)multisymplectic geometry} \cite{kijowski1,kijowski2} of the multivelocity and multimomentum phase spaces. 
(Pre)multisymplectic geometry in classical field theory is the geometric multivariable extension of (pre)symplectic geometry in classical mechanics where the physical observables of interest (position, velocity, momentum) depend on a single variable (time). 
Similarly to how to the simple (single-variable) variational calculus can be performed on (pre)symplectic phase spaces as done in classical mechanics, the multiple variational calculus can be posed geometrically on the (pre)multisymplectic phase spaces.

The De Donder--Weyl formalism, however, was hindered at the time of its development due to a lack of an understanding of constraints.
A constraint analysis is needed to obtain the full dynamics of a field theory with a Lagrangian whose Legendre map is singular. In the De Donder--Weyl approach, the Legendre map is singular when the multi-Hessian constructed from the multivelocities is singular. 
The appropriate constraint analysis that is needed for such singular field theories was not understood at the time when the De Donder--Weyl approach was first developed. 
Instead, field theories were studied using the canonical Hamiltonian formalism which arises from performing a space+time splitting where, given an initial value Cauchy data surface, one solves a set of partial differential equations for dynamical functions which describe the time evolution of the classical fields of interest. 
This approach sacrifices the manifest Lorentz covariance exhibited by Lorentz invariant field theories in the Lagrangian formalism. However, the advantage at the time was that the (non-covariant) Dirac--Bergmann constraint analysis \cite{constraints1,constraints2} gave the full dynamics of singular field theories
(``singular" with respect to the Legendre map that gives the canonical momenta rather than the multimomenta used in the De Donder--Weyl approach). 
Furthermore, the canonical formalism gave a straightforward method for performing quantization, making the canonical formalism for field theories exceptionally useful. 

The success of the canonical formalism in quantum field theory made it the standard Hamiltonian field theory formalism in theoretical physics. However, understanding how to develop the canonical formulation of field theories in a manifestly Lorentz covariant manner remained a mystery for many years until Crnkovi\'c and Witten discovered how to construct a classical covariant phase space in which the canonical formulation (on-shell) can be carried out without having to perform a space + time splitting \cite{wittencovariant}. 
Soon after, Lee and Wald generalized this new covariant phase space formalism to also work off-shell in \cite{leewaldcov}. The Dirac--Bergmann constraint analysis makes use of Poisson brackets which are defined on symplectic phase space manifolds; the covariant phase space manifolds presented by Crnkovi\'c, Witten, Lee, and Wald are infinite-dimensional (pre)symplectic manifolds it is therefore possible to carry out the geometric Gotay--Nester--Hinds constraint algorithm \cite{presymp3} which is equivalent to the Dirac--Bergmann constraint analysis. In fact, Lee and Wald specify in \cite{leewaldcov} how the covariant Poisson brackets of constraints corresponding to a local symmetry are related to the Lie algebra of the local symmetry under inspection. For a treatment of boundary terms in the covariant phase space formalism presented by Lee and Wald, see \cite{newcov2,harlow}.

The difficulty with performing an analogous constraint analysis {\it a la} Dirac on the multivelocity and multimomentum phase spaces on which the De Donder--Weyl formalism takes place arises fundamentally from 
the fact that Poisson brackets on jet and dual jet bundles are not yet fully understood.
A geometric algorithm for performing the desired covariant constraint analysis in the De Donder--Weyl formalism was developed in \cite{LMMMR-2005} where the premultisymplectic geometry of the multivelocity and multimomentum phase spaces is used to describe the constraint submanifolds on which field theories are physically relevant.
This analysis is geometrically analogous to how the presymplectic geometry of the phase spaces in classical mechanics is used to perform the geometric constraint analysis of mechanical systems equivalent to the Dirac--Bergmann constraint procedure for mechanics developed in \cite{presymp3,presymp1,LMMMR-2002,presymp4,MMT-97,presymp2}.
For classical field theories constructed on $m$-dimensional spacetimes, the jet and dual jet bundles, i.e. the multivelocity and multimomentum phase spaces, are (pre)multisymplectic manifolds equipped with (pre)multisymplectic $(m+1)$-forms. These differential forms {\sl multisymplectic} for regular field theories and are {\sl premultisymplectic} for singular field theories; the precise mathematical definitions of these forms will be given later.
Furthermore, the symmetries of a field theory are Lie groups which act on the phase spaces in such a way that their (pre)multisymplectic forms are preserved and Noether's theorem yields conserved currents called {\sl covariant momentum maps} which are $(m-1)$-forms on the (pre)multisymplectic phase spaces.
A general overview of (pre)multisymplectic geometry, including new geometry developed in this work, is provided in Section \ref{section:DiffGeo} of this paper.
For reviews of the previously known aspects of the multisymplectic formulation of field theories see also, for example,  \cite{GIMMSY,art:Roman09,book:Saunders89}.
The symplectic form on the covariant phase space introduced by Crnkovi\'c and Witten is derived from the multisymplectic form on the multimomentum phase space in \cite{multisympcov}. 
Furthermore, in \cite{Gimmsy2,Gimmsy3} it is shown how performing a space + time splitting of the multimomentum phase space gives the standard canonical formalism for field theories along with a canonical constraint analysis (equivalent to the constraint analysis {\it a la} Dirac) in which covariant momentum maps play a crucial role.

In this paper, the (pre)multisymplectic geometry of the multivelocity and multimomentum phase spaces is used in a variety of field theories to develop: 
the field equations both in the Lagrangian and in the De Donder--Weyl Hamiltonian formulations, the geometric constraint analysis \cite{LMMMR-2005}, 
and the covariant momentum maps \cite{GIMMSY} which give the conserved currents associated with Noether symmetries. 
The premultisymplectic constraint algorithm originally formalised in \cite{LMMMR-2005} is performed here using a local-coordinate approach which significantly simplifies the entire constraint algorithm;
this exposition of the constraint algorithm along with some new properties of the constraints are presented in Section \ref{constalg}.

The specific field theories developed in this paper were chosen to point out certain subtle features of the geometric constraint analysis and the investigation of symmetries. 
This paper presents the first premultisymplectic treatment of Carrollian field theories \cite{CarrollParticle,Gomis:2022spp,CarrollScalar,NL-Primer}.
It should be noted that the De Donder--Weyl formalism and its (pre)multisymplectic interpretation are not restricted to work only for Lorentz invariant theories; in fact, the Carrollian scalar field theories investigated in the premultisymplectic formalism in this paper are not Lorentz invariant. Carrollian field theories arise from taking the limit $c\rightarrow 0$ for the speed of light as first introduced in \cite{NL-1965, NL-1966}.
Furthermore, the multisymplectic formulation of string theory and $p$-branes on the multimomentum phase space found in \cite{Baez} is extended in this work by additionally providing the multisymplectic formulation on the multivelocity phase space and discussing the relevant symmetries on the multisymplectic phase spaces. The De Donder--Weyl formulation of string theory (and $p$-branes in general) can be found in \cite{ngbranes} without mention of the underlying multisymplectic structure.
Finally, this work reproduces the results found in the premultisymplectic treatment of Chern--Simons gravity done in \cite{Vey} where the primary constraint submanifold of the multimomentum phase space arising from the De Donder--Weyl Legendre map is presented. 
This work additionally provides the full premultisymplectic constraint analysis of Chern--Simons gravity, on both the multimomentum and multivelocity phase spaces, where it is found that the field equations define
the constraint submanifolds of the so-called {\sl Second Order Partial Differential Equation ({\sc sopde})} type;
these constraint submanifolds arise from imposing the {\sl {\sc sopde} condition} which guarantees that the final field equations are second order partial differential equations.
The treatment of the symmetries of Chern--Simons gravity in the premultisymplectic context is also given.
For additional references on the multisymplectic treatment of General Relativity, see  \cite{art:Capriotti2,first,GR1,GR2};
for an alternative geometric description of the De Donder--Weyl formalism 
applied to gravity see, for example, \cite{polysymplectic}
in which vector-valued forms called {\sl polysymplectic forms}
are used instead of the (pre)multisymplectic forms.
The existence of boundaries on the multisymplectic phase spaces will not be considered in this work. 
For a detailed treatment of boundaries in the multisymplectic and other formalisms
of field theories see, for instance, \cite{boundaries1,boundaries2,boundaries4}.

The organization of the paper is as follows: Section \ref{section:DiffGeo} provides a general overview of the (pre)multisymplectic setting used to formulate first-order classical field theories, including the local description of the constraint algorithm.
Additionally, symmetries and conservation laws in the (pre)multisymplectic framework are reviewed in this section.
Section \ref{section:ScalarFields} contains the premultisymplectic analysis of the canonical Klein Gordon Lagrangian and of the electric and magnetic Carrollian scalar field theories. All three scalar field theories exhibit different constraint and rigid symmetry structures. 
In section \ref{section:StringTheory}, the Nambu--Goto Lagrangian for bosonic strings and $p$-branes is found to be regular in the De Donder--Weyl sense and the symmetries of the Nambu--Goto action are presented in multisymplectic framework. 
Finally, the premultisymplectic treatment of gravity in $2+1$ dimensions as a Chern--Simons theory \cite{WittenGravity} is presented in Section \ref{section:2+1Gravity}. 

All the manifolds are real, second countable, and of class $\Cinfty$.
Manifolds and mappings are assumed to be smooth.
Sum over crossed repeated indices is understood.


\section{Preliminary Differential Geometry}
\label{section:DiffGeo}

This section reviews the foundations of the multisymplectic formulation of first-order field theories and basic concepts of symmetries and conservation laws; the constraint algorithm is also described in the premultisymplectic context. 
  
\subsection{Multivector Fields}
\label{ap:multivector}

Let  $E$ be an $n$-dimensional manifold and $m<n$; 
sections of $\Lambda^m (\Tan E):=\overbrace{\Tan E\wedge\ldots\wedge\Tan E}^{(m)}$ 
(the $m$th exterior product of $\Tan E$) are called {\sl\textbf{$m$-multivector fields}}
on $E$ and they are the contravariant skew-symmetric tensor fields of order $m$ on $E$.
(For more details on multivector fields and all the related topics presented here see, for instance, \cite{Ca96a,art:Echeverria_Munoz_Roman98,EMR-99b,IEMR-2012}). 
The set of $m$-multivector fields on $E$ is denoted $\mathfrak{X}^m(E)$.

For every $m$-multivector field $\textbf{X}\in \mathfrak{X}^m(E)$ and $p\in E$, there exists an open neighborhood $U_p \subset Y$ and $X_1,...,X_r \in \mathfrak{X}(E)$ such that
\begin{equation*}
\textbf{X}\vert_{U_p}= \sum_{1\leq i_1 \ldots i_m \leq r} f^{i_1...i_m}X_{i_1}\wedge\ldots \wedge X_{i_m},
\end{equation*}
where $f^{i_1...i_m} \in C^{\infty}(U_p)$ and $m \leq r \leq \text{dim}E$.
Then, $\textbf{X}\in \mathfrak{X}^m(E)$ is {\sl\textbf{locally decomposable}} if, for every $p\in E$, there is an open neighborhood $U_p \subset Y$ 
and $X_1,...X_m \in \mathfrak{X}(U_p)$ such that 
$\left.\textbf{E}\right\vert_{U_p}=X_1\wedge...\wedge X_m$. 
All multivector fields used in this paper
are assumed to be locally decomposable. 
The contraction between multivector fields 
$\textbf{X}\in \mathfrak{X}^m(E)$ and differentiable $k$-forms 
$\xi\in \df^k(E)$ is
\begin{equation*}
i(\textbf{X})\xi\vert_{U_p}= \sum_{1\leq i_1 \ldots i_m \leq r} f^{i_1...i_m}i(X_{i_1}\wedge ... \wedge X_{i_m})\xi = \sum_{1\leq i_1 \ldots i_m \leq r} f^{i_1...i_m}\,\xi(X_{i_1}),\ldots ,X_{i_m}),
\end{equation*}
if $k\geq m$, and $i(\textbf{X})\xi\vert_{U_p}=0$, if $k< m$. 
A {\sl\textbf{distribution}} $\mathpzc{D}$ of rank $k$ on $E$ is a subbundle of $\Tan E$ of rank $k$ (and an $m$-dimensional distribution on $E$ is an $m$-dimensional subbundle of $\Tan E$). 
For every $p \in E$ there is a linear subspace $\mathpzc{D}_p \subseteq\Tan_pY$ such that $\mathpzc{D} =  \bigcup_{p \in Y}  \mathpzc{D}_p$.
Given a distribution $\mathpzc{D}\subseteq\Tan E$, a nonempty immersed submanifold $W\subseteq E$ is a an {\sl\textbf{integral manifold of}} $\mathpzc{D}$ if $\Tan_pW=\mathpzc{D}_p$.
The distribution $\mathpzc{D}\subseteq\Tan E$ is an {\sl\textbf{integrable distribution}} if at every $p\in E$ there is an integral manifold of $\mathpzc{D}$ containing $p$.
A multivector field  $\textbf{X} \in \mathfrak{X}^m(E)$ and an $m$-dimensional distribution $\mathpzc{D}$ on $E$ are {\sl\textbf{locally associated}} if there exists an open set $U\subseteq E$ such that $\left.\textbf{X}\right\vert_U$ is a section of $\left.\Lambda^m\mathpzc{D}\right\vert_U$.

Locally decomposable $m$-multivector fields are locally associated to $m$-dimensional distributions. 
Multivector fields associated with the same distribution form an equivalence class $\{ \textbf{X} \}$ in $\mathfrak{X}^m(E)$. 
If $\textbf{X}_1, \ \textbf{X}_2 \in \mathfrak{X}^m(E)$ are two nonvanishing multivector fields, both locally associated with the same distribution on the open set $U \subseteq E$, then there exists 
$f \in C^{\infty}(U)$ such that $\textbf{X}_1=f\textbf{X}_2$ (on $U$),
and this is precisely the equivalence relation with equivalence class denoted $\{ \textbf{X}\}_U$.
A multivector field is {\sl\textbf{integrable}} if its locally associated distribution is integrable. 
An $m$-dimensional submanifold $W\hookrightarrow E$ is an {\sl\textbf{integral manifold}} of $\textbf{Z} \in \mathfrak{X}^m(E)$ if, and only if, for every point $p \in W$, the multivector $\textbf{X}_p$ spans $\Lambda^m\Tan_pW$.

Now, let $\pi\colon E\rightarrow M$ be a fiber bundle
with $\dim{M}=m$ and $\dim{E}=m+n$.
Vector fields on $E$ are called {\sl \textbf{$\pi$-vertical vector fields}} when they are tangent to the fibres of $\pi$. The set of such vector fields will be denoted as $\vf^{{\rm V}(\pi)}(E)$.
A multivector field $\textbf{Z} \in \mathfrak{X}^m(E)$  is $\pi$-{\sl\textbf{transverse}} if at every point $p \in Y$, $\left.i(\textbf{X})(\pi^*\beta)\right\vert_p \neq 0$
for every  $\beta \in \df^m(Y)$ with $\beta(\pi(p))\neq 0$.
Consider a local section of $\pi$, $\phi\colon U_x\subset M \rightarrow E: x \mapsto y=\phi(x)$ where $x \in M$ and $y \in E$;
 if $\phi(U_x)$ is the integral manifold of the multivector field $\textbf{X} \in \mathfrak{X}^m(E)$ at $y$, then $\phi$ is called an {\sl\textbf{integral section}} of \textbf{X}.
 If a multivector field $\textbf{X} \in \mathfrak{X}^m(E)$ is integrable, then it is $\pi$-transverse if, and only if, its integral manifolds are local sections of $\pi$; that is,
the local sections of $\pi$ are integral sections of  $\textbf{X}$. 
If $M$ is an orientable manifold,
the condition that a multivector field $\textbf{X} \in \mathfrak{X}^m(E)$ is $\pi-$transverse can be written as 
$i(\textbf{X})\text{d}^mx \neq 0$,
where $\d^mx\equiv\d x^1\wedge\ldots\wedge\d x^m$
is the local coordinate expression 
$(x^\mu)$ on $M$ $(\mu=0,...,m-1)$ of the volume form.
In particular, it is possible to take a representative $\textbf{X}$
in the class of $\pi$-transverse multivector fields 
 such that $f=1$ so that
\beq
 \inn(\textbf{X})\text{d}^mx=1 \ .
 \label{trans}
\eeq

If $x\in M$, two sections $\phi,\widetilde{\phi}$ of $\pi$ are \emph{($1st$-order) equivalent} at $x$ if $\phi(x)=\widetilde{\phi}(x)$ and $\partial_\mu\phi\vert_x=\partial_\mu\widetilde{\phi}\vert_x$, where $\displaystyle\partial_\mu\phi\equiv\derpar{\phi}{x^\mu}$; i.e. $T_x\phi=T_x\widetilde{\phi}$. The corresponding equivalence classes  are called the {\sl\textbf{$1$-jets}} of $\phi$ at $x$, denoted $j^1_x\phi$.
Then, the {\sl\textbf{first-order jet bundle}} $J^1E$ of $E$ is defined as
$J^1E=\{j^1_x\phi:x\in M, \phi\in\Gamma(\pi)\}$.
It is a bundle over $E$ and $M$
whose natural projections are denoted $\pi^1\colon J^1E\rightarrow E$ and $\bar\pi^1\colon J^1E\rightarrow M$, and $\text{dim}J^1E=m+n+mn$.
The set of sections of $\pi$ and $\bar\pi^1$ are denoted by $\Gamma(\pi)$ and $\Gamma(\bar\pi^1)$ respectively.
A section $\psi\in\Gamma(\bar{\pi}^1)$ is said to be a {\sl\textbf{holonomic section}} if $\psi=j^1\phi$; that is,
$\psi$ is is a holonomic section if it is the {\sl\textbf{first jet prolongation}} to $J^1E$ of a section $\phi\in\Gamma(\pi)$.
A multivector field $\textbf{X} \in \mathfrak{X}^m(J^1E)$ 
is said to be a {\sl\textbf{holonomic multivector field}} if it is integrable and its integral sections are holonomic sections of $\bar{\pi}^1$.
Observe that a multivector field $\textbf{X} \in \mathfrak{X}^m(J^1E)$  is a holonomic multivector field if, and only if,
(i) $\textbf{X}$ is integrable,
(ii) $\textbf{X}$ is $\bar{\pi}^1-$transverse, and
(iii) the integral sections of $\textbf{X}$ are holonomic.

Given coordinates $(x^\mu, y^A)$ $(A=1,...,n)$ on $E$ adapted to the bundle structure, the induced coordinates on $J^1E$ are $(x^\mu, y^A, y_\mu^A)$.
Then holonomic sections are written as
$\psi=j^1\phi(x)=\left(x^\mu(x),y^A(x),\partial_\mu y^A(x)\right)$.
Furthermore, the local expression for $\textbf{X}\in\mathfrak{X}^m(J^1E)$ is 
$$
\textbf{X}= \bigwedge^{m}_{\mu=1} \textbf{X}_\mu = \bigwedge^{m}_{\mu=1} f \left( \frac{\partial}{\partial x^\mu} + D_\mu^A\frac{\partial}{\partial y^A} + H_{\mu\nu}^A\frac{\partial}{\partial y_\nu^A}\right), \ \ \ \ f\in C^{\infty}(J^1E) \ ,
$$
which defines equivalence classes of multivector fields on $J^1E$. 

Integral sections of $\textbf{X}$ satisfy $\displaystyle\derpar{y^A}{x^\mu} = D_\mu^A$,
$\displaystyle\derpar{y_\nu^A}{x^\mu}= H^A_{\mu\nu}$. Furthermore, if $\textbf{X}$ is holonomic, then $D_\mu^A=y_\mu^A$ and hence
$\displaystyle\frac{\partial^2y^A}{\partial x^\mu \partial x^\nu}= H^A_{\mu\nu}$. Holonomic multivector fields are also called
{\sl\textbf{Second Order Partial Differential Equations}} ({\sc sopde}).

\subsection{Multisymplectic Lagrangian Field Theory}

Multisymplectic geometry can be viewed as the field theoretic extension of the symplectic geometrization of classical mechanics
(for more details on the multisymplectic Lagrangian formulation  of first-order field theories
see, for instance, \cite{GIMMSY,art:Roman09,book:Saunders89,art:Echeverria_Munoz_Roman98,art:Aldaya_Azcarraga78_2,EMR-96,Gc-73,GS-73}). 
The analysis begins by considering an orientable $m$-dimensional pseudo-Riemannian manifold $M$. 
The {\sl \textbf{configuration manifold}} is taken to be a fiber bundle $E$ over $M$ with $n$-dimensional fibers with (surjective) projection map $\pi:E\rightarrow M:(x^\mu, y^A)\mapsto x^\mu$. 
The fields under consideration are denoted $y^A(x)$ and are given by the local sections $\phi:M\rightarrow E:x^\mu\mapsto (x^\mu, y^A(x))$ of $\pi$. 
The {\sl \textbf{multivelocity phase space}} on which the Lagrangian formalism takes place is the \textsl{first-order jet bundle} $J^1E$ of $\pi$
which has natural coordinates $(x^\mu,y^A,y_\mu^A)$ and hence $\text{dim}J^1E=n+m+nm$; $J^1E$ is a fiber bundle over $E$ with projection map $\pi^1:J^1E\rightarrow E:(x^\mu,y^A,y_\mu^A)\mapsto (x^\mu,y^A)$ 
and also a fiber bundle over $M$ with projection $\bar{\pi}^1:J^1E\rightarrow M:(x^\mu,y^A,y^A_\mu)\mapsto x^\mu$. 

Densities on $J^1E$ can be obtained by lifting the volume form $\text{d}^mx$ on $M$ to $J^1E$.
The {\sl\textbf{Lagrangian density}} is then written as $\textbf{L}=\mathscr{L}(x^\mu,y^A,y_\mu^A)\text{d}^mx$ and $\mathscr{L}(x^\mu,y^A,y_\mu^A)\in C^\infty(J^1E)$ is referred to as the {\sl\textbf{Lagrangian function}}. The {\sl\textbf{Lagrangian energy}} $E_\mathscr{L}\in C^\infty(J^1E)$ is defined as 
\begin{equation*}
E_\mathscr{L}\equiv\frac{\partial\mathscr{L}}{\partial y^A_\mu}y^A_\mu-\mathscr{L}(x^\mu,y^A,y^A_\mu) \ .
\end{equation*}
The Lagrangian is said to be \textsl{regular} 
when the generalized Hessian matrix
\begin{equation}
\label{multiHessian}
H_{AB}^{\mu\nu}=\frac{\partial^2\mathscr{L}}{\partial y^A_\mu \partial y^B_\nu}
\end{equation}
is non-singular everywhere. 
Furthermore, 
the bundle $J^1E$ comes equipped with an $m$-form called the {\sl\textbf{$m$-Poincar\'e--Cartan form}} $\Theta_\mathscr{L}\in\df^m(J^1E)$, given by 
\begin{equation*}
\Theta_\mathscr{L}=\frac{\partial\mathscr{L}}{\partial y^A_\mu}\text{d}y^A\wedge\text{d}^{m-1}x_\mu-E_\mathscr{L}\text{d}^mx \ ,
\end{equation*}
and a closed (and hence locally exact) $(m+1)$-form $\Omega_\mathscr{L}\in \df^{m+1}(J^1E)$,
which is called the {\sl\textbf{$(m+1)$-Poincar\'e--Cartan form}}, given by
\begin{equation*}
\Omega_\mathscr{L}=-\text{d}\Theta_\mathscr{L}=-\text{d}\left(\frac{\partial\mathscr{L}}{\partial y^A_\mu}\right)\wedge\text{d}y^A\wedge\text{d}^{m-1}x_\mu+\text{d}E_\mathscr{L}\wedge\text{d}^mx \ ,
\end{equation*}
where
$\text{d}^{m-1}x_\mu\equiv i(\partial_\mu)\text{d}^mx=
\displaystyle\frac{1}{(m-1)!}\epsilon_{\mu\mu_2\dotsm\mu_m}\text{d}x^{\mu_2}\wedge\dotsc\wedge\text{d}x^{\mu_n}$,
$\text{d}^{m-2}x_{\mu\nu}\equiv i(\partial_{\nu}) i(\partial_\mu)\text{d}^mx=
\displaystyle\frac{1}{(m-2)!}\epsilon_{\mu\nu\mu_3\dotsm\mu_m}\text{d}x^{\mu_3}\wedge\dotsc\wedge\text{d}x^{\mu_m}$,
and so on.

The form $\Omega_\mathscr{L}$ is \textsl{$j$-nondegenerate} ($1\leq j \leq m)$ if, for every $p\in J^1E$ and $\textbf{Y}\in \mathfrak{X}^j(J^1E)$,
it follows that $\left.i(\textbf{Y})\Omega_\mathscr{L}\right\vert_p=0\ \Longleftrightarrow\ \left.\textbf{Y}\right\vert_p=0$.
When $\Omega_\mathscr{L}$ is $1$-nondegenerate, it is referred to as being a {\sl\textbf{multisymplectic form}}. This occurs when $\mathscr{L}(x^\mu,y^A,y_\mu^A)$ is regular. Otherwise, when $\mathscr{L}$ is singular, $\Omega_\mathscr{L}$ is $1$-degenerate and is referred to as being a {\sl\textbf{premultisymplectic form}}
(although here it will also be referred to as a \textsl{degenerate multisymplectic form} when this is the case). 

The couple $(J^1E,\ \Omega_{\mathscr{L}})$
is called a {\sl\textbf{Lagrangian system}}.
The field equations for this system are obtained from a 
variational action principle posed on $J^1E$.
The {\sl\textbf{action}}, denoted
$$
\displaystyle S\left[(j^1\phi)^*x^\mu,(j^1\phi)^* y^A, (j^1\phi)^*y^A_\mu\right]=\int_{\Sigma} (j^1\phi)^*\Theta_\mathscr{L}\ ,
$$
is a functional on the set of sections $\Gamma(\pi)$ given by
$\displaystyle 
\Gamma(\Sigma,E)\rightarrow \mathds{R}:\phi \mapsto \int_{M} (j^1\phi)^*\Theta_\mathscr{L}$.
The variational problem consists of finding {\sl critical}
({\sl stationary}) sections of $\Gamma(\pi)$ which satisfy 
\begin{equation*}
\left.\frac{\text{d}}{\text{d}s}\right\vert _{s=0}\int_{\Sigma} (j^1\phi_s)^*\Theta_{\mathscr{L}}=0 \ ,
\end{equation*}
for all variations $\phi_s=\eta_s\circ \phi$ of $\phi$. 
Here $\eta_s$ is the flow of a vertical vector field on $E$ compactly supported on $\Sigma$. 
For more details see \cite{art:Echeverria_Munoz_Roman98,Gc-73,GS-73}.
Critical sections which are solutions to this variational problem can be characterized in the following equivalent ways:
\ben
\item
$(j^1\phi)^*i(\mathpzc{X})\Omega_{\mathscr{L}}=0$,
for every $\mathpzc{X} \in \mathfrak{X}(J^1E)$.
\item
$j^1\phi$ is an integral section of a class of 
$\bar\pi^1$-transverse, locally decomposable, holonomic, multivector fields 
$\{\textbf{X}_{\mathscr{L}}\} \subset \mathfrak{X}^m(J^1E)$ which satisfy
\begin{equation}
\label{equation:LEOM}
\inn(\textbf{X}_{\mathscr{L}})\Omega_{\mathscr{L}}=0
 \quad , \quad 
 \mbox{\rm  for every $\textbf{X}_{\mathscr{L}} \in  \{\textbf{X}_{\mathscr{L}}\}$} \ .
\end{equation}
Furthermore, it is possible to choose a representative of the class $\{\textbf{X}_{\mathscr{L}}\}$
which satisfies the normalized transverse condition \eqref{trans}:
\begin{equation}
\label{equation:LEOM1}
\inn(\textbf{X}_{\mathscr{L}})\d^mx=1 \ .
\end{equation}
\item
Given a natural system of coordinates $(U;x^\mu, y^A,y_\mu^A)$ on $J^1E$, the first jet prolongations $j^1\phi=\left(x^\mu, y^A(x^\nu),\partial_\mu y^A(x^\nu)\right) \in U \subset J^1E$ satisfy the 
{\sl Euler--Lagrange equations}:
\begin{equation*}
\frac{\partial \mathscr{L}}{\partial y^A} \circ j^1\phi -\frac{\partial}{\partial x^\mu} \left( \frac{\partial \mathscr{L}}{\partial y_\mu^A} \circ j^1\phi \right) = 0 \ .
\end{equation*}
\een

When the Lagrangian is regular, the field equations are compatible (and have solutions) on all of $J^1E$.
In the singular case they are, in general, not compatible on all of $J^1E$ and it is therefore necessary to implement a {\sl constraint algorithm}
in order to find a submanifold of $J^1E$ on which the field equations are compatible where consistent solutions exist.
This constraint algorithm is explained in detail in Section \ref{constalg}.

\subsection{Multisymplectic Hamiltonian Field Theory}

For the De Donder--Weyl Hamiltonian formulation of first-order field theories 
(see, for instance, \cite{GIMMSY,art:Roman09,EMR-99b,
HK-04,CCI-91,
LMM-96,MS-98,EMR-00b} for details),
let ${\cal M}\pi\equiv\Lambda_2^m\Tan^*E$
be the bundle of $m$-forms on
$E$ vanishing by the action of two $\pi$-vertical vector fields.
The manifold ${\cal M}\pi$ is called the {\sl \textbf{extended multimomentum bundle}}
and has local coordinates $(x^\mu,\,y^A,p_A^\mu, p)$, hence $\text{dim}{\cal M}\pi=m+n+mn+1$.
Then, consider the quotient bundle $J^1E^* := {\cal M}\pi / \Lambda^m_1(\Tan^*{\rm E})$
(where $\Lambda_1^m(\Tan^*{\rm E})$ is the bundle of $\pi$-semibasic $m$-forms on ${\rm E}$),
which is called the {\sl\textbf{multimomentum bundle}} of $E$.
The bundle $J^1E^*$ has natural local coordinates  $(x^\mu,y^A,p^\mu_A)$ and natural projections 
\beann
\tau:J^1E^*\rightarrow E&:&(x^\mu,y^A,p^\mu_A)\mapsto (x^\mu,y^A) 
\ ,\\
\bar{\tau}:J^1E^*\rightarrow M&:&(x^\mu,y^A,p^\mu_A)\mapsto x^\mu 
\ ,\\
 \sigma:{\cal M}\pi\rightarrow J^1E^*&:&(x^\mu,y^A,p^\mu_A,p)\mapsto (x^\mu,y^A,p^\mu_A) \ . 
\eeann
As ${\cal M}\pi$ is a subbundle of $\Lambda^m\Tan^*E$
(the multicotangent bundle of $E$ of order $m$), ${\cal M}\pi$ is endowed with canonical forms
which are known as the {\sl \textbf{multimomentum Liouville $m$ and $(m+1)$-forms}}, whose local expressions are
$$
  \Theta = p_A^\alpha\d y^A\wedge\d^{m-1}x_\alpha+p\d^mx
  \quad , \quad
  \Omega = -\d p_A^\alpha\wedge\d y^A\wedge\d^{m-1}x_\alpha-\d p\wedge\d^mx \ .
$$
If $\Lag\in C^\infty(J^1E)$ is a Lagrangian function,
then the {\sl \textbf{extended Legendre map}}, $\widetilde{\mathscr{FL}}\colon J^1E\to {\cal M}\pi$,
and the {\sl \textbf{Legendre map}}, $\mathscr{FL} :=\sigma\circ\widetilde{\mathscr{FL}}\colon J^1E\to J^1E^*$, 
are locally given as
 $$
 \begin{array}{ccccccc}
 \widetilde{\mathscr{FL}}^*x^\nu = x^\nu &, &
  \widetilde{{\mathscr FL}}^*y^A = y^A &\quad , \quad&
 \widetilde{\mathscr{FL}}^*p_A^\nu =\displaystyle\frac{\partial\mathscr{L}}{\partial y^A_\nu}
 &\quad , \quad&
 \widetilde{\mathscr{FL}}^*p =\mathscr{L}-\displaystyle v^A_\nu\frac{\partial\mathscr{L}}{\partial y^A_\nu}
 \\
 \mathscr{FL}^*x^\nu = x^\nu &\quad , \quad&
 \mathscr{FL}^*y^A = y^A &\quad , \quad&
 \mathscr{FL}^*p_A^\nu =\displaystyle\frac{\partial\mathscr{L}}{\partial y^A_\nu}\quad . & &
 \end{array}
 $$
The Lagrangian $\Lag$ is {\sl \textbf{regular}}
if, and only if, $\mathscr{FL}$ is a local diffeomorphism
(this definition is equivalent to that given above).
As a particular case, $\Lag$ is {\sl \textbf{hyper-regular}}
if $\mathscr{FL}$ is a global diffeomorphism.
A singular Lagrangian $\Lag$ is {\sl \textbf{almost-regular}} if
$P_0:=\mathscr{FL} (J^1E)$ is a submanifold of $J^1E^*$
(where $\jmath_0\colon P_0\hookrightarrow J^1E^*$
denotes the natural embedding),
${\mathscr{FL}}$ is a submersion onto its image, and
for every $p\in J^1E$, the fibres
${\mathscr{FL}}^{-1}({\mathscr{FL}}(p))$
are connected submanifolds of $J^1E$.
 
If $\Lag$ is a hyper-regular Lagrangian, then the following diagram illustrates the full structure:
 \[
 \begin{array}{ccccc}
\begin{picture}(15,52)(0,0)
\put(0,0){\mbox{$J^1E$}}
\end{picture}
&
\begin{picture}(65,52)(0,0)
 \put(20,27){\mbox{$\widetilde{\mathscr{FL}}$}}
 \put(25,6){\mbox{${\mathscr{FL}}$}}
 \put(0,7){\vector(2,1){65}}
 \put(0,4){\vector(1,0){65}}
\end{picture}
&
\begin{picture}(15,52)(0,0)
 \put(0,0){\mbox{$J^1E^*$}}
 \put(0,41){\mbox{${\cal M}\pi$}}
 \put(5,38){\vector(0,-1){25}}
 \put(-5,22){\mbox{$\sigma$}}
 \put(10,13){\vector(0,1){25}}
 \put(15,22){\mbox{$h$}}
\end{picture}
\end{array}
 \]
The map constructed as
 $h:=\widetilde{\mathscr{FL}}\circ\mathscr{FL}^{-1}$ is called the {\sl Hamiltonian section} associated with the
hyper-regular Lagrangian $\Lag$.
The differential forms defined as
\begin{equation}
\label{defThetaOmegaH}
 \Theta_{\mathscr H}:=h^*\Theta\in\df^m(J^1E^*)   ,\qquad
 \Omega_{\mathscr H}:=-\d\Theta_{h}=h^*\Omega\in\df^{m+1}(J^1E^*) \ ,
\end{equation}
are called the {\sl\textbf{Hamilton--Cartan $m$ and $(m+1)$ forms}} of $J^1E^*$
associated with the Hamiltonian section $h$.
If the form $\Omega_{\mathscr H}$ is $1$-nondegenerate (and hence multisymplectic), then the Hamiltonian section
$h(x^\nu,y^A,p^\nu_A)=(x^\nu,y^A,p^\nu_A,p=-{\mathscr H}(x^\gamma,y^B,p_B^\gamma))$
 is locally specified by the {\sl Hamiltonian function}
$\mathscr{H}=p^\nu_A (\mathscr{FL}^{-1})^*y_\nu^A-(\mathscr{FL}^{-1})^*\Lag$
known as the {\sl\textbf{De Donder--Weyl Hamiltonian function}} and it follows that $\mathscr{FL}^*\mathscr{H}=E_\Lag$.
The local expressions for the Hamilton--Cartan forms defined in (\ref{defThetaOmegaH}) are
\begin{equation*}
 \Theta_\mathscr{H} = p_A^\nu\d y^A\wedge\d^{m-1}x_\nu -\mathscr{ H}\d^mx
\quad  , \quad
 \Omega_\mathscr{H} = -\d p_A^\nu\wedge\d y^A\wedge\d^{m-1}x_\nu +
 \d \mathscr{H}\wedge\d^mx \ ,
\end{equation*}
so ${\mathscr F}\Lag^*\Theta_{\mathscr H}=\Theta_{\Lag}$
and ${\mathscr F}\Lag^*\Omega_\mathscr{H}=\Omega_{\Lag}$.
For regular, but not hyper-regular, Lagrangians the construction is the same, but on the open sets 
${\mathscr F}\Lag(J^1E)\subset J^1E^*$.

The couple $(J^1E^*,\Omega_{\mathscr{H}})$
is called a {\sl \textbf{regular Hamiltonian system}} when $\Omega_\mathscr{H}$ is multisymplectic 
and its corresponding field equations are obtained from a variational action principle on $J^1E^*$. 
The {\sl\textbf{action}}, denoted
$$\displaystyle 
S\left[ \psi^*x^\mu, \psi^* y^A,  \psi^*p^\mu_A\right]=\int_{M} \psi^*\Theta_{\mathscr{H}}\ ,$$
is a functional on the set of sections $\Gamma(\bar\tau)$ given by
$\displaystyle 
\Gamma(M,J^1E^*)\rightarrow \mathds{R}:\psi \mapsto \int_{M}\psi^*\Theta_\mathscr{H}$.
As in the Lagrangian formalism, the objective is to find sections that are critical (stationary):
\begin{equation*}
\left.\frac{\text{d}}{\text{d}s}\right\vert_{s=0}\int_{M} \psi^*_s\Theta_{\mathscr{H}}=0 \ ,
\end{equation*}
for variations $\psi_s=\eta_s\circ \psi$ of $\psi$.
Critical sections are characterized in the following equivalent ways:
\ben
\item
$\psi^*i(\mathpzc{X})\Omega_{\mathscr{H}}=0$,
for every $\mathpzc{X} \in \mathfrak{X}(J^1E^*)$.
\item
$\psi$ is an integral section of a class of integrable,
locally decomposable, $\bar{\tau}$-transverse multivector fields 
$\{\textbf{X}_{\mathscr{H}}\} \subset \mathfrak{X}^n(J^1E^*)$ which satisfy 
\begin{equation}
\label{equationHEOM}
i(\textbf{X}_{\mathscr{H}})\Omega_{\mathscr{H}}=0
\quad , \quad \mbox{\rm  for every $\textbf{X}_{\mathscr{H}} \in  \{\textbf{X}_{\mathscr{H}}\}$} \ .
\end{equation}
Furthermore, we can take as a representative of the class $\{\textbf{X}_{\mathscr{L}}\}$
a normalized $\bar\tau^1$-transverse multivector field:
\begin{equation}
\label{equationHEOM1}
i(\textbf{X}_{\mathscr{H}})\d^mx=1 \ .
\end{equation}
\item
If $(U;x^\mu, y^A,p_A^\mu)$ is a natural system of coordinates in $J^1E^*$, then $\psi \in U \subset J^1E^*$ satisfies the {\sl \textbf{Hamilton--De Donder--Weyl equations}}:
$$
\frac{\partial (y^A\circ \psi)}{\partial x^\mu} = \frac{\partial \mathscr{H}}{\partial p_A^\mu} \circ \psi
\quad , \quad \frac{\partial (p^A_\mu \circ \psi)}{\partial x^\mu}=-\frac{\partial \mathscr{H}}{\partial y^A} \circ \psi \ .
$$
\een

For the almost-regular case,
$P_0$ is a fibre bundle over $E$ and $M$ with natural
projections $\tau_0\colon P_0\to E$ and
$\bar\tau_0:=\pi\circ\tau_0\colon P_0\to M$.
The $\sigma$-transverse submanifold
$\tilde P_0 = \widetilde{\mathscr F \Lag}(J^1E) \hookrightarrow{\cal M}\pi$ with natural embedding $\tilde\jmath_0\colon\tilde P_0\hookrightarrow{\cal M}\pi$
is diffeomorphic to $P_0$ according to the restriction of the projection $\sigma$ to $\tilde P_0$ given as
$\tilde\sigma\colon\tilde P_0\to P_0$.
Taking $\tilde h:=\tilde\sigma^{-1}$, the Hamilton--Cartan $(m+1)$-form is defined as
$\Omega^0_{\mathscr H}=(\tilde\jmath_0\circ\tilde h)^*\Omega\in\df^{m+1}(P)$,
which verifies that
${\mathscr F}\Lag_0^*\Omega^0_{\mathscr H}=\Omega_{\Lag}$
where ${\mathscr F}\Lag_0$ is the restriction map of
${\mathscr F}\Lag$ onto $P_0$ given as
${\mathscr F}\Lag=\jmath_0\circ{\mathscr F}\Lag_0$.
Furthermore, there exists a De Donder--Weyl Hamiltonian
function $\mathscr{H}_0\in\Cinfty(P_0)$
such that
$E_\mathscr{L}=\mathscr{FL}_0^*\mathscr{H}_0$.
This case is illustrated in the following diagram:
 \[
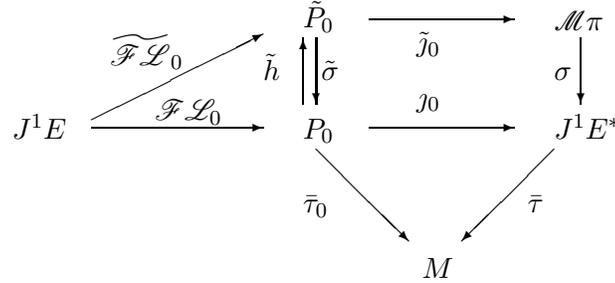
 \begin{array}{cccc}
\begin{picture}(20,52)(0,0)
\put(0,0){\mbox{$J^1E$}}
\end{picture}
&
\begin{picture}(65,52)(0,0)
 \put(7,28){\mbox{$\widetilde{{\mathscr F}\Lag}_0$}}
 \put(24,7){\mbox{${\mathscr F}\Lag_0$}}
 \put(0,7){\vector(2,1){65}}
 \put(0,4){\vector(1,0){65}}
\end{picture}
&
\begin{picture}(90,52)(0,0)
 \put(5,0){\mbox{$P_0$}}
 \put(5,42){\mbox{$\tilde P_0$}}
 \put(5,13){\vector(0,1){25}}
 \put(10,38){\vector(0,-1){25}}
 \put(-10,22){\mbox{$\tilde h$}}
 \put(12,22){\mbox{$\tilde\sigma$}}
 \put(30,45){\vector(1,0){55}}
 \put(30,4){\vector(1,0){55}}
 \put(48,12){\mbox{$\jmath_0$}}
 \put(48,33){\mbox{$\tilde\jmath_0$}}
 \end{picture}
&
\begin{picture}(15,52)(0,0)
 \put(0,0){\mbox{$J^1E^*$}}
 \put(0,41){\mbox{${\cal M}\pi$}}
 \put(10,38){\vector(0,-1){25}}
 \put(0,22){\mbox{$\sigma$}}
\end{picture}
\\
& &
\begin{picture}(90,35)(0,0)
 \put(10,35){\vector(1,-1){35}}
 \put(5,11){\mbox{$\bar\tau_0$}}
 \put(90,11){\mbox{$\bar\tau$}}
 \put(100,35){\vector(-1,-1){35}}
\end{picture}
 &
\\
& & \qquad M &
 \end{array}
 \]
In general, the couple $(P_0,\Omega_{\mathscr{H}}^0)$ (where $\Omega_{\mathscr{H}}^0$ is a pre-multisymplectic form)
is a {\sl \textbf{non-regular Hamiltonian system}}.
Now, the variational principle and the field equations are stated as in the regular case, but for sections
$\Gamma(\bar\tau_0)$ and the analog to the field equations \eqref{equationHEOM} read
\beq
i(\textbf{X}^0_\mathscr{H})\Omega^0_\mathscr{H}=0
\quad , \quad \mbox{\rm  for every $\textbf{X}^0_{\mathscr{H}} \in  \{\textbf{X}^0_{\mathscr{H}}\}
\subset\vf^m(P_0)$} \ .
\label{dimgravmvfh}
\eeq
As in the Lagrangian singular case, when $(P_0,\Omega_{\mathscr{H}}^0)$ is a non-regular Hamiltonian system,
in the best situations, the Hamiltonian field equations have consistent solutions only on a submanifold of $P_0$
which is obtained by using the constraint algorithm described in the next section \ref{constalg}.

The general geometric structure on which the field theories is developed is depicted in the diagram below:

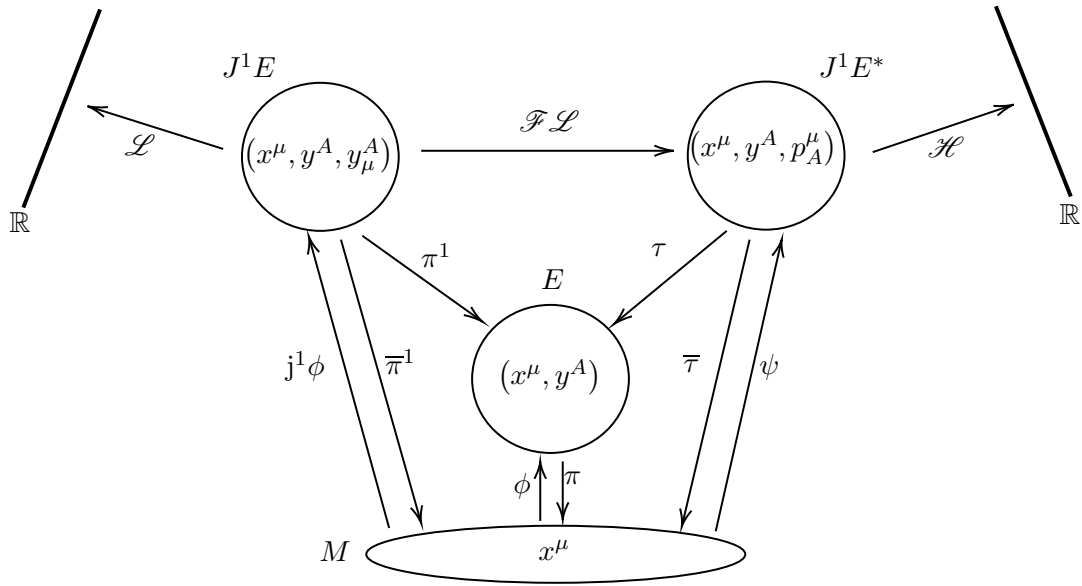
\tikzset{every picture/.style={line width=0.75pt}} 
\begin{tikzpicture}[x=0.75pt,y=0.75pt,yscale=-.8,xscale=.9]
\draw    (176,838) -- (101.88,811.67) ;
\draw [shift={(100,811)}, rotate = 379.56] [color={rgb, 255:red, 0; green, 0; blue, 0 }  ][line width=0.75]    (10.93,-3.29) .. controls (6.95,-1.4) and (3.31,-0.3) .. (0,0) .. controls (3.31,0.3) and (6.95,1.4) .. (10.93,3.29)   ;
\draw    (540,840) -- (617.13,810.71) ;
\draw [shift={(619,810)}, rotate = 519.21] [color={rgb, 255:red, 0; green, 0; blue, 0 }  ][line width=0.75]    (10.93,-3.29) .. controls (6.95,-1.4) and (3.31,-0.3) .. (0,0) .. controls (3.31,0.3) and (6.95,1.4) .. (10.93,3.29)   ;
\draw    (269,1077.15) -- (224.48,895.17) ;
\draw [shift={(224,893.23)}, rotate = 436.25] [color={rgb, 255:red, 0; green, 0; blue, 0 }  ][line width=0.75]    (10.93,-3.29) .. controls (6.95,-1.4) and (3.31,-0.3) .. (0,0) .. controls (3.31,0.3) and (6.95,1.4) .. (10.93,3.29)   ;
\draw   (186.52,842.98) .. controls (186.52,817.18) and (206.18,796.27) .. (230.44,796.27) .. controls (254.7,796.27) and (274.36,817.18) .. (274.36,842.98) .. controls (274.36,868.77) and (254.7,889.68) .. (230.44,889.68) .. controls (206.18,889.68) and (186.52,868.77) .. (186.52,842.98) -- cycle ;
\draw   (436.34,842.21) .. controls (436.34,816.42) and (456,795.51) .. (480.26,795.51) .. controls (504.51,795.51) and (524.18,816.42) .. (524.18,842.21) .. controls (524.18,868.01) and (504.51,888.92) .. (480.26,888.92) .. controls (456,888.92) and (436.34,868.01) .. (436.34,842.21) -- cycle ;
\draw   (256.2,1093.73) .. controls (256.2,1083.8) and (303.74,1075.74) .. (362.39,1075.74) .. controls (421.04,1075.74) and (468.58,1083.8) .. (468.58,1093.73) .. controls (468.58,1103.67) and (421.04,1111.73) .. (362.39,1111.73) .. controls (303.74,1111.73) and (256.2,1103.67) .. (256.2,1093.73) -- cycle ;
\draw   (315.5,983.1) .. controls (315.5,957.3) and (335.17,936.39) .. (359.43,936.39) .. controls (383.68,936.39) and (403.35,957.3) .. (403.35,983.1) .. controls (403.35,1008.89) and (383.68,1029.8) .. (359.43,1029.8) .. controls (335.17,1029.8) and (315.5,1008.89) .. (315.5,983.1) -- cycle ;
\draw    (253.98,892.75) -- (319.88,945.85) ;
\draw [shift={(321.43,947.11)}, rotate = 218.86] [color={rgb, 255:red, 0; green, 0; blue, 0 }  ][line width=0.75]    (10.93,-3.29) .. controls (6.95,-1.4) and (3.31,-0.3) .. (0,0) .. controls (3.31,0.3) and (6.95,1.4) .. (10.93,3.29)   ;
\draw    (458.2,889.68) -- (396.29,947.28) ;
\draw [shift={(394.82,948.64)}, rotate = 317.07] [color={rgb, 255:red, 0; green, 0; blue, 0 }  ][line width=0.75]    (10.93,-3.29) .. controls (6.95,-1.4) and (3.31,-0.3) .. (0,0) .. controls (3.31,0.3) and (6.95,1.4) .. (10.93,3.29)   ;
\draw    (366.28,1035.16) -- (366.28,1069.91) ;
\draw [shift={(366.28,1071.91)}, rotate = 270] [color={rgb, 255:red, 0; green, 0; blue, 0 }  ][line width=0.75]    (10.93,-3.29) .. controls (6.95,-1.4) and (3.31,-0.3) .. (0,0) .. controls (3.31,0.3) and (6.95,1.4) .. (10.93,3.29)   ;
\draw    (353.68,1072.68) -- (353.68,1037.16) ;
\draw [shift={(353.68,1035.16)}, rotate = 450] [color={rgb, 255:red, 0; green, 0; blue, 0 }  ][line width=0.75]    (10.93,-3.29) .. controls (6.95,-1.4) and (3.31,-0.3) .. (0,0) .. controls (3.31,0.3) and (6.95,1.4) .. (10.93,3.29)   ;
\draw    (286.96,839.15) -- (426.55,839.15) ;
\draw [shift={(428.55,839.15)}, rotate = 180] [color={rgb, 255:red, 0; green, 0; blue, 0 }  ][line width=0.75]    (10.93,-3.29) .. controls (6.95,-1.4) and (3.31,-0.3) .. (0,0) .. controls (3.31,0.3) and (6.95,1.4) .. (10.93,3.29)   ;
\draw    (452,1079.23) -- (488.61,897.19) ;
\draw [shift={(489,895.23)}, rotate = 461.37] [color={rgb, 255:red, 0; green, 0; blue, 0 }  ][line width=0.75]    (10.93,-3.29) .. controls (6.95,-1.4) and (3.31,-0.3) .. (0,0) .. controls (3.31,0.3) and (6.95,1.4) .. (10.93,3.29)   ;
\draw    (242,895.23) -- (286.51,1072.29) ;
\draw [shift={(287,1074.23)}, rotate = 255.89] [color={rgb, 255:red, 0; green, 0; blue, 0 }  ][line width=0.75]    (10.93,-3.29) .. controls (6.95,-1.4) and (3.31,-0.3) .. (0,0) .. controls (3.31,0.3) and (6.95,1.4) .. (10.93,3.29)   ;
\draw    (471,895.23) -- (433.41,1074.27) ;
\draw [shift={(433,1076.23)}, rotate = 281.86] [color={rgb, 255:red, 0; green, 0; blue, 0 }  ][line width=0.75]    (10.93,-3.29) .. controls (6.95,-1.4) and (3.31,-0.3) .. (0,0) .. controls (3.31,0.3) and (6.95,1.4) .. (10.93,3.29)   ;
\draw [line width=1.5]    (609,748) -- (651,868) ;
\draw [line width=1.5]    (107,750) -- (64,875) ;
\draw (54.55,875.96) node [anchor=north west][inner sep=0.75pt]    {$\mathbb{R} \ $};
\draw (119.24,827.34) node [anchor=north west][inner sep=0.75pt]    {$\mathscr{L}$};
\draw (569.24,828.34) node [anchor=north west][inner sep=0.75pt]    {$\mathscr{H}$};
\draw (173.55,775.69) node [anchor=north west][inner sep=0.75pt]    {$J^{1}E$};
\draw (209.54,963.56) node [anchor=north west][inner sep=0.75pt]    {$\mathrm{j^{1} \phi }$};
\draw (508.06,775.75) node [anchor=north west][inner sep=0.75pt]    {$J^{1} E^{*}$};
\draw (228.34,1083.88) node [anchor=north west][inner sep=0.75pt]    {$M$};
\draw (337.03,1040.24) node [anchor=north west][inner sep=0.75pt]    {$\mathrm{\phi }$};
\draw (365.2,1039.47) node [anchor=north west][inner sep=0.75pt]    {$\pi $};
\draw (265.36,962.65) node [anchor=north west][inner sep=0.75pt]    {$\overline{\pi }^{1}$};
\draw (414.25,895.53) node [anchor=north west][inner sep=0.75pt]    {$\tau $};
\draw (474.95,965.44) node [anchor=north west][inner sep=0.75pt]    {$\psi $};
\draw (352.46,911.9) node [anchor=north west][inner sep=0.75pt]    {$E$};
\draw (340.24,812.07) node [anchor=north west][inner sep=0.75pt]    {$\mathscr{FL}$};
\draw (185.61,829.27) node [anchor=north west][inner sep=0.75pt]    {$\left( x^{\mu } ,y^{A} ,y^{A}_{\mu }\right)$};
\draw (434.55,824.56) node [anchor=north west][inner sep=0.75pt]    {$\left( x^{\mu } ,y^{A} ,p^{\mu }_{A}\right)$};
\draw (327.98,968.98) node [anchor=north west][inner sep=0.75pt]    {$\left( x^{\mu } ,y^{A}\right)$};
\draw (351.08,1083) node [anchor=north west][inner sep=0.75pt]    {$x^{\mu }$};
\draw (285.36,895.65) node [anchor=north west][inner sep=0.75pt]    {$\pi ^{1}$};
\draw (432.36,965.65) node [anchor=north west][inner sep=0.75pt]    {$\overline{\tau }$};
\draw (643,871) node [anchor=north west][inner sep=0.75pt]    {$\mathbb{R} \ $};
\end{tikzpicture}

\subsection{Geometric Constraint Algorithm}
\label{constalg}

As stated in the previous sections, it is compulsory to implement a constraint algorithm in the Lagrangian and the Hamiltonian formalisms of singular (almost-regular) field theories in order to find the submanifolds of the multivelocity and multimomentum phase spaces
on which the field equations are compatible and where consistent solutions exist.
The geometric algorithm for uncovering the intrinsic constraint structure 
in premultisymplectic systems is developed in detail in \cite{LMMMR-2005}
and is a generalization of the algorithms stated for non-autonomous singular mechanics \cite{presymp1,LMMMR-2002}.
The constraint algorithm presented in \cite{LMMMR-2005} uses the characterization of the variational principle and the resulting field equations via multivector fields as discussed in the previous sections
(i.e. equations \eqref{equation:LEOM} and \eqref{equation:LEOM1} in the Lagrangian case,
and \eqref{equationHEOM} and \eqref{equationHEOM1} in the Hamiltonian case),
as the multivector field characterization is the most suitable for its implementation.

This section presents a procedure for obtaining the constraints which define the submanifolds arising at each step of the constraint algorithm in a more practical manner than what is given in previous literature.
It should be noted that the procedure described here is done in local charts as opposed to the intrinsic global description given in \cite{LMMMR-2005}.
The coordinate description of the constraint algorithm will be carried out explicitly in each example field theory studied in this paper.

The general geometrical setting for the Lagrangian and Hamiltonian formalisms
for singular field theories consists of taking
a fiber bundle $\kappa\colon F\to M$, where
$\dim\,M=m>1$ and $\dim\,F=N+m$ ($N>m$), and $M$ is an
orientable manifold with a volume form whose pull-back to $F$
is denoted  $\omega\in\df^m(F)$.
Let $\Omega\in\df^{m+1}(F)$ be a premultisymplectic form on $F$
satisfying the so-called {\sl variational condition} given as
\beq
\inn(Y_1)\inn(Y_2)\inn(Y_3)\Omega=0 \ , \
\mbox{\rm for every $Y_1,Y_2,Y_3\in\vf^{{\rm V}(\kappa)}(F)$} \ .
\label{triplecont}
\eeq

Now, consider the problem of 
finding a submanifold $\jmath_C\colon C\hookrightarrow F$
and a locally decomposable $m$-vector field
${\bf X}\in\vf^m(F)$ along $C$ such that
\beq
\inn({\bf X})\Omega\vert_C = 0 \quad , \quad 
 \inn({\bf X})\omega\vert_C\not=0 \ ,
\label{genfieldeq}
\eeq
where the second equation is the condition that ${\bf X}$ is $\kappa$-transverse, and can be written simply as $\inn({\bf X})\omega\vert_C=1$.

The algorithmic procedure consists of the following steps:
\bit
\item
{\sl Compatibility conditions\/}:
First, look for the conditions called {\sl compatibility constraints} which define the set $C_1\subset F$
where the field equations \eqref{genfieldeq} have solutions.
This set is assumed to be a submanifold $C_1\hookrightarrow F$.
The compatibility constraints can be obtained from direct inspection of the local expression of the field equations \eqref{genfieldeq}.
\item
{\sl Tangency conditions\/}:
Now, given locally decomposable multivector fields 
$\textbf{X}$ which are
solutions to the field equations \eqref{genfieldeq} on $C_1$, ensure the stability of these solutions by imposing the {\sl stability} or {\sl tangency condition} of these multivector fields on $C_1$. This is done by requiring that the components of $\textbf{X}$ are tangent to $C_1$;
that is, if $\textbf{X}=\bigwedge_\mu X_\mu$, for every compatibility constraints defining $C_1$, $\zeta_j=\inn(Z_j)\alpha\in\Cinfty(F)$,
the stability condition is given by 
$$
\mathpzc{L}_{X_\mu}\zeta\vert_{C_1}=0 \ .
$$
This procedure may produce a new constraint submanifold
$C_2\hookrightarrow C_1\hookrightarrow F$; if this is the case, then it is necessary to impose the tangency of $\textbf{X}$ to $C_2$ as well. This procedure is reiterated until, in the most favorable cases (which include most physical field theories), no new constraints are produced and hence, a {\sl final constraint submanifold} $C\hookrightarrow\ldots C_1\hookrightarrow F$
is found such that the solutions $\textbf{X}$ to the field equations are tangent to $C$. 
However, note that this procedure does not in general terminate for all field theories, but these 
cases will not be studied in this work. 
\item 
{\bf Remark}:
Recall that the variational principle in the Lagrangian formalism requires that multivector field solutions to the field equations on $J^1E$ are holonomic multivector fields. However, multivector fields which solve the field equations are not, in general, holonomic and it is therefore sometimes necessary to impose this condition after finding the compatibility constraint submanifold $C_1\hookrightarrow J^1E$.
Imposing the {\sl holonomy ({\sc sopde}) condition} may give rise to additional constraints which define a new constraint submanifold $S_1\hookrightarrow C_1\hookrightarrow J^1E$; such constraints will be referred to here as {\sl {\sc sopde} constraints\/}.
Then, as described above, the tangency condition for the {\sc sopde} multivector fields $\textbf{X}_\Lag\in\vf^m(J^1E)$ is imposed
on all the constraints defining $S_1$ and subsequently also on any other resulting constraints until (in well behaved field theories) a final constraint submanifold is found. 

It is important to note that the {\sc sopde} condition does not apply in the Hamiltonian formalism and the algorithm is simply applied as described in the previous points. 
\eit

Integrability conditions for ${\bf X}$ are typically imposed after the constraint analysis,
however, this analysis will be unnecessary in the field theories studied in this work since they are integrable.   

Now that the geometric constraint algorithm has been described in detail, it is possible to establish various properties of constraints by means of a local analysis. 
The first of such properties presented here concerns compatibility constraints.
As a consequence of the variational condition \eqref{triplecont}, the premultisymplectic form can be expressed as the following splitting:
\beq
\label{Omegasplit}
\Omega=\widetilde\Omega+\alpha\wedge\omega \ ,
\eeq
where $\widetilde\Omega\in\df^{m+1}(F)$ and $\alpha\in\df^1(F)$ are closed forms.
In order to perform such a splitting globally,
it is necessary to use an {\sl Ehresmann connection} on the bundle $F\to M$
(although all the results are independent of this choice)  \cite{LMMMR-2005,LMMMR-2002,EMR-96};
however, since the field theories in this work are studied only locally, the connection on each local chart $U\subset F$ induced by the natural connection on $\Real^N\times\Real^m\to\Real^m$ will always be used.
It thereby follows that, in the Lagrangian and the Hamiltonian formulations 
of classical field theories, the (pre)multisymplectic forms are split in the following manner:
\beann
\Omega_\mathscr{L}=\widetilde{\Omega}_\mathscr{L}+\alpha_\mathscr{L}\wedge\omega\in \df^{m+1}(J^1E)
&\Rightarrow& \widetilde{\Omega}_\mathscr{L}=
\displaystyle -\text{d}\left(\frac{\partial\mathscr{L}}{\partial y^A_\mu}\right)\wedge\text{d}y^A\wedge\text{d}^{m-1}x_\mu \ , \\
& & \alpha_\Lag\wedge\omega=\d E_\mathscr{L}\wedge\d^mx \ , \\
\Omega_\mathscr{H}=\widetilde{\Omega}_\mathscr{H}+\alpha_\mathscr{H}\wedge\omega\in \df^{m+1}(P_0)
&\Rightarrow& \widetilde{\Omega}_\mathscr{H}=
-\text{d}p_A^\mu\wedge\text{d}y^A\wedge\text{d}^{m-1}x_\mu \ ,\\
& & \alpha_\mathscr{H}\wedge\omega=\d \mathscr{H}\wedge\d^mx 
\ ,
\eeann
where $F=J^1E$ is the multivelocity phase space for the Lagrangian formulation of any generic field theory and $F=P_0\subset J^1E^*$ is the multimomentum phase space for the De Donder--Weyl Hamiltonian formulation of singular field theories.
Then, as a consequence of splitting the (pre)multisymplectic forms as shown above, the following lemma and subsequent propositions hold:

\begin{lemma}
If ${\bf X}\in\vf^m(U)$ is a locally decomposable and $\kappa$-transverse multivector field on $U$,
then
\beq
\label{lema1}
\inn({\bf X})\Omega=\inn({\bf X})\widetilde\Omega+
\widetilde\alpha_{\bf X}+(-1)^mf\,\alpha  \ ,
\eeq
where $f=\inn({\bf X})\omega\in\Cinfty(U)$ is a nonvanishing function and
$\widetilde\alpha_{\bf X}\in\df^1(U)$
is a $\kappa$-semibasic $1$-form
(that is, $\inn(Y)\widetilde\alpha_{\bf X}=0$,
for every $Y\in\vf^{{\rm V}(\kappa)}(F)$).
\end{lemma}
\proof
If ${\bf X}=X_0\wedge\ldots\wedge X_{m-1}$ in $U$,
it follows from \eqref{Omegasplit} that
\beann
\inn({\bf X})\Omega&=&
\inn({\bf X})\widetilde\Omega+\inn({\bf X})(\alpha\wedge\omega)
\\ &=&
\inn({\bf X})\widetilde\Omega+
\sum_{\mu=0}^{m-1}(-1)^{m-1}\inn(X_\mu)\alpha\,\inn(X_0\wedge\ldots \wedge X_{\mu-1}\wedge X_{\mu+1}\wedge\ldots\wedge X_{m-1})\omega+
(-1)^m\alpha\,\inn({\bf X})\omega
\\ &=&
\inn({\bf X})\widetilde\Omega+
\sum_{\mu=0}^{m-1}(-1)^{m-1}g_\mu\,\inn(X_\mu)\alpha\,\d x^\mu+
(-1)^m\alpha\,\inn({\bf X})\omega \, ;
\eeann
then, denoting \, $f\equiv\inn({\bf X})\omega$ \, and
$$
\widetilde\alpha_{\bf X}=
\sum_{\mu=0}^{m-1}(-1)^{m-1}\inn(X_\mu)\alpha\,\inn(X_0\wedge\ldots \wedge X_{\mu-1}\wedge X_{\mu+1}\wedge\ldots\wedge X_{m-1})\omega 
\equiv \sum_{\mu=0}^{m-1}(-1)^{m-1}\inn(X_\mu)\alpha\,g_\mu\,\d x^\mu\ ,
$$
where $f,g_\mu\in\Cinfty(U)$ are non-vanishing functions
and the result follows.
\qed

In particular, taking $\inn({\bf X})\omega=1$, it follows that $f=1=g_\mu$ and
$\inn({\bf X})\Omega=\inn({\bf X})\widetilde\Omega+
\widetilde\alpha_{\bf X}+(-1)^m\alpha$,
with
$\widetilde\alpha_{\bf X}=(-1)^{m-1}\inn(X_\mu)\alpha\,\d x^\mu$.

\begin{prop}
\label{maintheor}
Let $(F,\Omega)$ be a premultisymplectic bundle $F\to M$ where $\Omega$ satisfies the variational condition \eqref{triplecont} and is written as the splitting \eqref{Omegasplit}.
If $C_1\hookrightarrow F$ is the maximal submanifold on which solutions to the field equations \eqref{genfieldeq} exist
(the {\sl compatibility constraint submanifold}),
then the following condition holds
at ${\rm p}\in U\subset C_1$:
\beq
\label{ncond}
[\inn(Z)\alpha]({\rm p})=0,\ \mbox{\rm for every $Z\in\text{ker}\,\widetilde\Omega\cap\text{ker}\,\omega$} \ .
\eeq
\end{prop}
\proof
As $C_1$ is the maximal submanifold where the field equations are compatible,
there exists a locally decomposable multivector ${\bf X}\in\vf^m(U)$
such that $\inn({\bf X})\Omega=0$ and $\inn({\bf X})\omega\not=0$. 
Then \eqref{lema1} holds and, 
as $\ker\omega=\vf^{{\rm V}(\kappa)}(F)$,
if $Z\in\ker\,\widetilde\Omega\cap\ker\,\omega$,
then $\inn(Z)\widetilde\alpha_{\bf X}=0$,
because $\widetilde\alpha_{\bf X}$ is $\kappa$-semibasic.
Therefore, for every $Z\in\text{ker}\,\widetilde\Omega\cap\text{ker}\,\omega$ and 
for ${\rm p}\in U$ it follows that
$$
0=\inn(Z_{\rm p})\inn({\bf X}_{\rm p})\Omega_{\rm p}=
\inn(Z_{\rm p})\left[\inn({\bf X}_{\rm p})\widetilde\Omega_{\rm p}+\widetilde\alpha_{{\bf X}{\rm p}}+(-1)^m f\,\alpha_{\rm p}\right]=(-1)^m f\,\inn(Z_{\rm p})\alpha_{\rm p}
\ \Longleftrightarrow\ \inn(Z_{\rm p})\alpha_{\rm p}=0\ .
$$
\qed

Thus, \eqref{ncond} is a necessary condition of compatibility of the field equations \eqref{genfieldeq}.
In the particular case where $m=1$
(that is, the non-autonomous mechanics)
this is also a sufficient condition and then
\eqref{ncond} is a geometrical characterization
of all the compatibility constraints.
Nevertheless, it remains under investigation to prove
the converse of Proposition \ref{maintheor} in the general case where $m>1$: the compatibility constraint submanifold $C_1$ is completely determined by \eqref{ncond}.

As a final remark, in the Lagrangian and Hamiltonian formalisms of field theories,
it is interesting to show the relation between the vector fields belonging to $\ker{\mathscr F}\Lag_*$
and the {\sl primary Hamiltonian constraints} arising from the Legendre map.
This result is analogous to that verified for singular dynamical systems
\cite{BGPR-86,BGPR-87}.

Let $(J^1E,\ \Omega_{\mathscr{L}})$ be an almost-regular Lagrangian system.
First observe that in natural coordinates of $J^1E$
the map ${\mathscr F}\Lag_*$ is represented by the matrix
$$
\Tan{{\mathscr F}\Lag}\equiv
\left(\begin{matrix}
({\rm Id})_{m\times m} & (0)_{m\times n} &  (0)_{m\times nm}
\\
(0)_{n\times m} & ({\rm Id})_{n\times n}& (0)_{n\times nm} 
\\
(0)_{nm\times m} & \left(\displaystyle\frac{\partial^2\Lag}{\partial y^B\partial y^A_\mu} \right) & \left(\displaystyle\frac{\partial^2\Lag}{\partial y^B_\nu \partial y^A_\mu}\right)
\end{matrix}\right) \ .
$$
It follows that $Y_i=\left(\gamma_i\right)_{\mu}^A\derpar{}{y^A_\mu}\in\ker{\mathscr F}\Lag_*$ 
with (local) component functions $\left(\gamma_i\right)_{\mu}^A\in\Cinfty(J^1E)$ which annihilate the Hessian,
$\displaystyle \left(\frac{\partial^2\mathscr{L}}{\partial y^B_\nu \partial y^A_\mu}\right)(\left(\gamma_i\right)_{\mu}^A)=(0)$, and hence
$\ker{\mathscr F}\Lag_*\subset\vf^{{\rm V}(\pi_1)}(J^1E)$.
Furthermore, is worth noting that
$$
\text{ker}\, \Omega_\mathscr{L}\cap\vf^{{\rm V}(\pi_1)}(J^1E)=\text{ker}\, \mathscr{FL}_*=\text{ker}\, \widetilde{\Omega}_\mathscr{L}\cap\vf^{{\rm V}(\pi_1)}(J^1E)
\subset\text{ker}\, \widetilde{\Omega}_\mathscr{L}\cap\vf^{{\rm V}(\bar{\pi}_1)}(J^1E)=\text{ker}\, \widetilde{\Omega}_\mathscr{L}\cap\text{ker}\, \omega .
$$
Finally:

\begin{prop}
\label{prop1}
For every constraint $\zeta_i\in\Cinfty(J^1E^*)$
which locally defines the ({\sl primary Hamiltonian constraint\/}) submanifold $P_0={\mathscr F}\Lag(J^1E)\hookrightarrow J^1E^*$, there exists a vector field 
$\displaystyle Y_i=\left(\gamma_i\right)_{\mu}^A\derpar{}{y^A_\mu}\in\ker{\mathscr F}\Lag_*$ such that
\beq
\label{gammarel}
\left(\gamma_i\right)_{\mu}^A={\mathscr F}\Lag^*\left(\derpar{\zeta_i}{p^\mu_A} \right) \ .
\eeq
\end{prop}
\proof
As $\zeta_i$ is a primary Hamiltonian constraint it follows that
${\mathscr F}\Lag^*\zeta_i=0$. Then,
\beann
0&=&\d{\mathscr F}\Lag^*\zeta_i=\d(\zeta_i\circ{\mathscr F}\Lag)
\\ &=&
\left(\derpar{\zeta_i}{p^\mu_A}\circ{\mathscr F}\Lag\right)\derpar{p^\mu_A}{y^B_\nu}\,\d y^B_\nu+
\left(\derpar{\zeta_i}{p^\mu_A}\circ{\mathscr F}\Lag\right)\derpar{p^\mu_A}{y^B}\,\d y^B+
\left(\derpar{\zeta_i}{y^B}\circ{\mathscr F}\Lag\right)\,\d y^B
\\ &=&
{\mathscr F}\Lag^*\left(\derpar{\zeta_i}{p^\mu_A}\right)\left(\frac{\partial^2\mathscr{L}}{\partial y^B_\nu \partial y^A_\mu}\right)\,\d y^B_\nu+
\left[{\mathscr F}\Lag^*\left(\derpar{\zeta_i}{p^\mu_A}\right)\left(\frac{\partial^2\mathscr{L}}{\partial y^B \partial y^A_\mu}\right)+
{\mathscr F}\Lag^*\left(\derpar{\zeta_i}{y^B}\right)\right]\,\d y^B
\\ & \Longleftrightarrow& 
{\mathscr F}\Lag^*\left(\derpar{\zeta_i}{p^\mu_A}\right)\left(\frac{\partial^2\mathscr{L}}{\partial y^B_\nu \partial y^A_\mu}\right)=0 \ , \
{\mathscr F}\Lag^*\left(\derpar{\zeta_i}{p^\mu_A}\right)\left(\frac{\partial^2\mathscr{L}}{\partial y^B \partial y^A_\mu}\right)+
{\mathscr F}\Lag^*\left(\derpar{\zeta_i}{y^B}\right)=0 \ ,
\eeann
and equation \eqref{gammarel} follows from the first equality above.
\qed


\subsection{Multisymplectic Symmetries and Noether Currents}

In the physics literature it is typical to work with the sections of the various fiber bundles that have been discussed. As mentioned earlier, the fields $y^A(x)$ are given by sections of the configuration bundle $\pi:E\rightarrow M$. Furthermore, the transformations which are used to investigate symmetries are also typically constructed on the sections of the configuration bundle in the physics literature. 
The Lagrangian function is typically defined on the jet prolongations $j^1\phi:M\rightarrow J^1E:x^\mu\mapsto (x^\mu,y^A(x),\partial_\mu y^A(x)) $ and the field transformations naturally produce transformations of the spacetime derivatives of the fields. 
The De Donder--Weyl formalism is similarly constructed on sections of $J^1E^*$ (or $P_0$). In this paper, the physics is being studied on the configuration manifold $E$, the jet manifold $J^1E$, and the dual jet manifold $J^1E^*$ (or $P_0$) as opposed to working on the sections of those manifolds as it is usual in most physical literature.
That is, one transforms $y^A$ which is a coordinate on some fiber bundle whose fibers give the field $y^A(x)$. 

Noether's theorem can be formulated on the multivelocity and multimomentum phase spaces. {\sl\textbf{Noether symmetries}} are regarded as transformations on the phase spaces which preserve the (pre)multisymplectic forms \cite{GIMMSY}. The conserved quantities on the multivelocity and multimomentum phase spaces associated with Noether symmetries are called {\sl covariant momentum maps} which will be discussed later in \ref{ap:covmomentummap}. 
{\sl\textbf{Gauge transformations}} are transformations on the configuration manifold $E$ which do not induce transformations on $M$; that is, the vector fields that generate gauge transformations are vector fields on $E$ which are $\pi$-vertical.
Gauge transformations whose liftings to $J^1E$ preserve the form $\Omega_\mathscr{L}$ are called
{\sl\textbf{gauge symmetries}}. 
Alternatively, lifted base space diffeomorphisms (such as spacetime diffeomorphisms or worldsheet diffeomorphisms in the case of string theory) are not referred to as gauge symmetries in this work. 

Since physical symmetries preserve $\Omega_\mathscr{L}$, and consequently $\Omega_\mathscr{H}$ as well, it is necessary to discuss how Lie groups act on the (pre-)multisymplectic multivelocity and multimomentum phase spaces. Afterwards, it will be shown how gauge transformations and base space diffeomorphisms are lifted to the multivelocity and multimomentum phase spaces.

\subsubsection{Covariant Momentum Maps and Noether Currents}
\label{ap:covmomentummap}

{\sl Covariant momentum maps} are the conserved quantities which result from the group action of a Lie group on a multisymplectic manifold. 
Consider a multisymplectic phase space $F$ with a multisymplectic $(m+1)$-form $\Omega=-\text{d}\Theta$ and a Lie group $G$ with group action $\Phi$ on $F$ written as
\begin{equation*}
\Phi:\mathpzc{G}\times F \rightarrow \text{Diff}(F):g\mapsto \Phi_g \quad ; \quad g \in G 
\ ,
\end{equation*}
and denote $\Phi_g:F \rightarrow F$ the corresponding diffeomorphisms induced by this action.
If $\mathfrak{g}$ is the Lie algebra of $G$, every $\xi\in\mathfrak{g}$ induces a vector field on $F$
\begin{equation*}
\left.\textbf{X}_{\xi}=\frac{\text{d}}{\text{d}\lambda}\right\vert_{\lambda=0}\Phi^*_{exp(\lambda\xi)} \ ,
\end{equation*}
with the use of some parameter $\lambda$ which is used to write $g\in G$ in terms of the exponential map
\begin{equation*}
exp:\textgoth{g}\rightarrow G:\xi\mapsto g=exp(\lambda\xi)\ .
\end{equation*}
The group $G$ is a {\sl\textbf{symmetry group}} of the multisymplectic phase space $F$ if every $g\in G$ generates a {\sl\textbf{multisymplectomorphism}}
(or a {\sl\textbf{covariant canonical transformation\/}}) on $F$; 
that is, a diffeomorphism $\Phi_g\colon F\to F$ such that $\Phi^*_g \Omega = \Omega$, or equivalently: 
\begin{equation}
\label{covtrans}
\mathpzc{L}_{\textbf{X}_{\xi}}\Omega = 0
\quad, \quad \mbox{\rm for every $\xi\in\mathfrak{g}$}\ ,
\end{equation}
where $\mathpzc{L}_{\textbf{X}_{\xi}}$ denotes the Lie derivative associated with ${\textbf{X}_{\xi}}$.
Furthermore, \textsl{\textbf{exact multisymplectomorphisms}} 
(or {\sl\textbf{special covariant canonical transformations\/}}) are those which also preserve the form $\Theta\in\df^m(F)$:
\begin{equation*}
\Phi^*_g \Theta = \Theta \Longleftrightarrow \mathpzc{L}_{\textbf{X}_{\xi}}\Theta = 0 \ .
\end{equation*}
Then, using of Cartan's identity, $\mathpzc{L}_{\textbf{X}_\xi}=i(\textbf{X}_\xi)\text{d}+\text{d}i(\textbf{X}_\xi)$, and as $\text{d}\Omega=0$, 
it follows from \eqref{covtrans} that $\text{d} i(\textbf{X}_{\xi})\Omega = 0$, so (locally) there exists some $(m-1)$-from $J_{\xi}\in \df^{m-1}(F)$ such that 
\begin{equation}
\label{equation:dJ}
i(\textbf{X}_{\xi})\Omega = -\text{d}J_{\xi} \ .
\end{equation}
Working with the infinitesimal generators of every $\xi\in\textgoth{g}$ 
amounts to $J_{\xi}$ to be linear in $\xi$ as will be the case in the theories that are investigated in this paper. This makes it possible to define a linear map, $\left.J\right\vert_{\rm p}$, by considering a point ${\rm p}\in F$ so that 
$J_{\rm p}(\xi) \equiv \left.J_{\xi}\right\vert_{\rm p} \in \Lambda^{m-1}(\Tan_{\rm p}F)$ gives 
$\left.J\right\vert_{\rm p}:\textgoth{g}\rightarrow \Lambda^{m-1}(\Tan_{\rm p}F):\xi \mapsto J_{\rm p}(\xi)$,
for every $\xi \in \textgoth{g}$. 
The linear map $\left.J\right\vert_{\rm p}$ is an element of $\textgoth{g}^*\otimes \Lambda^{m-1}(\Tan_zF)$. 
This structure makes it possible to construct another map, $J$, using $J({\rm p}) \equiv \left.J\right\vert_{\rm p}$ so that 
$J:F \rightarrow \textgoth{g}^*\otimes\df^{m-1}(F): {\rm p} \mapsto J({\rm p})$, for every ${\rm p}\in F$. 
The map $J$ is called the {\sl\textbf{covariant momentum map}}, although this terminology is used to refer to both $J$ and $\left.J\right\vert_{\xi}$. 
This setting can be concisely specified through the natural pairing $\left<\cdot,\cdot\right>$ between $\textgoth{g}^*$ and $\textgoth{g}$ written as 
\begin{equation}
\left.J_{\xi}\right\vert_{z}= \left<J(z),\xi \right> \ .
\end{equation}
Working directly with equation (\ref{equation:dJ}) and Cartan's identity gives 
\begin{equation}
\label{equation:dJcartan}
\text{d}J_{\xi}=-i(\textbf{X}_{\xi})\Omega=i(\textbf{X}_{\xi})\text{d}\Theta=\mathpzc{L}_{\textbf{X}_{\xi}}\Theta-\text{d}i(\textbf{X}_{\xi})\Theta \ ,
\end{equation}
but also
\begin{equation*}
0=\mathpzc{L}_{\textbf{X}_{\xi}}\text{d}\Theta=\text{d}\mathpzc{L}_{\textbf{X}_{\xi}}\Theta \ ,
\end{equation*}
so (at least locally) $\mathpzc{L}_{\textbf{X}_{\xi}}\Theta=\text{d}\alpha_{\xi}$ for some $\alpha_{\xi} \in \df^{m-1}(F)$. Then, equation (\ref{equation:dJcartan}) yields
\begin{equation*}
J_{\xi}=-i(\textbf{X}_{\xi})\Theta + \alpha_{\xi} \ .
\end{equation*}
However, equation (\ref{equation:dJ}) only determines $J_{\xi}$ up to some exact form $\text{d}\beta_{\xi}$ with $\beta_{\xi} \in \df^{m-2}(F)$. 
That is, equation (\ref{equation:dJ}) is still satisfied with the redefinition 
$J_{\xi}\rightarrow J_{\xi} - \text{d}\beta_{\xi}$;
so, in order to keep full generality, the momentum map $J_{\xi}$ takes the form 
\begin{equation}
\label{equation:mmap}
J_{\xi}=-i(\textbf{X}_{\xi})\Theta + \alpha_{\xi} + \text{d}\beta_{\xi} \ .
\end{equation}
It is worth noting that, for symmetries generated by a vector field $\textbf{X}_\xi$ which is a lift from $M$ to $F$, it follows that
$\mathpzc{L}_{\textbf{X}_{\xi}}\Theta=0 \Rightarrow \alpha_{\xi} = \text{d}\gamma_\xi$ for some $\gamma_\xi\in\Omega^{m-2}(F)$. Exact forms such as $\text{d}\beta_\xi$ and $\text{d}\gamma_\xi$ will be ignored in this paper as boundary terms will not be considered.

Furthermore, in the Lagrangian formulation of field theories where $F=J^1E$,
the covariant momentum maps on $J^1E$ are said to be in the \textsl{Lagrangian representation} and the \textsl{Noether current} $j=j^\mu\text{d}^{m-1}x_\mu\in \df^{m-1}(M)$ of a symmetry is obtained by pulling back the associated momentum map $J_\xi\in\df^{m-1}(J^1E)$ to $M$, 
using the first jet prolongations $j^1\phi$ as follows:
$j=j^1\phi^*J_\xi$.

\subsubsection{Lifting Transformations to $J^1E$ and $J^1E^*$}
\label{liftingvf}

Consider infinitesimal spacetime diffeomorphisms $M\rightarrow M$ produced by the coordinate transformation $x'^\mu=x^\mu+\xi^\mu(x)$. Such spacetime diffeomorphisms are generated by the vector field 
$\displaystyle\xi=-\xi^\mu(x)\frac{\partial}{\partial x^\mu}\in\vf(M)$. Then, the $\pi$-projectable vector field $\xi_E\in\vf(E)$, which generates the corresponding transformations $(x^\mu,y^A)\rightarrow (x^\mu+\xi^\mu(x),y^A+\xi^A(x,y))$ on the configuration manifold $E$ with $\xi^A(x,y)\in\Cinfty(E)$ can be written as
\begin{equation}
\label{diffsEvector}
\xi_E=-\xi^\mu(x)\frac{\partial}{\partial x^\mu}-\xi^A(x,y)\frac{\partial}{\partial y^A}\in\vf(E)\ .
\end{equation}
Furthermore, the canonical lift of $\xi_E\in\mathfrak{X}(E)$ to $J^1E$ is given by the following expression \cite{GIMMSY,book:Saunders89}:
\begin{equation}
X_\xi = -\xi^\mu\derpar{}{x^\mu}-\xi^A\frac{\partial}{\partial y^A}-\left(\derpar{\xi^A}{x^\mu}-y^A_\nu\derpar{\xi^\nu}{x^\mu}+y^B_\mu\frac{\partial\xi^A}{\partial y^B}\right)\frac{\partial}{\partial y^A_\mu}\in\mathfrak{X}(J^1E)\ .
\label{diffsEvector2}
\end{equation}

The field transformations $\delta y^A(x)$, which are sometimes referred to as the {\sl \textbf{local variation}} of the fields $y^A(x)$, are defined as the Lie derivatives of the fields $y^A(x)$ 
by the vector field $\xi$ on $M$ and take the form,
\begin{equation}
\label{liederivative}
\delta y^A(x)\equiv\mathpzc{L}_\xi y^A(x)=y'^A(x)-y^A(x)=-\xi^\mu(x)\derpar{y^A}{x^\mu}(x) +\widetilde{\xi}^A(x)\ ,
\end{equation}
where $\widetilde{\xi}^A(x)=\xi^A\circ\phi$.
The first term in (\ref{liederivative}), 
$-\xi^\mu(x)\derpar{y^A}{x^\mu}(x)$, is sometimes referred to as the {\sl \textbf{transport term}} while the second term, $\widetilde{\xi}^A(x)$, is referred to as the {\sl \textbf{global variation}} of the fields $y^A(x)$ and is given as
\begin{equation*}
\widetilde{\xi}^A(x)=\Delta_{g}y^A(x)\equiv y'^A(x')-y^A(x)\ .
\end{equation*}
Since the fields $y^A(x)$ are given by the local sections $\phi:M\rightarrow E$, the field transformations $\delta y^A(x)$ in (\ref{liederivative}) 
can be interpreted geometrically as the Lie derivative of the local sections $\phi$ defined as \cite{GIMMSY}:
\begin{equation}
\label{LieDerivSections1}
\mathpzc{L}_\xi\phi=T\phi\circ\xi-\xi_E\circ\phi \ .
\end{equation}

The variation of the spacetime derivatives of the fields, $\delta y^A_\mu(x)$, can also be characterized by the Lie derivative of local sections as follows: given holonomic local sections $\psi=j^1\phi:M\rightarrow J^1E$, the Lie derivative of $\psi$ with respect to the vector field $\displaystyle\xi=-\xi^\mu(x)\frac{\partial}{\partial x^\mu}\in\vf(M)$ is defined as
\begin{equation}
\label{LieDerivSections2}
\mathpzc{L}_\xi\psi=T\psi\circ \xi-X_\xi\circ\psi,
\end{equation}
where $X_\xi \in \vf(J^1E)$ is the canonical lift of $\xi\in\vf(M)$ to $J^1E$ given by (\ref{diffsEvector2}). It thereby follows that
\begin{equation*}
\mathpzc{L}_\xi\psi^A_\mu=\delta y^A_\mu(x)=-\xi^\nu\derpar{y^A_\mu}{x^\nu}-\derpar{\xi^\nu}{x^\mu}\derpar{y^A}{x^\nu}+\derpar{\widetilde{\xi}^A}{x^\mu}+
\derpar{y^B}{x^\mu}\frac{\partial\widetilde{\xi}^A}{\partial y^B}\ ,
\end{equation*}
where $\widetilde{\xi}^A(x)=\xi^A\circ\psi$.
It is important to point out that, sometimes, it is only possible to interpret the field variations $\delta y^A(x)$ as Lie derivatives of the jet prolongations $j^1\phi$; this may occur when the vector fields which generate Noether symmetries on the (pre)multisymplectic phase spaces are not projectable onto $E$ and are not produced from canonical lifts from $E$ to $J^1E$. Such examples will come up in sections \ref{kgsymmetries} and \ref{carrollsymmetries0}.

In the case of a gauge transformation, the vector field $\xi_E$ only has the component along $\partial/\partial y^A$,
\begin{equation*}
\xi_E=-\xi^A(x,y)\frac{\partial}{\partial y^A}\in\vf(E)\ ,
\end{equation*}
so the canonical lift to $J^1E$ is given by
\begin{equation*}
X_\xi = -\xi^A\frac{\partial}{\partial y^A}-\left(\derpar{\xi^A}{x^\mu}+y^B_\mu\frac{\partial\xi^A}{\partial y^B}\right)\frac{\partial}{\partial y^A_\mu}\in \vf(J^1E)\ .
\end{equation*}

Canonical lifts from $E$ to $J^1E$ satisfy the property that they preserve the Lagrangian density if, and only if, they preserve the Poincar\'e--Cartan form, i.e. $\mathpzc{L}_{X_\xi}\textbf{L}=0\ \Leftrightarrow \mathpzc{L}_{X_\xi}\Theta_\mathscr{L}=0$ 
\cite{EMR-96,book:Saunders89}. 
The vector fields on $J^1E$ used to construct the covariant momentum maps on $J^1E$ are precisely canonical lifts and, for regular Lagrangians (i.e. when $\mathscr{FL}$
is a diffeomorphism), the vector fields used to construct the covariant momentum maps on $J^1E^*$ are the push-forward by the Legendre map of the canonical lifts: $Y_\xi=\mathscr{FL}_*X_\xi\in\vf(J^1E^*)$. 
When the Lagrangian is singular (in particular, almost-regular), it is sometimes the case that the vector field $X_\xi\in\vf(J^1E)$ is only projectable onto the multimomentum phase space on some constraint submanifold of the multivelocity phase space: $Y_\xi=\mathscr{FL}_{0*}X_\xi\vert_{S\subset J^1E}\in\vf(P_0)$. This scenario occurs in both the Klein--Gordon field theory in Section \ref{canonicalKG} and the magnetic Carrollian scalar field theory in Section \ref{CarrollMagnetic}.


\section{Scalar Field Theories}\label{section:ScalarFields}

The scalar field theories presented in this section are formulated on an $m$-dimensional spacetime $M$ and are constructed from two scalar fields, $\phi(x)$ and $\pi(x)$, given by the sections of the configuration bundle $E\rightarrow M: (\phi, \pi) \mapsto x^\mu$ with local sections denoted as $\phi:M\rightarrow E: x^\mu\mapsto (x^\mu,\phi(x), \pi(x))$. Three different scalar field theories will be constructed on this geometric setup while the second and third field theories come from Carroll contractions of the canonical Klein--Gordon Lagrangian. For more details on Carrollian field theories, see for example \cite{CarrollParticle,Gomis:2022spp,CarrollScalar,NL-Primer}.

\subsection{Klein--Gordon From Geometric Constraints}
\label{canonicalKG}

In this section, spacetime, $M$, is taken to be $m-$dimensional Minkowski space with the Minkowski metric denoted as $\eta_{\mu\nu}$ with signature $(-+\dotsc +)$.

\subsubsection{Lagrangian Formulation}

The scalar field theory investigated in this section is given by the canonical Klein--Gordon Lagrangian which originates from the canonical construction of the Klein--Gordon field on the canonical momentum phase space where the canonical momentum, $\pi(x)$, is given by the Legendre map $\pi(x)=\partial\mathscr{L}(x)/\partial\dot{\phi}(x)$. 
Conversely, in the construction of the canonical Klein--Gordon Lagrangian given here, $\pi(x)$ is proposed simply as a second variational field while the notion of the canonical Legendre map is abandoned and the relation that is typically obtained from the canonical Legendre map instead shows up as a {\sc sopde} constraint. 
The jet bundle which serves as the multivelocity phase space for this specific field theory has coordinates $(x^\mu,\phi,\pi,\phi_\mu,\pi_\mu)$ and it thereby follows that the first-order jet prolongations of sections $\phi:M\to E$ have the expression $j^1\phi:M\rightarrow J^1E:x^\mu\mapsto (x^\mu,\phi(x), \pi(x),\partial_\mu\phi(x), \partial_\mu \pi(x))$.

Now, instead of writing the Lagrangian in terms of the fields themselves, the Lagrangian is written as a function on the multivelocity phase space,
\begin{equation}
\label{scalarLag}
\mathscr{L}=\pi\phi_0-\frac{1}{2}\pi^2-\frac{1}{2}\phi_i\phi^i \in C^\infty (J^1E)\ .
\end{equation}
This Lagrangian is singular as it is directly evident from the Hessian matrix (with respect to the multivelocities), whose components are
$$
\frac{\partial^2\mathscr{L}}{\partial\phi_0\partial\phi_0}=0\ , \
\frac{\partial^2\mathscr{L}}{\partial\phi_0\partial\phi_i}=0 \ , \
\frac{\partial^2\mathscr{L}}{\partial\phi_i\partial\phi_j}=-\delta^{ij}\ ,\
\frac{\partial^2\mathscr{L}}{\partial\pi_\mu\partial\pi_\nu}=0 \ ,\
\frac{\partial^2\mathscr{L}}{\partial\pi_\mu\partial\phi_\nu}=0 \ .
$$
The null vectors associated with $\derpar{^2\mathscr{L}}{\phi_\mu\partial\phi_\nu}$ and $\derpar{^2\mathscr{L}}{\pi_\mu\partial\pi_\nu}$ are
\begin{equation}
\label{KGnullvectors}
\gamma^{(\phi)}_\mu= 
\begin{pmatrix}
1\\
0\\
\vdots\\
0
\end{pmatrix}
\quad , \quad
\gamma^{(\pi)}_\mu= 
\begin{pmatrix}
1\\
1\\
\vdots\\
1
\end{pmatrix}\ ,
\end{equation}
respectively.
The Lagrangian energy is 
$$
E_\mathscr{L}=\frac{\partial\mathscr{L}}{\partial\phi_\mu}\phi_\mu+\frac{\partial\mathscr{L}}{\partial \pi_\mu}\pi_\mu-\mathscr{L}=
\frac{1}{2}(\pi^2-\phi_i\phi^i)\in C^\infty (J^1E) \ ,
$$
and the Poincar\'e--Cartan forms are
\bea
\Theta_\mathscr{L}&=&\frac{\partial\mathscr{L}}{\partial\phi_\mu}\text{d}\phi\wedge\text{d}^{m-1}x_\mu+\frac{\partial\mathscr{L}}{\partial \pi_\mu}\text{d}p\wedge\text{d}^{m-1}x_\mu-E_\mathscr{L}\text{d}^mx 
\nonumber \\ &=&
\pi\text{d}\phi\wedge\text{d}^{m-1}x_0-\phi^i\text{d}\phi\wedge\text{d}^{m-1}x_i- \frac{1}{2}(\pi^2-\phi_i\phi^i)\text{d}^mx\in\df^m(J^1E) \ ,
\label{thetatilde} \\
\Omega_\mathscr{L}&=&
-\text{d}\Theta_\mathscr{L}=
\text{d}\phi\wedge\text{d}\pi\wedge\text{d}^{m-1}x_0-\text{d}\phi^i\wedge\text{d}\phi\wedge\text{d}^{m-1}x_i+(\pi\text{d}\pi-\phi^i\text{d}\phi_i)\wedge\text{d}^mx
\nonumber \\ &\equiv&
\widetilde{\Omega}_\mathscr{L} + \text{d}E_\mathscr{L}\wedge\text{d}^mx 
\in\df^{m+1}(J^1E)\ .
\label{omegatilde}
\eea
As $\Lag$ is a singular Lagrangian, 
${\Omega}_\mathscr{L}$ is a premultisymplectic form.
Now, take a locally decomposable multivector field $\textbf{X}_\Lag=X_0\wedge X_1\wedge\dotsc\wedge X_{m-1}\in\vf^m(J^1E)$, with components
\begin{equation*}
X_\mu=\frac{\partial}{\partial x^\mu}+A_\mu\frac{\partial}{\partial\phi}+B_\mu\frac{\partial}{\partial \pi}+C_{\mu\nu}\frac{\partial}{\partial\phi_\nu}+D_{\mu\nu}\frac{\partial}{\partial \pi_\nu} \ ,
\end{equation*}
which satisfies $i(\textbf{X}_\Lag)\d^mx=1$.
Then, the field equation \eqref{equation:LEOM} gives
\beann
0=i(\textbf{X}_\Lag)\Omega_\mathscr{L}&=&-B_0\text{d}\phi+A_0\text{d}\pi+C_i^i\text{d}\phi-A^i\text{d}\phi_i-\pi \text{d}\pi+\phi^i\text{d}\phi_i \\
& &+\left(B_0 A_\lambda-B_\lambda A_0+C_{\lambda i}A_i-C_{ii}A_\lambda +\pi B_\lambda -\phi_i C_\lambda^i\right)\text{d}x^\lambda \ ,
\eeann
which, by setting differential forms separately equal to zero, produces the following equations
\begin{subequations}
\begin{align}
&C_i^i-B_0=0 \ ,  \label{scalareq1} 
\\
&A_0-\pi=0 \ , \label{scalareq2} \\
&\phi_i-A_i=0 \ ,\label{scalareq3} \\
&B_0 A_\lambda-B_\lambda A_0 +C_{\lambda i}A_i-C_{ii}A_\lambda +\pi B_\lambda -\phi_i C_\lambda^i=0 \ .\label{scalareq4}
\end{align}
\end{subequations}
Equation (\ref{scalareq4}) is a combination of the first three equations and is thereby and identity. 
Equation (\ref{scalareq2}) determines the coefficient $A_0$,
equation \eqref{scalareq1} is a relation among the coefficients $B_0$ and $C_i^i$,
and equation (\ref{scalareq3}) is the spatial part of the holonomic or {\sc sopde} condition for the 
$\partial/\partial\phi$ piece of the multivector field. 
As these equations show, no compatibility constraints appear.
As it has been proven in Proposition \ref{maintheor},
this can also be analyzed geometrically by enforcing 
\begin{equation*}
i(Z)\text{d}E_\mathscr{L}=0 \quad ,\quad
\mbox{\rm for every $Z\in\text{ker}\,\widetilde{\Omega}_\mathscr{L}\cap\text{ker}\,\text{d}^mx$} \ ,
\end{equation*}
where,
$\widetilde\Omega_\Lag=
\text{d}\phi\wedge\text{d}\pi\wedge\text{d}^{m-1}x_0-\text{d}\phi^i\wedge\text{d}\phi\wedge\text{d}^{m-1}x_i$,
as it is shown in \eqref{omegatilde}. In this case
\begin{equation*}
\text{ker},\widetilde{\Omega}_\mathscr{L}\cap\text{ker}\, \text{d}^mx=
\left\langle \frac{\partial}{\partial\phi_0},
\frac{\partial}{\partial \pi_\mu}\right\rangle \ ,
\end{equation*}
and it follows that 
\begin{equation*}
i\left(\frac{\partial}{\partial\phi_0}\right)\text{d}E_\mathscr{L}=0=
i\left(\frac{\partial}{\partial \pi_\mu}\right)\text{d}E_\mathscr{L} \ ,
\end{equation*}
which agrees with the fact that no compatibility constraints are produced in this system.

Now, imposing the {\sc sopde} condition
$A_\mu=\phi_\mu$, $B_\mu=\pi_\mu$ on $\textbf{X}_\mathscr{L}$
(i.e. demanding that the multivector fields $\textbf{X}_\Lag$ are holonomic),
it follows that the relevant field equations left over are
\begin{subequations}
\begin{align}
&C_i^i-\pi_0=0 \ , \label{scalareq2.1} 
\\
&\phi_0-\pi=0 \ . \label{scalareq2.2}
\end{align}
\end{subequations}
At this stage, the first equation \eqref{scalareq2.1} gives
another relation for the coefficients $C_i^i$.
Equation \eqref{scalareq2.2} is a {\sc sopde} constraint which defines the constraint submanifold $S_1\hookrightarrow J^1E$; this is a constraint giving a relation between the fields and multivelocities.
Notice that the Lagrangian function (\ref{scalarLag}) on the submanifold $S_1$ takes the form of the standard Klein--Gordon Lagrangian:
\begin{equation*}
\left.\mathscr{L}\right\vert_{S_1}=-\frac{1}{2}\phi_\mu\phi^\mu\ .
\end{equation*}
As a final step, it is necessary to impose the tangency condition of the multivector fields ${\bf X}_\Lag$
to the submanifold $S_1$:
\beq
\mathpzc{L}_{X_\mu}(\phi_0-\pi)=C_{\mu0}-B_\mu=0  \quad \mbox{\rm (on $S_1$)} \ .
\label{holcon0}
\eeq
These are new relations among the coefficients of ${\bf X}_\Lag$ but are not constraints, therefore $S_1$ is the final constraint submanifold on which the constraint algorithm terminates.

Now, upon taking
$(x^\mu,\phi(x),\pi(x),\phi_\nu(x),\pi_\nu(x))$
to be integral sections of the multivector field $\textbf{X}_\Lag$ by setting
\beq
A_\mu=\derpar{\phi}{x^\mu} \quad ,\quad
B_\mu=\derpar{\pi}{x^\mu} \quad ,\quad
C_{\mu\nu}=\derpar{\phi_\nu}{x^\mu} \quad ,\quad 
D_{\mu\nu}=\derpar{\pi_\nu}{x^\mu} \ ,
\label{intconds}
\eeq
the field equations on $S_1$ become 
\begin{subequations}
\begin{align}
\frac{\partial^2\phi}{\partial x^i\partial x_i}-\dot \pi\equiv
\partial_i\partial^i\phi-\dot \pi&=0\ ,\label{KG}\\
\dot\phi-\pi&=0\ .\label{scalarconstraint}
\end{align}
\end{subequations}
The combination of equations (\ref{KG}) and (\ref{scalarconstraint}) gives the Klein--Gordon equation: 
\begin{equation*}
\frac{\partial^2\phi}{\partial x^\mu\partial x_\mu}\equiv
\partial_\mu\partial^\mu\phi=0 \ .
\end{equation*}
Furthermore, the integral sections \eqref{intconds} which satisfy the field equations (\ref{KG}) and (\ref{scalarconstraint}) satisfy the tangency condition on $S_1$ given by \eqref{holcon0} particularly as a consequence of \eqref{scalarconstraint}.

\subsubsection{Hamiltonian Formulation}

The De Donder--Weyl Hamiltonian formulation takes place on the bundle $J^1E^*$ which serves as the multimomentum phase space and has coordinates $(x^\mu,\phi,\pi,p_\phi^\mu,p^\mu_\pi)$, where the multimomenta $p_\phi^\mu$, $p^\mu_\pi$ are obtained from the Legendre map $\mathscr{FL}:J^1E\rightarrow J^1E^*$,
\beq
\mathscr{FL}^*p_\phi^i=\frac{\partial\mathscr{L}}{\partial\phi_i}=-\phi^i \quad ,\quad
\mathscr{FL}^* p_\phi^0=\frac{\partial\mathscr{L}}{\partial\phi_0}=\pi \quad , \quad
\mathscr{FL}^*p_\pi^\mu=\frac{\partial\mathscr{L}}{\partial \pi_\mu}=0 \ .
\label{legtrans1}
\eeq
The first equation above is invertible and gives the relation among the multimomenta $\pi^i$ and the multivelocities $\phi^i$ and is therefore not a constraint.
The second and third equations give the primary constraints
$ p_\phi^0-\pi=0$, $p^\mu_\pi=0$,
which define the submanifold $\jmath_0:P_0\hookrightarrow J^1E^*$
with local coordinates $(x^\mu,\phi,\pi,p_\phi^i)$. 
Note that, by Proposition \ref{prop1}, the null vectors \eqref{KGnullvectors} are given as
$$
\gamma^{(\phi)}_\mu=\mathscr{FL}^*\derpar{}{p^\mu_\phi}(p^0_\phi-\pi)\quad , \quad
\gamma^{(\pi)}_\mu=\mathscr{FL}^*\derpar{}{p^\mu_\phi}\left(p^\mu_\pi\right)\ .
$$
Furthermore, $\mathscr{FL}$ maps onto the primary constraint submanifold $P_0\subset J^1E^*$ as a consequence of the fact that the Lagrangian function $\Lag$ is almost-regular and the restricted Legendre map $\mathscr{FL}_0$ is given by $\mathscr{FL}=\jmath_0\circ\mathscr{FL}_0$.

The De Donder--Weyl Hamiltonian $\mathscr{H}_0\in\Cinfty(P_0)$ obtained from $E_\mathscr{L}=\mathscr{FL}_0^*\mathscr{H}_0$ is given as, 
\begin{equation*}
\mathscr{H}_0=\frac{1}{2}(\pi^2-p_\phi^i p_{\phi i}) \ .
\end{equation*}
The Hamilton--Cartan forms obtained from \eqref{defThetaOmegaH} are
\beann
\Theta_\mathscr{H}^0&=&\pi\text{d}\phi\wedge\text{d}^{m-1}x_0+p_\pi^i\text{d}\phi\wedge\text{d}^{m-1}x_i-\frac{1}{2}(\pi^2-p_\phi^i p_{\phi i})\text{d}^mx\in \df^m(P_0) \ , \\
\Omega_\mathscr{H}^0=-\text{d}\Theta_\mathscr{H}^0&=&
\text{d}\phi\wedge\text{d}\pi\wedge\text{d}^{m-1}x_0+\text{d}\phi\wedge\text{d}p_\phi^i\wedge\text{d}^{m-1}x_i+(\pi\text{d}\pi- p_{\phi i}\text{d}p_\phi^i)\wedge\text{d}^mx \\
&\equiv& \widetilde\Omega_\mathscr{H}^0+\d\mathscr{ H}_0\wedge\text{d}^mx \in \df^{m+1}(P_0) \ .
\eeann
Now, in order to produce the Hamiltonian field equations, take a locally decomposable multivector field 
$\textbf{X}_\mathscr{ H}^0=X_0\wedge X_1\wedge\dotsb\wedge X_{m-1}\in \mathfrak{X}^m(P_0)$
satisfying the normalized transverse condition $i(\textbf{X}_\mathscr{ H}^0)\d^mx=1$, 
whose components are written as
\begin{equation*}
X_\mu=\frac{\partial}{\partial x^\mu}+A_\mu\frac{\partial}{\partial\phi}+B_\mu\frac{\partial}{\partial \pi}+C^i_{\mu}\frac{\partial}{\partial p_\phi^i} \ .
\end{equation*}
Then, the field equation $i(\textbf{X}_\mathscr{ H}^0)\Omega_\mathscr{H}^0=0$
yields
\begin{subequations}
\begin{align}
&B_0+C^i_i=0 \ , \label{Hscalareq1} \\
&A_0-\pi=0 \ , \label{Hscalareq2} \\
&A_i+p_{\phi i}=0 \ , \label{Hscalareq3} \\
&\pi B_\lambda-p_{\phi i}C_\lambda^i+A_\lambda B_0-A_0B_\lambda+A_\lambda C^i_i-A_iC_\lambda^i=0\ . \label{Hscalareq4}
\end{align}
\end{subequations}
Similarly to the Lagrangian formalism,
equation (\ref{Hscalareq4}) is a combination of the first three equations and is thereby an identity.
Equations (\ref{Hscalareq2}) and (\ref{Hscalareq3}) determine the coefficients 
$A_0$ and $A_i$ respectively
and equation \eqref{Hscalareq1} is a relation among the coefficients $B_0$ and $C_i^i$.
As before, no compatibility constraints appear;
this can also be shown geometrically as in the Lagrangian formulation
by searching for compatibility constraints via the application of Proposition \ref{maintheor}:
set
$i(Z)\text{d}\mathscr{H}_0=0$ for every 
$Z\in\text{ker}\,\widetilde{\Omega}_\mathscr{H}^0\cap\text{ker}\,\text{d}^mx$. 
However, as 
\begin{equation}
\text{ker}\,\widetilde{\Omega}_\mathscr{H}^0\cap\text{ker}\,\text{d}^mx=\{ 0\}\ , 
\end{equation}
it follows that condition \eqref{ncond} holds
in agreement with the fact that there are no compatibility constraints (as one would expect from the Lagrangian formulation).

Now, taking $(x^\mu,\phi(x),\pi(x),p^i_\phi(x))$ to be integral sections of $\textbf{X}_\mathscr{ H}^0$ given by, and hence
$$
A_\mu=\derpar{\phi}{x^\mu} \quad ,\quad
B_\mu=\derpar{\pi}{x^\mu} \quad ,\quad
C^i_\mu=\derpar{p^i_\phi}{x^\mu} \ ,
$$
the field equations above become
\begin{subequations}
\begin{align}
&\dot{\pi}+\derpar{p^i_\phi}{x^i}=0 \ , \label{Hintsec1} \\
&\dot{\phi}-\pi=0 \ , \label{Hintsec2} \\
&\derpar{\phi}{x^i}+p_\phi^i=0 \ .  \label{Hintsec3}
\end{align}
\end{subequations}
Plugging \eqref{Hintsec2} and \eqref{Hintsec3} into \eqref{Hintsec1} yields 
\begin{equation*}
\ddot{\phi}-\partial_i\partial^i\phi=0 \ ,
\end{equation*}
which is again the Klein--Gordon equation as expected and thereby displaying the equivalence between the Lagrangian and the Hamiltonian formalisms.

\subsubsection{Symmetries}\label{kgsymmetries}

The Lagrangian density associated with the Lagrangian function (\ref{scalarLag}),
\begin{equation*}
\textbf{L}=\mathscr{L}\text{d}^mx=\left(\pi\phi_0-\frac{1}{2}\pi^2-\frac{1}{2}\phi_i\phi^i\right)\text{d}^mx\ ,
\end{equation*}
is invariant under Lorentz transformations 
$\Lambda^\mu_{\ \nu}$:
\begin{equation*}
x'^\mu=\Lambda^\mu_{\ \nu} x^\nu \ ,\
\phi'(x')=\phi( x) \ ,\
\pi'(x')=\pi( x) \ ,\
\phi'_\mu(x')=\frac{\partial x^\nu}{\partial x'^\mu} \phi_\nu(x) \ ,\
\pi'_\mu(x')=\frac{\partial x^\nu}{\partial x'^\mu} \pi_\nu(x) \ .
\end{equation*}
Let the Lorentz transformations on $M$ be infinitesimal so that the Lorentz matrices take the form $\Lambda^\mu_{\ \nu}=\delta^\mu_{\ \nu}+\epsilon^\mu_{\ \nu}$, 
where $\epsilon_{\mu\nu}$ is the infinitesimal antisymmetric Lorentz matrix. 
It follows that the spacetime coordinates transform as $x'^\mu=x^\mu+\xi^\mu(x)$ with $\xi^\mu(x)=\epsilon^\mu_{\ \nu}x^\nu$
and the resulting infinitesimal Lorentz transformations of the fields and their spacetime derivatives are given by the Lie derivatives of the first--jet prolongations $j^1\phi:M\rightarrow J^1E$ with respect to the vector field 
$\xi=-\xi^\mu(x)\derpar{}{x^\mu}\in\vf(M)$ 
as defined in \eqref{LieDerivSections2}:
\begin{align}
\label{KGtransformations}
\begin{split}
\delta\phi(x)&\equiv \phi'(x)-\phi(x)=-\epsilon^\mu_{\ \nu}x^\nu\derpar{\phi}{x^\mu} \ ,\\
\delta \pi(x)&\equiv \pi'(x)-\pi(x)=-\epsilon^\mu_{\ \nu}x^\nu\derpar{\pi}{x^\mu}-\epsilon^\alpha_{\ 0}\phi_\alpha \ ,\\
\delta\phi_\mu(x)&\equiv \phi'_\mu(x)-\phi_\mu(x)= -\epsilon^\alpha_{\ \beta}x^\beta\derpar{\phi_\mu}{x^\alpha}-\epsilon^\alpha_{\ \mu}\phi_\alpha\ , \\
\delta \pi_\mu(x)&\equiv \pi'_\mu(x)-\pi_\mu(x)= -\epsilon^\alpha_{\ \beta}x^\beta\derpar{\pi_\mu}{x^\alpha}-\epsilon^\alpha_{\ \mu}\pi_\alpha\ ,
\end{split}
\end{align}
which give $\displaystyle\delta\mathscr{L}(x)=-\derpar{}{x^\alpha}\left(\epsilon^\alpha_{\ \beta}x^\beta\mathscr{L}(x)\right)$.
The vector field which generates the Lorentz transformations above on $J^1E$ is
\begin{align*}
\begin{split}
X_\Lambda&=-\epsilon^\mu_{\ \nu}x^\nu\frac{\partial}{\partial x^\mu}+\epsilon^\nu_{\ 0}\phi_\nu\frac{\partial}{\partial \pi}+\epsilon^\nu_{\ \mu}\phi_\nu\frac{\partial}{\partial\phi_\mu}+\epsilon^\nu_{\ \mu}\pi_\nu\frac{\partial}{\partial \pi_\mu}\\
&=-\epsilon^\mu_{\ \nu}x^\nu\frac{\partial}{\partial x^\mu}+\epsilon^i_{\ 0}\phi_i\frac{\partial}{\partial \pi}+\epsilon^i_{\ 0}\phi_i\frac{\partial}{\partial\phi_0}+\left(\epsilon^0_{\ i}\phi_0+\epsilon^j_{\ i}\phi_j\right)\frac{\partial}{\partial\phi_i}+\epsilon^\nu_{\ \mu}\pi_\nu\frac{\partial}{\partial \pi_\mu}\in\mathfrak{X}(J^1E)\ .
\end{split}
\end{align*}
It follows that $\mathpzc{L}_{X_\Lambda}\textbf{L}=0$ holds everywhere on $J^1E$ while
$$
\mathpzc{L}_{X_\Lambda}\Theta_\mathscr{L}=\epsilon^{0i}(\pi-\phi_0)(\text{d}\phi\wedge\text{d}^{m-1}x_i-\phi_i\text{d}^mx)\ .
$$
Also, observe that $X_\Lambda$ is tangent to the constraint submanifold $S_1\subset J^1E$ given by the constraint $\phi_0-\pi=0$ as desired
since $\mathpzc{L}_{X_\Lambda}(\phi_0-\pi)=0$. 
However, $X_\Lambda$ is not projectable onto $E$ as a result of the component attached to $\derpar{}{\pi}$; 
it is for this reason that the transformations \eqref{KGtransformations} can be interpreted as the Lie derivatives of $j^1\phi$ only (as the sections $\phi$ do not provide the term associated with the second term of $X_\Lambda$ that is not projectable onto $E$).
Now $\mathpzc{L}_{X_\Lambda}\Theta_\mathscr{L}\vert_{S_1}=0$,
so the exact Noether symmetry exists only on the constraint submanifold $S_1$.
Then, the corresponding momentum map $J_{\mathscr{L}}(X_\Lambda)=-i(X_\Lambda)\Theta_\mathscr{L}$ also exists only on the points of $S_1$ and is given by
$$
J_{\mathscr{L}}(X_\Lambda)\vert_{S_1}=
-\epsilon^i_{\ \beta}x^\beta \pi\text{d}\phi\wedge\text{d}^{m-2}x_{0i}+\epsilon^j_{\ \beta}x^\beta\phi^i\text{d}\phi\wedge\text{d}^{m-2}x_{ij}-\frac{1}{2}\epsilon^i_{\ \beta}x^\beta(\pi^2-\phi_j\phi^j)\text{d}^{m-1}x_i\ .
$$
In the Hamiltonian formalism on $P_0\subset J^1E^*$, the symmetry transformations are generated by $Y_\Lambda=\mathscr{FL}_{0*}\left.X_\Lambda\right\vert_{S_1}\in\mathfrak{X}(P_0)$ which is given as
\begin{equation*}
Y_\Lambda=-\epsilon^\mu_{\ \nu}x^\nu\frac{\partial}{\partial x^\mu}-\epsilon_{0i}p_\phi^i\frac{\partial}{\partial \pi}-\left(\epsilon^i_{\ 0}\pi+\epsilon^i_{\ j}p_\phi^j\right)\frac{\partial}{\partial p_\phi^i}\ .
\end{equation*}
The momentum map $J_{\mathscr{H}_0}(Y_\Lambda)=-i(Y_\Lambda)\Theta_\mathscr{H}^0\in\df^{m-1}(P_0)$ is given by
\beann
J_{\mathscr{H}_0}(Y_\Lambda)=-\epsilon^i_{\ \beta}x^\beta \pi\text{d}\phi\wedge\text{d}^{m-2}x_{0i}-\epsilon^j_{\ \beta}x^\beta p_\phi^i\text{d}\phi\wedge\text{d}^{m-2}x_{ij}-\frac{1}{2}\epsilon^i_{\ \beta}x^\beta(\pi^2-p_{\phi j}p_\phi^j)\text{d}^{m-1}x_i\ .
\eeann

\subsection{Carrollian Electric Scalar Field Theory}
\label{carrollsymmetries1}

The {\sl electric Carrollian contraction} \cite{CarrollScalar} of the canonical Klein--Gordon Lagrangian is performed by making the field redefinition given by $\phi(x)\rightarrow c\phi(x)$, $\pi(x)\rightarrow \frac{1}{c}\pi(x)$, and taking the limit $c\rightarrow 0$ for the speed of light. 
It follows that the Minkowski metric becomes degenerate as 
\begin{equation}
\label{carrollian metric}
\text{d}s^2=\eta_{\mu\nu}\text{d}x^\mu\otimes\text{d}x^\nu=-c^2\text{d}x^0\otimes\text{d}x^0+\delta_{ij}\text{d}x^i\otimes\text{d}x^j\ \longrightarrow\  \delta_{ij}\text{d}x^i\otimes\text{d}x^j\ .
\end{equation}

\subsubsection{Lagrangian Formulation}

The Lagrangian function $\mathscr{L}\in C^\infty (J^1E)$ obtained from the electric Carrollian contraction of the canonical Klein--Gordon Lagrangian is
\begin{equation}
\label{electricL}
\mathscr{L}=\pi\phi_0-\frac{1}{2}\pi^2\ ,
\end{equation}
which is clearly singular since
\begin{equation*}
\frac{\partial^2\mathscr{L}}{\partial\phi_\mu\partial\phi_\nu}=0\quad ,  \quad 
\frac{\partial^2\mathscr{L}}{\partial\pi_\mu\partial\pi_\nu}=0\quad ,  \quad 
\frac{\partial^2\mathscr{L}}{\partial\pi_\mu\partial\phi_\nu}=0\ .
\end{equation*}
The null vectors of the Hessians $\derpar{^2\mathscr{L}}{\phi_\mu\partial\phi_\nu}$ and $\derpar{^2\mathscr{L}}{\pi_\mu\partial\pi_\nu}$ are
\begin{equation}
\label{Enullvectors}
\gamma^{(\phi)}_\mu= 
\begin{pmatrix}
1\\
1\\
\vdots\\
1
\end{pmatrix}
\quad , \quad
\gamma^{(\pi)}_\mu= 
\begin{pmatrix}
1\\
1\\
\vdots\\
1
\end{pmatrix}\ ,
\end{equation}
respectively.
The Lagrangian energy function and the Poincar\'e--Cartan forms are
\beann
E_\mathscr{L}&=&\frac{1}{2}\pi^2\in C^\infty(J^1E) \ , \\
\Theta_\mathscr{L}&=&\pi\text{d}\phi\wedge\text{d}^{m-1}x_0-\frac{1}{2}\pi^2\text{d}^mx\in \Omega^m(J^1E) \ , \\
\Omega_\mathscr{L}&=&\text{d}\phi\wedge\text{d}\pi\wedge\text{d}^{m-1}x_0+\pi\text{d}\pi\wedge\text{d}^mx\equiv
\widetilde{\Omega}_\mathscr{L} + \text{d}E_\mathscr{L}\wedge\text{d}^mx \in \Omega^{m+1}(J^1E)\ .
\eeann
Since the Lagrangian function (\ref{electricL}) is singular, $\Omega_\mathscr{L}$ is a premultisymplectic form.
As before, the field equations are obtained by using locally decomposable multivector fields 
$\textbf{X}_\Lag=X_0\wedge X_1\wedge\dotsb\wedge X_{m-1}\in\vf^m(J^1E)$
such that $i(\textbf{X}_\Lag)\d^mx=1$, 
so it follows that their components are given by 
\begin{equation*}
X_\mu=\frac{\partial}{\partial x^\mu}+A_\mu\frac{\partial}{\partial\phi}+B_\mu\frac{\partial}{\partial \pi}+C_{\mu 0}\frac{\partial}{\partial\phi_0}+C_{\mu i}\frac{\partial}{\partial\phi_i}+D_{\mu\nu}\frac{\partial}{\partial \pi_\nu} \ .
\end{equation*}
Then the field equation is
\begin{equation*}
0=i(\textbf{X}_\Lag)\Omega_\mathscr{L}=A_0\text{d}\pi-B_0\text{d}\phi-\pi\text{d}\pi+(A_\lambda B_0-B_\lambda A_0+\pi B_\lambda)\text{d}x^\lambda \ .
\end{equation*}
which leads to
\begin{subequations}
\begin{align}
&-B_0=0 \ , \label{Leq1} \\
&A_0-\pi=0 \ , \label{Leq2} \\
&A_\lambda B_0-B_\lambda A_0+\pi B_\lambda=0 \label{Leq3} \ .
\end{align}
\end{subequations}
Equations \eqref{Leq1} and \eqref{Leq2} determine the coefficients $B_0$ and $A_0$; equation \eqref{Leq3} is an identity as a result of the previous equations as usual.
There are no compatibility constraints,
which is corroborated 
by applying Proposition \ref{maintheor}:
\begin{equation*}
\text{ker}\ \widetilde{\Omega}_\mathscr{L}\cap\text{ker}\ \text{d}^mx=\left\langle\frac{\partial}{\partial\phi_\mu}, \frac{\partial}{\partial \pi_\mu}\right\rangle,
\end{equation*}
and $\displaystyle i\left(\frac{\partial}{\partial\phi_\mu}\right)\text{d}E_\mathscr{L}=0=
i\left(\frac{\partial}{\partial \pi_\mu}\right)\text{d}E_\mathscr{L}$ is satisfied {\sl a priori} without producing any constraints.

Imposing the {\sc sopde} condition, $A_\mu=\phi_\mu$, $B_\mu=\pi_\mu$, the field
equations \eqref{Leq1} and \eqref{Leq2} produce two {\sc sopde} constraints
\beq
\pi_0=0 \quad ,\quad \phi_0-\pi=0 \ ,
\label{electcons}
\eeq
which define the constraint submanifold $S_1\hookrightarrow J^1E$.
The tangency condition for ${\bf X}_\Lag$ on $S_1$ is imposec by 
\beq
\mathpzc{L}_{X_\mu}(\pi_0)=D_{\mu 0}=0  \quad , \quad
\mathpzc{L}_{X_\mu}(\phi_0-\pi)=C_{\mu0}-B_\mu=0 \quad
\mbox{\rm (on $S_1$)} \ .
\label{tangent2}
\eeq
It is evident above that no new constraints arise, hence, $S_1$ is the final constraint submanifold in the Lagrangian formalism.

Consider now $(x^\mu,\phi(x),\pi(x),\phi_\nu(x),\pi_\nu(x))$ which are integral sections 
of the multivector fields $\textbf{X}_\Lag$ and therefore
satisfy \eqref{intconds}.
Then, the {\sc sopde} constraints \eqref{electcons} yield
\beq
\dot \pi=0 \quad , 
\quad \dot\phi-\pi=0 \ , \label{electric2}
\eeq
which can be combined giving
\begin{equation*}
\ddot\phi=0 \ ,
\end{equation*}
as in \cite{CarrollParticle}.
The tangency conditions \eqref{tangent2} lead to
\begin{equation*}
\derpar{\pi_0}{x^\mu}\equiv\derpar{\dot\pi}{x^\mu}=0 \quad , \quad 
\derpar{\pi}{x^\mu}=\derpar{\phi_0}{x^\mu}\equiv\derpar{\dot\phi}{x^\mu}
 \quad ; \quad\mbox{\rm (on $S_1$)} \ .
\end{equation*}
These equations are fulfilled by the solutions to the field equations,
in particular, as a consequence of the second equation in
\eqref{electric2}.

\subsubsection{Hamiltonian Formulation}

The multimomenta obtained from the Legendre map are
\begin{equation*}
\mathscr{FL}^*p_\phi^0\equiv \frac{\partial\mathscr{L}}{\partial\phi_0}=\pi
\quad ,\quad
\mathscr{FL}^*p_\phi^i=\frac{\partial\mathscr{L}}{\partial\phi_i}=0
\quad ,\quad
\mathscr{FL}^*p^\mu_\pi=\frac{\partial\mathscr{L}}{\partial \pi_\mu}=0 \ ,
\end{equation*}
and give the primary constraints
$p_\phi^0-\pi=0$,
$p^i_\phi=0$, and $p^\mu_\pi=0$,
which define the primary constraint submanifold $\jmath_0:P_0\hookrightarrow  J^1E^*$ with local coordinates $(x^\mu,\phi,\pi)$ and the restricted Legendre map $\mathscr{FL}_0$ is given by $\mathscr{FL}=\jmath_0\circ\mathscr{FL}_0$ as before. Once again, the primary constraints can be used to compute the null vectors \eqref{Enullvectors} by using equation \eqref{gammarel} given in Proposition \ref{prop1}.
The Lagrangian function $\Lag$ is almost-regular
and the De Donder--Weyl Hamiltonian $\mathscr{H}_0\in\Cinfty(P_0)$ obtained from $E_\mathscr{L}=\mathscr{FL}_0^*\mathscr{H}_0$ is 
\begin{equation*}
\mathscr{H}_0=\frac{1}{2}\pi^2 \ ,
\end{equation*}
and the Hamilton-Cartan forms are 
\beann
\Theta_\mathscr{H}^0&=&\pi\text{d}\phi\wedge\text{d}^{m-1}x_0-\frac{1}{2}\pi^2\text{d}^mx\in\df^m(P_0) \ , \\
\Omega_\mathscr{H}^0&=&\text{d}\phi\wedge\text{d}\pi\wedge\text{d}^{m-1}x_0+\pi\text{d}\pi\wedge\text{d}^mx
\equiv \widetilde\Omega_\mathscr{H}^0+\d\mathscr{H}_0\wedge\text{d}^mx\in\df^{m+1}(P_0) \ .
\eeann
The locally decomposable multivector fields $\textbf{X}_\mathscr{H}^0=X_0\wedge X_1\wedge\dotsb\wedge X_{m-1}\in \mathfrak{X}^m(P_0)$
satisfying the normalized transverse condition $i(\textbf{X}_\mathscr{ H}^0)\d^mx=1$ used to produce the field equations have the following components:
\begin{equation*}
X_\mu=\frac{\partial}{\partial x^\mu}+A_\mu\frac{\partial}{\partial\phi}+B_\mu\frac{\partial}{\partial \pi}\ .
\end{equation*}
The field equations obtained from 
$i(\textbf{X}_\mathscr{H}^0)\Omega_\mathscr{H}^0=0$ are
\begin{subequations}
\begin{align}
&B_0=0 \ , \label{Heq1} \\
&A_0-\pi=0 \ , \label{Heq2} \\
&A_\lambda B_0-A_0B_\lambda +\pi B_\lambda=0 \ . \label{Heq3}
\end{align}
\end{subequations}
Once again, using \eqref{Heq1} and \eqref{Heq2}, the third equation \eqref{Heq3} is an identity.
These equations do not produce any compatibility constraints. This again is in agreement with Proposition \ref{maintheor} which dictates that
compatibility constraints may be produced by 
$i(Z)\text{d}\mathscr{H}_0=0$, for every $Z\in \text{ker}\,\widetilde{\Omega}_\mathscr{H}^0\cap\text{ker}\, \text{d}^mx$, and as
\begin{equation}
\text{ker}\, \widetilde{\Omega}_\mathscr{H}^0\cap\text{ker}\, \text{d}^mx=\{ 0\}
\end{equation}
is obtained computationally, it is confirmed that no compatibility constraints exist.

Upon working on $(x^\mu,\phi(x),\pi(x))$ which are integral sections 
of the multivector fields $\textbf{X}_\mathscr{H}^0$,
and hence
\begin{equation*}
A_\mu=\derpar{\phi}{x^\mu} \quad , \quad
B_\mu=\derpar{\pi}{x^\mu} \ ,
\end{equation*}
it follows that the field equations \eqref{Heq1} and  \eqref{Heq2} take the form
\begin{equation*}
\dot{\pi}=0 \quad , \quad \dot{\phi}-\pi=0 \ .
\end{equation*}
Notice that plugging in the second equation into the first above gives
\begin{equation*}
\ddot{\phi}=0 \ .
\end{equation*}
The equivalence between the Lagrangian and the Hamiltonian formalisms again is evident.

\subsubsection{Symmetries} 
\label{carrollsymmetries}

The spacetime symmetry of the Carroll spacetime geometry are the Carroll transformations \cite{CarrollScalar}:

\begin{equation}
\label{carroll1}
x'^0=x^0+b_kx^k \quad, \quad x'^i=x^i+\epsilon^i_{\ j}x^j
\quad \Rightarrow \quad x'^{\mu}=C^\mu_{\ \nu}x^\nu \ ,
\end{equation}
where $C^0_{\ 0}=1$, $C^0_{\ k}=b_k$, $C^k_{\ 0}=0$, $(C^i_{\ j})\in O(d)$.
These transformations can be represented infinitesimally by
\begin{equation}
\label{carroll2}
C^\mu_{\ \nu}= \delta^\mu_{\ \nu}+\epsilon^\mu_{\ \nu}\ ,
\end{equation}
so that now $b_k=\epsilon^0_{\ k}$ is infinitesimal. 
Up to linear terms in $\phi'(x')=\phi(x)$ and $\pi'(x')=\pi(x)$, the infinitesimal transformations of the fields, $\delta\phi(x)\equiv \phi'(x)-\phi(x)$ and $\delta \pi(x)\equiv \pi'(x)-\pi(x)$, are
\begin{align*}
\delta \phi(x)&=-\epsilon_{\ \beta}^{\alpha}x^{\beta}\derpar{\phi}{x^\alpha}=- b_kx^k\derpar{\phi}{x^0}-\epsilon^i_{\ j}x^j\derpar{\phi}{x^i} \ , \\
\delta \pi(x)&=-\epsilon_{\ \beta}^{\alpha}x^\beta\derpar{\pi}{x^\alpha}= -b_kx^k\derpar{\pi}{x^0}-\epsilon^i_{\ j}x^j\derpar{\pi}{x^i} \ .
\end{align*}
Furthermore 
\begin{equation*}
\phi'_\mu(x')=\frac{\partial x^\nu}{\partial x'^\mu}\phi_\nu(x)\ \Longleftrightarrow\ \phi'_\mu(C\cdot x) =(C^{-1})_{\ \mu}^{\nu}\phi_\nu(x)\ ,
\end{equation*}
where $C\equiv(C_{\ \nu}^\mu)$ and $C\cdot x\equiv C_{\ \nu}^\mu x^\nu$; 
hence,
\begin{align*}
\delta\phi_\mu(x)&\equiv \phi'_\mu(x)-\phi_\mu(x)= -\epsilon^\nu_{\ \mu}\phi_\nu-\epsilon_{\ \beta}^{\alpha}x^{\beta}\derpar{\phi_\mu}{x^\alpha} \ ,\\ 
\delta \pi_\mu(x)&\equiv \pi'_\mu(x)-\pi_\mu=-\epsilon^\nu_{\ \mu}\pi_\nu -\epsilon_{\ \beta}^{\alpha}x^{\beta}\derpar{\pi_\mu}{x^\alpha} \ .
\end{align*}
Component-wise, 
\begin{align*}
\delta\phi_0(x)&=-\epsilon_{\ \beta}^{\alpha}x^{\beta}\derpar{\phi_0}{x^\alpha}=-b_kx^k\derpar{\phi_0}{x^0}-\epsilon^k_{\ j}x^j\derpar{\phi_0}{x^k} \ ,\\ 
\delta\phi_i(x)&=-\epsilon^\nu_{\ i}\phi_\nu-\epsilon_{\ \beta}^{\alpha}x^{\beta}\derpar{\phi_i}{x^\alpha}=-b_i\phi_0-\epsilon^k_{\ i}\phi_k- b_kx^k\derpar{\phi_i}{x^0}-\epsilon^k_{\ j}x^j\derpar{\phi_i}{x^k} \ ,
\end{align*}
and similarly for the components of $\delta \pi_\mu(x)$. Under the transformations presented above, 
the Lagrangian transforms as $\displaystyle\delta\mathscr{L}(x)=-\derpar{}{x^\alpha}\left(\epsilon^\alpha_{\ \beta}x^\beta\mathscr{L}(x)\right)$. 
These transformations are produced by the Lie derivatives of the local sections $\phi:M\rightarrow E$ 
and $j^1\phi:M\rightarrow J^1E$ in the direction of a vector field $\xi=-\xi^\mu(x)\derpar{}{x^\mu}\in\vf(M)$ 
which generates the Carroll transformations, where now $\xi^\mu(x)=\epsilon^\mu_{\ \nu}x^\nu$ is given by (\ref{carroll1}) and (\ref{carroll2}). 
The vector field which generates the Carroll transformations on the configuration bundle $E$, obtained from \eqref{LieDerivSections1}, is
\begin{equation*}
\xi_E=-\epsilon^\mu_{\ \nu}x^\nu\frac{\partial}{\partial x^\mu}\in \vf(E)\ ,
\end{equation*}
while the canonical lift of $\xi_E$ to $J^1E$ which generates the Carroll transformations on the multivelocity phase space is given by
\begin{align*}
\begin{split}
X&=-\epsilon^\mu_{\ \nu}x^\nu\frac{\partial}{\partial x^\mu} +\epsilon^\nu_{\ \mu}\phi_\nu\frac{\partial}{\partial\phi_\mu}+\epsilon^\nu_{\ \mu}\pi_\nu\frac{\partial}{\partial \pi_\mu} \\
&=-b_ix^i\frac{\partial}{\partial x^0}-\epsilon^i_{\ j}x^j\frac{\partial}{\partial x^i}+\left(b_i\phi_0+\epsilon_{\ i}^{ j}\phi_j\right)\frac{\partial}{\partial \phi_i}+\left(b_i\pi_0+\epsilon_{\ i}^{ j}\pi_j\right)\frac{\partial}{\partial \pi_i}\in\vf(J^1E) \ .
\end{split}
\end{align*}
Evidently $X$ is tangent to the constraint submanifold $S_1$ since $\mathpzc{L}_X(\pi-\phi_0)=0$ and $\mathpzc{L}_X \pi_0=0$.
Finally, since $\mathpzc{L}_X\textbf{L}=0 \Rightarrow \mathpzc{L}_X\Theta_\mathscr{L}=0$, 
the momentum map $J_\mathscr{L}(X)$ corresponding to the Carroll spacetime transformations in the Lagrangian setting is given by
\begin{equation*}
J_\mathscr{L}(X)=-i(X)\Theta_\mathscr{L}=-\epsilon^i_{\ j}x^j\pi\text{d}\phi\wedge\text{d}^{m-2}x_{0i}-\frac{1}{2}\pi^2\left(b_ix^i\text{d}^{m-1}x_0+\epsilon^i_{\ j}x^j\text{d}^{m-1}x_i\right)\in\df^{m-1}(J^1E) \ .
\end{equation*}

In the Hamiltonian setting, which takes place on $P_0\subset J^1E^*$, the momentum map $J_{\mathscr{H}_0}(Y)\in\df^{m-1}(P_0)$ is constructed using the vector field $Y\in\vf(P_0)$ given by
\begin{equation*}
Y=\mathscr{FL}_{0*}X=-\epsilon^\mu_{\ \nu}x^\nu\derpar{}{x^\mu}=-b_ix^i\frac{\partial}{\partial x^0}-\epsilon^i_{\ j}x^j\frac{\partial}{\partial x^i}\ .
\end{equation*}
As in the Lagrangian setting, $\mathpzc{L}_Y\Theta^0_\mathscr{H}=0$, 
it follows that $J_{\mathscr{H}_0}(Y)$ is given by,
\begin{equation*}
J_{\mathscr{H}_0}(Y)=-i(Y)\Theta_\mathscr{H}^0=-\epsilon^i_{\ j}x^j\pi\,\text{d}\phi\wedge\text{d}^{m-2}x_{0i}-\frac{1}{2}\pi^2\left(b_ix^i\text{d}^{m-1}x_0+\epsilon^i_{\ j}x^j\text{d}^{m-1}x_i\right)\ .
\end{equation*}

\subsection{Carrollian Magnetic Scalar Field Theory}
\label{CarrollMagnetic}

The {\sl magnetic Carrollian contraction} \cite{CarrollScalar} of the canonical Klein--Gordon Lagrangian is performed by reinserting the factors of $c$ (speed of light) 
into the Lagrangian (\ref{scalarLag}) and taking the limit $c\rightarrow 0$. The Minkowski metric (\ref{carrollian metric}) becomes degenerate as in the electric Carrollian scalar field theory.

\subsubsection{Lagrangian Formulation}

The Lagrangian function $\mathscr{L}\in C^\infty (J^1E)$ obtained from taking the magnetic Carrollian contraction of the canonical Klein--Gordon Lagragian (\ref{scalarLag}) is
\begin{equation*}
\mathscr{L}=\pi\phi_0-\frac{1}{2}\phi_i\phi^i \ .
\end{equation*}
This Lagrangian is singular since the components of the Hessian
matrix (with respect to the multivelocities) are
\begin{equation*}
\frac{\partial^2\mathscr{L}}{\partial\phi_0\partial\phi_0}=0\quad , \quad 
\frac{\partial^2\mathscr{L}}{\partial\phi_0\partial\phi_i}=0\quad , \quad 
\frac{\partial^2\mathscr{L}}{\partial\phi_i\partial\phi_j}=-\delta^{ij}\quad , \quad 
\frac{\partial^2\mathscr{L}}{\partial\pi_\mu\partial\pi_\nu}=0\quad , \quad
\frac{\partial^2\mathscr{L}}{\partial\pi_\mu\partial\phi_\nu}=0\ . 
\end{equation*}
The null vectors of the Hessians $\derpar{^2\mathscr{L}}{\phi_\mu\partial\phi_\nu}$ and $\derpar{^2\mathscr{L}}{\pi_\mu\partial\pi_\nu}$ are
\begin{equation}
\label{Mnullvectors}
\gamma^{(\phi)}_\mu= 
\begin{pmatrix}
1\\
0\\
\vdots\\
0
\end{pmatrix}
\quad , \quad
\gamma^{(\pi)}_\mu= 
\begin{pmatrix}
1\\
1\\
\vdots\\
1
\end{pmatrix}\ ,
\end{equation}
respectively. 
The Lagrangian energy function and the Poincar\'e--Cartan forms are now
\beann
E_\mathscr{L}&=&-\frac{1}{2}\phi_i\phi^i\in C^\infty(J^1E)  \ ,
\\
\Theta_\mathscr{L}&=&\pi\text{d}\phi\wedge\text{d}^{m-1}x_0-\phi^i\text{d}\phi\wedge\text{d}^{m-1}x_i+\frac{1}{2}\phi_i\phi^i\text{d}^mx \in \Omega^m(J^1E)  \ ,
\\
\Omega_\mathscr{L}&=&\text{d}\phi\wedge\text{d}\pi\wedge\text{d}^{m-1}x_0+\text{d}\phi^i\wedge\text{d}\phi\wedge\text{d}^{m-1}x_i+\phi^i\text{d}\phi_i\wedge\text{d}^mx
\equiv\widetilde{\Omega}_\mathscr{L} + \text{d}E_\mathscr{L}\wedge\text{d}^mx\in \Omega^{m+1}(J^1E) ,
\eeann
and $\Omega_\mathscr{L}$ is a premultisymplectic form.
Taking locally decomposable  multivector fields $\textbf{X}_\Lag=X_0\wedge X_1\wedge\dotsb\wedge X_{m-1}\in\vf^m(J^1E)$
satisfying $i(\textbf{X}_\Lag)\d^mx=1$, their components are given as
\begin{equation*}
X_\mu=\frac{\partial}{\partial x^\mu}+A_\mu\frac{\partial}{\partial\phi}+B_\mu\frac{\partial}{\partial \pi}+C_{\mu 0}\frac{\partial}{\partial\phi_0}+C_{\mu i}\frac{\partial}{\partial\phi_i}+D_{\mu\nu}\frac{\partial}{\partial \pi_\nu} \ ,
\end{equation*}
and the field equations obtained from $i(\textbf{X}_\Lag)\Omega_\mathscr{L}=0$ are
\begin{subequations}
\begin{align}
&C_i^i-B_0=0 \ , \label{Lageq1} \\
&A_0=0  \ , \label{Lageq2} \\
&A_\lambda B_0-B_\lambda A_0+C_\lambda^i A_0+A_\lambda C_0^i +\phi^i C_{\lambda i}=0 \ . \label{Lageq3}
\end{align}
\end{subequations}
Equations \eqref{Lageq1} and \eqref{Lageq2} are relations for the coefficients
of the multivector fields
while equation \eqref{Lageq3} is an identity following from \eqref{Lageq1} and \eqref{Lageq2} as usual.
There are no compatibility constraints 
in agreement with the geometric analysis 
presented in Proposition \ref{maintheor}, which shows that
\begin{equation*}
\text{ker}\, \widetilde{\Omega}_\mathscr{L}\cap\text{ker}\, \text{d}^mx=\left\langle \frac{\partial}{\partial\phi_0}, \frac{\partial}{\partial \pi_\mu}\right\rangle \ ,
\end{equation*}
and $\displaystyle\inn\left(\frac{\partial}{\partial\phi_0}\right)\text{d}E_\mathscr{L}=0=
i\left(\frac{\partial}{\partial \pi_\mu}\right)\text{d}E_\mathscr{L}$ is satisfied without producing any constraints. 

Then, upon imposing the {\sc sopde} condition $A_\mu=\phi_\mu$, $B_\mu=\pi_\mu$,  equations  \eqref{Lageq1} and \eqref{Lageq2} become
\begin{equation*}
C_i^i-\pi_0=0 \quad ,\quad \phi_0=0 \ .
\end{equation*}
The first equation gives a relation for the coefficients $C_i^i$,
and the second one is a {\sc sopde}-constraint 
which defines the constraint submanifold $S_1\hookrightarrow J^1E$.

Now impose the tangency condition of the multivector fields ${\bf X}_\Lag$ to $S_1$,
\beq
\mathpzc{L}_{X_\mu}(\phi_0)=C_{\mu0}=0  \quad \mbox{\rm (on $S_1$)} \ .
\label{holcon}
\eeq
These are not constraints, but new relations among the coefficient of ${\bf X}_\Lag$
and hence the final constraint submanifold in the Lagrangian formalism is $S_1$.
Then, on $(x^\mu,\phi(x),\pi(x),\phi_\nu(x),\pi_\nu(x))$,
the integral sections of the multivector fields $\textbf{X}_\Lag$ satisfy equations \eqref{intconds} and 
the field equations take the form
\beq
\partial_i\partial^i\phi-\dot{\pi}=0 \quad , \quad \dot{\phi}=0  \quad ; \quad\mbox{\rm (on $S_1$)}\ .
\label{intsec0}
\eeq
For further details on these equations see \cite{CarrollParticle, CarrollScalar}.
Furthermore, from \eqref{holcon} it follows that, on $S_1$,
\begin{equation*}
\partial_\mu\phi_0\equiv\partial_\mu\dot\phi=0 \ .
\end{equation*}
The equations above are fulfilled by the solutions to the field equations, in particular, as a consequence of the second equation in
\eqref{intsec0}.

\subsubsection{Hamiltonian Formulation}

The Legendre map $\mathscr{FL}:J^1E\rightarrow J^1E^*$ gives the multimomenta
\begin{equation*}
\mathscr{FL}^*p_\phi^0= \frac{\partial\mathscr{L}}{\partial\phi_0}=\pi
\quad , \quad
\mathscr{FL}^*p_\phi^i=\frac{\partial\mathscr{L}}{\partial\phi_i}=-\phi^i
\quad , \quad
\mathscr{FL}^*p^\mu_\pi=\frac{\partial\mathscr{L}}{\partial \pi_\mu}=0 \ .
\end{equation*}
It follows that $p_\phi^0-\pi=0$ and $p_\pi^\mu=0$ are primary constraints which define the primary constraint submanifold $P_0\hookrightarrow J^1E^*$
with local coordinates $(x^\mu,\phi,\pi,p_\phi^i)$; consequently, the Lagrangian is again almost-regular. 
As above, the null vectors \eqref{Mnullvectors} are given by the partial derivatives of the primary constraints with respect to the multimomenta according to \eqref{gammarel} in Proposition \ref{prop1}.
The restricted Legendre map $\mathscr{FL}_0$ is given by $\mathscr{FL}=\jmath_0\circ\mathscr{FL}_0$ as usual. 

The De Donder--Weyl Hamiltonian $\mathscr{H}_0\in C^\infty(P_0)$ is given by
\begin{equation*}
\mathscr{H}_0=-\frac{1}{2}p_{\phi i}p_\phi^i \ ,
\end{equation*}
and the Hamilton--Cartan forms on $P_0$ are 
\beann
\Theta_\mathscr{H}^0&=&\pi\text{d}\phi\wedge\text{d}^{m-1}x_0+p_\phi^i\text{d}\phi\wedge\text{d}^{m-1}x_i+\frac{1}{2}p_{\phi i}p_\phi^i\text{d}^mx\in \df^m(P_0) \ , \\
\Omega_\mathscr{H}^0&=&\text{d}\phi\wedge\text{d}\pi\wedge\text{d}^{m-1}x_0+\text{d}\phi\wedge\text{d}p_\phi^i\wedge\text{d}^{m-1}x_i-p_{\phi i}\text{d}p_\phi^i\wedge\text{d}^mx
\equiv \widetilde\Omega_\mathscr{H}^0+\d\mathscr{ H}_0\wedge\text{d}^mx\in\df^{m+1}(P_0)\,.
\eeann
Now take a locally decomposable multivector field 
$\textbf{X}_\mathscr{H}^0=X_0\wedge X_1\wedge\dotsb\wedge X_{m-1}\in \mathfrak{X}^m(P_0)$ 
satisfying $i(\textbf{X}_\mathscr{ H}^0)\d^mx=1$,
so their components are given as
\begin{equation*}
X_\mu=\frac{\partial}{\partial x^\mu}+A_\mu\frac{\partial}{\partial\phi}+B_\mu\frac{\partial}{\partial \pi}+C^i_{\mu}\frac{\partial}{\partial p_\phi^i} \ .
\end{equation*}
The field equations obtained from $i(\textbf{X}_\mathscr{H}^0)\Omega_\mathscr{H}^0=0$ are:
\begin{subequations}
\begin{align}
&B_0+C_i^i=0 \ , \label{HFeq1} \\
&A_0=0 \ , \label{HFeq2} \\
&A_i+p_{\phi i}=0 \ , \label{HFeq3} \\
&A_\lambda B_0-A_0B_\lambda +A_\lambda C_i^i-A_iC_\lambda^i -p_{\phi i}C_\lambda^i=0 \ . \label{HFeq4}
\end{align}
\end{subequations}
Equation (\ref{HFeq4}) is a combination of the equations 
(\ref{HFeq1}), (\ref{HFeq2}), and (\ref{HFeq3}) which give relations or determine some coefficients.
Then, no compatibility constraints appear.
Again, this is in agreement with the procedure given in Proposition \ref{maintheor}:
set $i(Z)\text{d}\mathscr{H}_0=0$
for every $Z\in \text{ker}\, \widetilde{\Omega}_\mathscr{H}^0\cap\text{ker}\, \text{d}^mx$;
however, 
\begin{equation*}
\text{ker}\, \widetilde{\Omega}_\mathscr{H}^0\cap\text{ker}\ \text{d}^mx=\{ 0\}\ , 
\end{equation*}
and condition \eqref{ncond} from Proposition \ref{maintheor} holds.

Working on $(x^\mu,\phi(x),\pi(x),p^i_0(x))$ taken to be integral sections 
of $\textbf{X}_\mathscr{H}^0$, 
\begin{equation*}
A_\mu=\derpar{\phi}{x^\mu} \quad , \quad
B_\mu=\derpar{\pi}{x^\mu} \quad , \quad
 C^i_\mu=\derpar{p_\phi^i}{x^\mu} \ ,
\end{equation*}
then (\ref{HFeq1}), (\ref{HFeq2}), and (\ref{HFeq3}) give
\begin{equation*}
\dot{\pi}+\derpar{p_\phi^i}{x^i} =0\quad , \quad
\dot{\phi}=0\quad , \quad
\derpar{\phi}{x^i} +p_{\phi i}=0 \ .
\end{equation*}
Plugging in the third equation into the first equation gives
\begin{equation*}
\dot{\pi}-\frac{\partial^2\phi}{\partial x^i\partial x_i}\equiv
\dot{\pi}-\partial_i\partial^i\phi=0 \quad , \quad \dot{\phi}=0 \ .
\end{equation*}
Once again, the equivalence between the Lagrangian and the Hamiltonian formalisms is evident as the resulting field equations are shown to be equivalent.

\subsubsection{Symmetries}
\label{carrollsymmetries0}

The Carroll transformations for the magnetic scalar field theory are written slightly differently than how they are written for the electric scalar field \ref{carrollsymmetries1}.
The Carroll transformations which are symmetries of the magnetic scalar field theory are
\begin{align*}
\delta \phi(x)&=-\epsilon_{\ \beta}^{\alpha}x^{\beta}\derpar{\phi}{x^\alpha}=- b_kx^k\derpar{\phi}{x^0}-\epsilon^i_{\ j}x^j\derpar{\phi}{x^i} \ , \\
\delta \pi(x)&=-\epsilon_{\ \beta}^{\alpha}x^\beta\derpar{\pi}{x^\alpha}-b_i\phi^i= -b_kx^k\derpar{\pi}{x^0}-\epsilon^i_{\ j}x^j\derpar{\pi}{x^i}-b_i\phi^i \ , \\
\delta\phi_0(x)&=-\epsilon_{\ \beta}^{\alpha}x^{\beta}\derpar{\phi_0}{x^\alpha}=-b_kx^k\derpar{\phi_0}{x^0}-\epsilon^k_{\ j}x^j\derpar{\phi_0}{x^k} \ ,\\ 
\delta\phi_i(x)&=-\epsilon^\nu_{\ i}\phi_\nu-\epsilon_{\ \beta}^{\alpha}x^{\beta}\derpar{\phi_i}{x^\alpha}=- b_i\phi_0-\epsilon^k_{\ i}\phi_k- b_kx^k\derpar{\phi_i}{x^0}-\epsilon^k_{\ j}x^j\derpar{\phi_i}{x^k} \ .
\end{align*}
Under these transformations the Lagrangian transforms as $\displaystyle\delta\mathscr{L}(x)=-\derpar{}{x^\alpha}\Big(\epsilon^\alpha_{\ \beta}x^\beta\mathscr{L}(x)\Big)$. 
Notice that the field $\pi(x)$ no longer transforms as a Carrollian scalar; instead, $\pi(x)$ transforms under a Carroll spacetime transformation $-\epsilon_{\ \beta}^{\alpha}x^\beta\derpar{\pi}{x^\alpha}$ plus a term $-b_i\phi^i(x)$ which cannot be represented on $E$. 
The field transformations behave geometrically as the Lie derivatives of the local sections $j^1\phi:M\rightarrow J^1E$.
The vector field generating the Carroll transformations on $J^1E$ is
\begin{align*}
\begin{split}
X&=-\epsilon^\mu_{\ \nu}x^\nu\frac{\partial}{\partial x^\mu} + b_i\phi^i\frac{\partial}{\partial \pi}+\epsilon^\nu_{\ \mu}\phi_\nu\frac{\partial}{\partial\phi_\mu}+\epsilon^\nu_{\ \mu}\pi_\nu\frac{\partial}{\partial \pi_\mu} \\
&=-b_ix^i\frac{\partial}{\partial x^0}-\epsilon^i_{\ j}x^j\frac{\partial}{\partial x^i}+ b_i\phi^i\frac{\partial}{\partial \pi}+\left(b_i\phi_0+\epsilon_{\ i}^{ j}\phi_j\right)\frac{\partial}{\partial \phi_i} +\epsilon^\nu_{\ \mu}\pi_\nu\frac{\partial}{\partial \pi_\mu}\in\vf(J^1E)\ ,
\end{split}
\end{align*}
which is tangent to the constraint submanifold $S_1\subset J^1E$, given by the constraint $\phi_0=0$,
since $\mathpzc{L}_X\phi_0=0$. 
Furthermore, although the vector field $X$ leaves the Lagrangian density invariant ($\mathpzc{L}_X\textbf{L}=0$), it does not produce an exact Noether symmetry as 
$$
\mathpzc{L}_{X}\Theta_\mathscr{L}=b^i\phi_0(\phi_i\text{d}^mx-\text{d}\phi\wedge\text{d}^{m-1}x_i) \ .
$$
Moreover, the vector field $X$ as written above on all of $J^1E$ is not $\mathscr{FL}$-projectable onto $P_0\subset J^1E^*$. Instead, $X$ on $S_1$ given by
\begin{equation}
\left.X\right\vert_{S_1}=-b_ix^i\frac{\partial}{\partial x^0}-\epsilon^i_{\ j}x^j\frac{\partial}{\partial x^i}+ b_i\phi^i\frac{\partial}{\partial \pi}+\epsilon_{\ i}^{ j}\phi_j\frac{\partial}{\partial \phi_i}+\epsilon^\nu_{\ \mu}\pi_\nu\frac{\partial}{\partial \pi_\mu}\ ,
\label{XS1}
\end{equation}
is $\mathscr{ FL}_0$-projectable onto $P_0$, giving
\begin{equation*}
Y=\mathscr{FL}_{0*}X\vert_{S_1}
=-b_ix^i\frac{\partial}{\partial x^0}-\epsilon^i_{\ j}x^j\frac{\partial}{\partial x^i}-b_ip_\phi^i\frac{\partial}{\partial \pi}+\epsilon^{ i}_{\ j}p_\phi^j\frac{\partial}{\partial p_\phi^i}\in\mathfrak{X}(P_0) \ .
\end{equation*}
Now, letting $\widetilde{X}$ be the local extension of $X\vert_{S_1}$ to $J^1E$
whose coordinate expression is given by \eqref{XS1};
then, it follows that
$\mathpzc{L}_{\widetilde{X}}\Theta_\mathscr{L}=0$ on $J^1E$.
It is thereby possible to define the momentum maps as $J_\mathscr{L}(\widetilde{X})=
-i(\widetilde{X})\Theta_\mathscr{L}\in\df^{m-1}(J^1E)$ and $J_{\mathscr{H}_0}(Y)=
-i(Y)\Theta^0_\mathscr{H}\in\df^{m-1}(P_0)$, giving:
\beann
J_\mathscr{L}(\widetilde{X})&=&-\left(\epsilon^i_{\ j}x^j\pi+b_jx^j\phi^i\right)\text{d}\phi\wedge\text{d}^{m-2}x_{0i}+\phi^i\epsilon^j_{\ k}x^k\text{d}\phi\wedge\text{d}^{m-2}x_{ij}\\
& &+\frac{1}{2}\phi_i\phi^i\left(b_kx^k\text{d}^{m-1}x_0+\epsilon^k_{\ j}x^j\text{d}^{m-1}x_k\right) \ ,
\\
J_{\mathscr{H}_0}(Y)&=&-\left(\epsilon^i_{\ j}x^j\pi-b_jx^jp_\phi^i\right)\text{d}\phi\wedge\text{d}^{m-2}x_{0i}-p_\phi^i\epsilon^j_{\ k}x^k\text{d}\phi\wedge\text{d}^{m-2}x_{ij}\\
& &+\frac{1}{2}p_{\phi i}p_\phi^i\left(b_kx^k\text{d}^{m-1}x_0+\epsilon^k_{\ j}x^j\text{d}^{m-1}x_k\right) \ .
\eeann

\section{Bosonic Strings and p-Branes}
\label{section:StringTheory}

In the next sections, the multisymplectic formalism for bosonic 
$p$-branes is given by working out explicitly the case for string theory $(p=1)$ and then discussing the generalization to $p>1$. 
The De Donder--Weyl Hamiltonian treatment for $p$-branes can be found in \cite{ngbranes}. 

The fields of interest are the embeddings of a brane worldvolume in spacetime. Spacetime $M$ is a smooth $D=(d+1)$-dimensional manifold with local coordinates $x^{\mu}$ ($\mu=0,1,...,d$) and spacetime metric $G_{\mu\nu}$ with signature $(-+\dotsb +)$.
The $p$-brane worldvolume $\Sigma$ is a smooth manifold with $\text{dim} \Sigma = m=p+1$ and local coordinates $\sigma^a$ (with  $a=0,1,...,p$). 
Given the embedding $X: \Sigma \rightarrow M: \sigma^a\mapsto x^{\mu}(\sigma)$, the embedding maps $x^\mu(\sigma)$ are taken to be fields on $\Sigma$. The configuration bundle $E$ over $\Sigma$ is taken to be the trivial bundle $E=\Sigma\times M$ and comes with a surjective projection map $\pi:E\rightarrow \Sigma$ as usual and sections of $E$ are given by $\phi:\Sigma\rightarrow\Sigma\times M: \sigma^a \mapsto (\sigma^a,x^{\mu}(\sigma))$.
The coordinates on $E$ are $(\sigma^a,x^\mu)$, 
while on $J^1E$ and $J^1E^*$ the local coordinates are $(\sigma^a,x^\mu,x_a^\mu)$ and $(\sigma^a,x^\mu,p^a_\mu)$ respectively. It follows that $\text{dim}E=m+D$ and $\text{dim}J^1E=\text{dim} J^1E^*=m+D+mD$. 
It is also worth noting that the first-order jet prolongations are given as $j^1\phi:\Sigma\rightarrow J^1E:\sigma^a\mapsto \Big(\sigma^a,x^\mu(\sigma), \derpar{x^\mu}{\sigma^a}(\sigma)\Big)$.

 The Lagrangian density is given by
\begin{equation}
\label{eq:NG-L}
\textbf{L}=\mathscr{L}(\sigma^a,x^{\mu},x^{\mu}_a) \text{d}^m\sigma= -T_p \sqrt{-\text{det} g} \ \text{d}^m\sigma= -T_p\sqrt{-\text{det}(G_{\mu\nu}x^{\mu}_a x^{\nu}_b)} \ \text{d}^m\sigma,
\end{equation}
where $T_p$ is the $p$-brane tension which has units of $[T_p]=(mass)/(length)^{p}$ and $g=\frac{1}{2}g_{ab} \text{d}\sigma^a \wedge \text{d}\sigma^b$ is a 2-form on $\Sigma$ on $J^1E$ whose pullback by jet prolongations give the induced metric on $\Sigma$: 
\begin{equation}
\label{stringmetric}
j^1\phi^* g=h\equiv \frac{1}{2}h_{ab}\text{d}\sigma^a\wedge\text{d}\sigma^b\ , \quad \text{where} \quad h_{ab}=G_{\mu\nu}\frac{\partial X^\mu}{\partial \sigma^a}\frac{\partial X^\nu}{\partial \sigma^b}\ .
\end{equation}
The 2-form $g$ can also be constructed on $J^1E^*$ by pushforward of the Legendre map, that is $\mathscr{FL}_*g=g\in\Omega^2(J^1E^*)$, so that $\psi^*g=h$.
The Legendre map $\mathscr{FL}: J^1E\rightarrow J^1E^*:(\sigma^a,x^{\mu},x^{\mu}_a)\mapsto (\sigma^a, x^{\mu},p^a_{\mu})$ gives
\begin{equation*}
\mathscr{FL}^*p^a_{\mu} = \frac{\partial\mathscr{L}}{\partial x^{\mu}_a} \ ;
\end{equation*}
the Hessian matrix
\begin{equation}
\label{pbraneHessian}
\frac{\partial^2\mathscr{L}}{\partial x_a^\mu \partial x_b^\nu}=-T_p\sqrt{-\text{det}g}\left[ G_{\mu\nu}g^{ba} -G_{\mu\alpha}G_{\beta\nu}x^\alpha_c x^\beta_i\left(g^{ba}g^{ci}+g^{cb}g^{ai}-g^{ca}g^{bi}\right)\right]\ .
\end{equation}
is non-singular and the Lagrangian is regular thusly.

The full geometric setting is illustrated in the following figure:

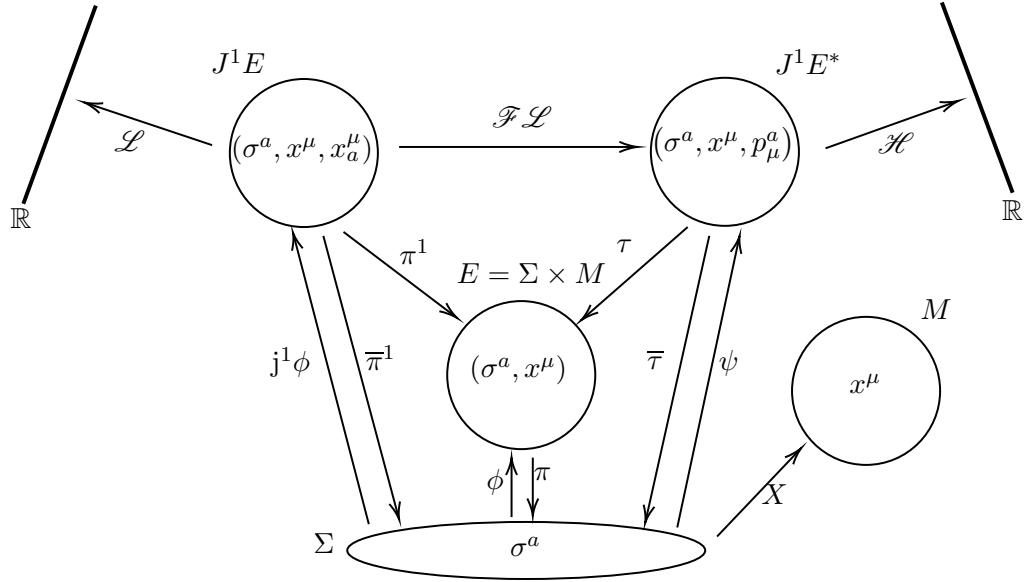
\tikzset{every picture/.style={line width=0.75pt}} 
\begin{tikzpicture}[x=0.75pt,y=0.75pt,yscale=-.8,xscale=.85]
\draw    (172,284) -- (97.88,257.67) ;
\draw [shift={(96,257)}, rotate = 379.56] [color={rgb, 255:red, 0; green, 0; blue, 0 }  ][line width=0.75]    (10.93,-3.29) .. controls (6.95,-1.4) and (3.31,-0.3) .. (0,0) .. controls (3.31,0.3) and (6.95,1.4) .. (10.93,3.29)   ;
\draw    (536,286) -- (613.13,256.71) ;
\draw [shift={(615,256)}, rotate = 519.21] [color={rgb, 255:red, 0; green, 0; blue, 0 }  ][line width=0.75]    (10.93,-3.29) .. controls (6.95,-1.4) and (3.31,-0.3) .. (0,0) .. controls (3.31,0.3) and (6.95,1.4) .. (10.93,3.29)   ;
\draw    (265,523.15) -- (220.48,341.17) ;
\draw [shift={(220,339.23)}, rotate = 436.25] [color={rgb, 255:red, 0; green, 0; blue, 0 }  ][line width=0.75]    (10.93,-3.29) .. controls (6.95,-1.4) and (3.31,-0.3) .. (0,0) .. controls (3.31,0.3) and (6.95,1.4) .. (10.93,3.29)   ;
\draw   (182.52,288.98) .. controls (182.52,263.18) and (202.18,242.27) .. (226.44,242.27) .. controls (250.7,242.27) and (270.36,263.18) .. (270.36,288.98) .. controls (270.36,314.77) and (250.7,335.68) .. (226.44,335.68) .. controls (202.18,335.68) and (182.52,314.77) .. (182.52,288.98) -- cycle ;
\draw   (432.34,288.21) .. controls (432.34,262.42) and (452,241.51) .. (476.26,241.51) .. controls (500.51,241.51) and (520.18,262.42) .. (520.18,288.21) .. controls (520.18,314.01) and (500.51,334.92) .. (476.26,334.92) .. controls (452,334.92) and (432.34,314.01) .. (432.34,288.21) -- cycle ;
\draw   (252.2,539.73) .. controls (252.2,529.8) and (299.74,521.74) .. (358.39,521.74) .. controls (417.04,521.74) and (464.58,529.8) .. (464.58,539.73) .. controls (464.58,549.67) and (417.04,557.73) .. (358.39,557.73) .. controls (299.74,557.73) and (252.2,549.67) .. (252.2,539.73) -- cycle ;
\draw   (311.5,429.1) .. controls (311.5,403.3) and (331.17,382.39) .. (355.43,382.39) .. controls (379.68,382.39) and (399.35,403.3) .. (399.35,429.1) .. controls (399.35,454.89) and (379.68,475.8) .. (355.43,475.8) .. controls (331.17,475.8) and (311.5,454.89) .. (311.5,429.1) -- cycle ;
\draw   (516.1,439.05) .. controls (516.1,413.25) and (535.77,392.34) .. (560.02,392.34) .. controls (584.28,392.34) and (603.95,413.25) .. (603.95,439.05) .. controls (603.95,464.84) and (584.28,485.75) .. (560.02,485.75) .. controls (535.77,485.75) and (516.1,464.84) .. (516.1,439.05) -- cycle ;
\draw    (249.98,338.75) -- (315.88,391.85) ;
\draw [shift={(317.43,393.11)}, rotate = 218.86] [color={rgb, 255:red, 0; green, 0; blue, 0 }  ][line width=0.75]    (10.93,-3.29) .. controls (6.95,-1.4) and (3.31,-0.3) .. (0,0) .. controls (3.31,0.3) and (6.95,1.4) .. (10.93,3.29)   ;
\draw    (454.2,335.68) -- (392.29,393.28) ;
\draw [shift={(390.82,394.64)}, rotate = 317.07] [color={rgb, 255:red, 0; green, 0; blue, 0 }  ][line width=0.75]    (10.93,-3.29) .. controls (6.95,-1.4) and (3.31,-0.3) .. (0,0) .. controls (3.31,0.3) and (6.95,1.4) .. (10.93,3.29)   ;
\draw    (362.28,481.16) -- (362.28,515.91) ;
\draw [shift={(362.28,517.91)}, rotate = 270] [color={rgb, 255:red, 0; green, 0; blue, 0 }  ][line width=0.75]    (10.93,-3.29) .. controls (6.95,-1.4) and (3.31,-0.3) .. (0,0) .. controls (3.31,0.3) and (6.95,1.4) .. (10.93,3.29)   ;
\draw    (349.68,518.68) -- (349.68,483.16) ;
\draw [shift={(349.68,481.16)}, rotate = 450] [color={rgb, 255:red, 0; green, 0; blue, 0 }  ][line width=0.75]    (10.93,-3.29) .. controls (6.95,-1.4) and (3.31,-0.3) .. (0,0) .. controls (3.31,0.3) and (6.95,1.4) .. (10.93,3.29)   ;
\draw    (282.96,285.15) -- (422.55,285.15) ;
\draw [shift={(424.55,285.15)}, rotate = 180] [color={rgb, 255:red, 0; green, 0; blue, 0 }  ][line width=0.75]    (10.93,-3.29) .. controls (6.95,-1.4) and (3.31,-0.3) .. (0,0) .. controls (3.31,0.3) and (6.95,1.4) .. (10.93,3.29)   ;
\draw    (448,525.23) -- (484.61,343.19) ;
\draw [shift={(485,341.23)}, rotate = 461.37] [color={rgb, 255:red, 0; green, 0; blue, 0 }  ][line width=0.75]    (10.93,-3.29) .. controls (6.95,-1.4) and (3.31,-0.3) .. (0,0) .. controls (3.31,0.3) and (6.95,1.4) .. (10.93,3.29)   ;
\draw    (471.62,530.93) -- (520.33,476.53) ;
\draw [shift={(521.66,475.04)}, rotate = 491.84] [color={rgb, 255:red, 0; green, 0; blue, 0 }  ][line width=0.75]    (10.93,-3.29) .. controls (6.95,-1.4) and (3.31,-0.3) .. (0,0) .. controls (3.31,0.3) and (6.95,1.4) .. (10.93,3.29)   ;
\draw    (238,341.23) -- (282.51,518.29) ;
\draw [shift={(283,520.23)}, rotate = 255.89] [color={rgb, 255:red, 0; green, 0; blue, 0 }  ][line width=0.75]    (10.93,-3.29) .. controls (6.95,-1.4) and (3.31,-0.3) .. (0,0) .. controls (3.31,0.3) and (6.95,1.4) .. (10.93,3.29)   ;
\draw    (467,341.23) -- (429.41,520.27) ;
\draw [shift={(429,522.23)}, rotate = 281.86] [color={rgb, 255:red, 0; green, 0; blue, 0 }  ][line width=0.75]    (10.93,-3.29) .. controls (6.95,-1.4) and (3.31,-0.3) .. (0,0) .. controls (3.31,0.3) and (6.95,1.4) .. (10.93,3.29)   ;
\draw [line width=1.5]    (605,194) -- (647,314) ;
\draw [line width=1.5]    (103,196) -- (60,321) ;
\draw (50.55,321.96) node [anchor=north west][inner sep=0.75pt]    {$\mathbb{R} \ $};
\draw (112.24,273.34) node [anchor=north west][inner sep=0.75pt]    {$\mathscr{L}$};
\draw (565.24,274.34) node [anchor=north west][inner sep=0.75pt]    {$\mathscr{H}$};
\draw (169.55,221.69) node [anchor=north west][inner sep=0.75pt]    {$J^{1} E$};
\draw (205.54,409.56) node [anchor=north west][inner sep=0.75pt]    {$\mathrm{j^{1} \phi }$};
\draw (504.06,221.75) node [anchor=north west][inner sep=0.75pt]    {$J^{1} E^{*}$};
\draw (590.58,379.05) node [anchor=north west][inner sep=0.75pt]    {$M$};
\draw (230.34,526.88) node [anchor=north west][inner sep=0.75pt]    {$\Sigma$};
\draw (333.03,486.24) node [anchor=north west][inner sep=0.75pt]    {$\mathrm{\phi }$};
\draw (361.2,485.47) node [anchor=north west][inner sep=0.75pt]    {$\pi $};
\draw (261.36,408.65) node [anchor=north west][inner sep=0.75pt]    {$\overline{\pi }^{1}$};
\draw (410.25,341.53) node [anchor=north west][inner sep=0.75pt]    {$\tau $};
\draw (470.95,411.44) node [anchor=north west][inner sep=0.75pt]    {$\psi $};
\draw (496.24,496.19) node [anchor=north west][inner sep=0.75pt]    {$X$};
\draw (315.46,356.9) node [anchor=north west][inner sep=0.75pt]    {$E=\Sigma \times M$};
\draw (336.24,258.07) node [anchor=north west][inner sep=0.75pt]    {$\mathscr{FL}$};
\draw (181.61,275.27) node [anchor=north west][inner sep=0.75pt]    {$\left( \sigma ^{a} ,x^{\mu } ,x^{\mu }_{a}\right)$};
\draw (430.55,270.56) node [anchor=north west][inner sep=0.75pt]    {$\left( \sigma ^{a} ,x^{\mu } ,p^{a}_{\mu }\right)$};
\draw (322.98,414.98) node [anchor=north west][inner sep=0.75pt]    {$\left( \sigma ^{a} ,x^{\mu }\right)$};
\draw (347.08,529) node [anchor=north west][inner sep=0.75pt]    {$\sigma ^{a}$};
\draw (548.45,426.4) node [anchor=north west][inner sep=0.75pt]    {$x^{\mu }$};
\draw (281.36,341.65) node [anchor=north west][inner sep=0.75pt]    {$\pi ^{1}$};
\draw (428.36,411.65) node [anchor=north west][inner sep=0.75pt]    {$\overline{\tau }$};
\draw (639,317) node [anchor=north west][inner sep=0.75pt]    {$\mathbb{R} \ $};
\end{tikzpicture}

\subsection{The String Lagrangian Formulation}

The Lagrangian for the Nambu--Goto string ($p=1$) is
\begin{equation*}
\textbf{L}=\mathscr{L}(\sigma^a, x^{\mu}, x^{\mu}_a) \text{d}^2\sigma= -T \sqrt{-\text{det} g} \ \text{d}^2\sigma \ ,
\end{equation*}
from which it follows that
\begin{equation}
\label{eq:dLdv}
\frac{\partial\mathscr{L}}{\partial x_a^\mu}=-T\sqrt{-\text{det}g}\ G_{\mu\nu}g^{ba}x_b^\nu=\frac{T}{\sqrt{-\text{det}g}}\ \epsilon^{bd}\epsilon^{ac}g_{dc}G_{\mu\nu}x^\nu_b \ ,
\end{equation}
where $g^{ba}\equiv(g^{-1})^{ba}= \frac{1}{\text{det}g}\epsilon^{bc}\epsilon^{ad}g_{dc}$ 
and so it follows that the Hessian can be written as
\begin{equation*}
\frac{\partial^2\mathscr{L}}{\partial x_a^\mu \partial x_b^\nu}=-T\sqrt{-\text{det}g}\left[ G_{\mu\nu}g^{ba} -G_{\mu\alpha}G_{\beta\nu}x^\alpha_c x^\beta_i\left(g^{ba}g^{ci}+g^{cb}g^{ai}-g^{ca}g^{bi}\right)\right]\ .
\end{equation*}
The Lagrangian energy is
\begin{equation}
\label{eq:stringLenergy}
E_{\mathscr{L}}\equiv \frac{\partial \mathscr{L}}{\partial x_a^\mu}x_a^\mu - \mathscr{L}=-T \sqrt{-\text{det} g}\ (g^{ba}g_{ab}-1)=-T \sqrt{-\text{det} g}\in C^\infty(J^1E)\ ,
\end{equation}
where the last equality follows as $g_{ab}$ is a $2\times2$ matrix in the case of the string.
Using (\ref{eq:dLdv}) and (\ref{eq:stringLenergy}),
the Poincar\'e--Cartan $2$-from and multisymplectic  Poincar\'e--Cartan $3$-form on $J^1E$ are
\beann
\Theta_\mathscr{L}&=&
\frac{\partial \mathscr{L}}{\partial x_a^\mu} \ \text{d}x^\mu\wedge \text{d}^{1}\sigma_a-E_\mathscr{L} \wedge\text{d}^2\sigma=
-T\sqrt{-\text{det}g} \left[G_{\mu\nu}g^{ba}x_b^\nu\text{d}x^\mu\wedge\text{d}^1\sigma_a-\text{d}^2\sigma\right] \ , 
\\
\Omega_\mathscr{L}&=&
T\Bigg\{\sqrt{-\text{det}g}\left[ G_{\mu\nu}g^{ba} -G_{\mu\alpha}G_{\beta\nu}x^\alpha_c x^\beta_i\left(g^{ba}g^{ci}+g^{cb}g^{ai}-g^{ca}g^{bi}\right)\right] \text{d}x^b_\nu\wedge\text{d}x^\mu\wedge\text{d}^1\sigma_a\\
& &+\derpar{}{x^\rho}\left(\sqrt{-\text{det}g}\, G_{\mu\nu}g^{ba}x^\nu_b\right)\text{d}x^\rho\wedge\text{d}x^\mu\wedge\text{d}^1\sigma_a\\
& &- \sqrt{-\text{det}g}\, \left[G_{\mu\nu}g^{ba} -G_{\mu\alpha}G_{\beta\nu}x^\alpha_c x^\beta_i\left(g^{ba}g^{ci}+g^{cb}g^{ai}-g^{ca}g^{bi}\right) \right]x^\mu_a\text{d}x^\nu_b\wedge\text{d}^2\sigma \\ 
& & -\left[
\derpar{}{\sigma^a}\left(\sqrt{-\text{det}g}\, G_{\mu\nu}g^{ba}x^\nu_b\right)
+\frac{1}{2}\sqrt{-\text{det}g}\,g^{ba} x^\alpha_ax^\beta_b\partial_{\mu} G_{\alpha\beta}
\right]\text{d}x^\mu\wedge\text{d}^2\sigma\Bigg\} \ .
\eeann

The field equations, $i(\textbf{X}_{\mathscr{L}})\Omega_{\mathscr{L}}=0$, can be computed using a representative of a class of $\bar{\pi}^1$-transverse and locally decomposable multivector fields $\textbf{X}_{\mathscr{L}}\in \mathfrak{X}^2(J^1E)$, 
written as $\textbf{X}_{\mathscr{L}}=X_0\wedge X_1$, with components given by the following local expression
\begin{equation*}
X_{a}=\frac{\partial}{\partial \sigma^a}+B_a^\mu\frac{\partial}{\partial x^{\mu}} + H_{ab}^\nu \frac{\partial}{\partial x_b^{\nu}},
\end{equation*}
from which it follows that $i(\textbf{X}_{\mathscr{L}})\text{d}^2\sigma=1$. Then, $i(\textbf{X}_{\mathscr{L}})\Omega_{\mathscr{L}}=0$ is given as
\begin{align*}
\begin{split}
i(\textbf{X}_{\mathscr{L}})\Omega_{\mathscr{L}}=\ &
T \Bigg\{\sqrt{-\text{det}g}\left[ G_{\mu\nu}g^{ba} -G_{\mu\alpha}G_{\beta\nu}x^\alpha_c x^\beta_i\left(g^{ba}g^{ci}+g^{cb}g^{ai}-g^{ca}g^{bi}\right)\right] \left(B_a^\mu - x_a^\mu \right)\text{d}x_b^\nu \\
&+ B^\rho_a\left[\derpar{}{x^\rho}\left(\sqrt{-\text{det}g}\, G_{\mu\nu}g^{ba}x^\nu_b\right)-\derpar{}{x^\mu}\left(\sqrt{-\text{det}g}\, G_{\rho\nu}g^{ba}x^\nu_b\right)\right]\text{d}x^\mu\\
&+\sqrt{-\text{det}g}\left[ G_{\mu\nu}g^{ba} -G_{\mu\alpha}G_{\beta\nu}x^\alpha_c x^\beta_i\left(g^{ba}g^{ci}+g^{cb}g^{ai}-g^{ca}g^{bi}\right)\right] H_{ab}^\nu\text{d}x^\mu\\
&+\left[\frac{1}{2}\sqrt{-\text{det}g}\, g^{ba}x_a^\alpha x_b^\beta\derpar{G_{\alpha\beta}}{x^\mu}+\derpar{}{\sigma^a}\left(\sqrt{-\text{det}g}\, G_{\rho\nu}g^{ba}x^\nu_b\right)x^\rho_a\right]\text{d}x^\mu \\
&-\bigg[\derpar{}{x^\rho}\left(\sqrt{-\text{det}g}\, G_{\mu\nu}g^{ba}x^\nu_bb\right)x_a^\mu B_k^\rho\\
&+ \frac{1}{2}\sqrt{-\text{det}g}\, g^{ba} x_a^\alpha x_b^\beta B_k^\rho\derpar{G_{\alpha\beta}}{x^\rho} +\derpar{}{\sigma^a}\left(\sqrt{-\text{det}g}\, G_{\rho\nu}g^{ba}x^\nu_b\right)B_k^\rho\\
&+\sqrt{-\text{det}g}\left[ G_{\mu\nu}g^{ba} -G_{\mu\alpha}G_{\beta\nu}x^\alpha_c x^\beta_i\left(g^{ba}g^{ci}+g^{cb}g^{ai}-g^{ca}g^{bi}\right)\right] \left(x_a^\mu H^\nu_{kb}-B_a^\mu H^\nu_{kb}+B_k^\mu H^\nu_{ab}\right)\\
&-\derpar{}{x^\rho}\left(\sqrt{-\text{det}g}\, G_{\mu\nu}g^{ba}x^\nu_b\right)\left(B_a^\mu B_k^\rho-B_k^\mu B_a^\rho\right)\bigg] \text{d}\sigma^k\Bigg\}=0\ .
\end{split}
\end{align*}
Setting differential forms separately equal to zero produces the following equations:
\bea
0&=&B^\rho_a\left[\derpar{}{x^\rho}\left(\sqrt{-\text{det}g}\, G_{\mu\nu}g^{ba}x^\nu_b\right)-\derpar{}{x^\mu}\left(\sqrt{-\text{det}g}\, G_{\rho\nu}g^{ba}x^\nu_b\right)\right]\
\nonumber \\
&&+\sqrt{-\text{det}g}\left[ G_{\mu\nu}g^{ba} -G_{\mu\alpha}G_{\beta\nu}x^\alpha_c x^\beta_i\left(g^{ba}g^{ci}+g^{cb}g^{ai}-g^{ca}g^{bi}\right)\right] H_{ab}^\nu
\nonumber \\
&&+\left[\frac{1}{2}\sqrt{-\text{det}g}\, g^{ba}x_a^\alpha x_b^\beta\derpar{G_{\alpha\beta}}{x^\mu}+\derpar{}{\sigma^a}\left(\sqrt{-\text{det}g}\, G_{\rho\nu}g^{ba}x^\nu_b\right)x^\rho_a\right]
\label{eq:stringfieldeqa} 
\\
0&=&\sqrt{-\text{det}g}\left[ G_{\mu\nu}g^{ba} -G_{\mu\alpha}G_{\beta\nu}x^\alpha_c x^\beta_i\left(g^{ba}g^{ci}+g^{cb}g^{ai}-g^{ca}g^{bi}\right)\right] \left(B_a^\mu - x_a^\mu \right) \ ,
 \label{eq:stringfieldeqb}
\\
0&=&\derpar{}{x^\rho}\left(\sqrt{-\text{det}g}\, G_{\mu\nu}g^{ba}x^\nu_bb\right)x_a^\mu B_k^\rho
\nonumber \\
&&+ \frac{1}{2}\sqrt{-\text{det}g}\, g^{ba} x_a^\alpha x_b^\beta B_k^\rho\derpar{G_{\alpha\beta}}{x^\rho} +\derpar{}{\sigma^a}\left(\sqrt{-\text{det}g}\, G_{\rho\nu}g^{ba}x^\nu_b\right)B_k^\rho
\nonumber \\
&&+\sqrt{-\text{det}g}\left[ G_{\mu\nu}g^{ba} -G_{\mu\alpha}G_{\beta\nu}x^\alpha_c x^\beta_i\left(g^{ba}g^{ci}+g^{cb}g^{ai}-g^{ca}g^{bi}\right)\right] \left(x_a^\mu H^\nu_{kb}-B_a^\mu H^\nu_{kb}+B_k^\mu H^\nu_{ab}\right)
\nonumber \\
&&-\derpar{}{x^\rho}\left(\sqrt{-\text{det}g}\, G_{\mu\nu}g^{ba}x^\nu_b\right)\left(B_a^\mu B_k^\rho-B_k^\mu B_a^\rho\right)\ .
 \label{eq:stringfieldeqc}
\eea
Notice that plugging  (\ref{eq:stringfieldeqb}) into  (\ref{eq:stringfieldeqc}) implies  (\ref{eq:stringfieldeqa}), which means that  (\ref{eq:stringfieldeqc}) is an identity as expected. Equation (\ref{eq:stringfieldeqb}) is the {\sc sopde} condition for $\textbf{X}_\mathscr{L}$, i.e. $B_a^\mu=x_a^\mu$. Furthermore, (\ref{eq:stringfieldeqa}) becomes the Euler--Lagrange equations when working with $j^1\phi=\left(\sigma^a,x^\mu(\sigma),\derpar{ x^\mu}{\sigma^a}\right)$ that are holonomic sections of $\textbf{X}_\mathscr{L}$ which satisfy
\begin{equation}
\label{eq:stringholointsec}
D_a^\mu=x_a^\mu=\derpar{x^\mu}{\sigma^a} \quad , \quad H_{ab}^\nu=\derpar{x_b^\nu}{\sigma^a}=
\frac{\partial^2x^\nu}{\partial\sigma^a\partial\sigma^b}\quad , \quad g_{ab}=h_{ab}= G_{\mu\nu}\derpar{x^\mu}{\sigma^a}\derpar{x^\nu}{\sigma^b} \ .
\end{equation}
Working with such sections that satisfy (\ref{eq:stringholointsec}) it follows that (\ref{eq:stringfieldeqa}) becomes
\beann
& &-\sqrt{-\text{det}h}\, h^{ba}\derpar{x^\alpha}{\sigma^a}\derpar{x^\beta}{\sigma^b}\derpar{G_{\alpha\beta}}{x^\mu}+\derpar{}{\sigma^a}\left(\sqrt{-\text{det}h}\, G_{\mu\nu}h^{ba}\derpar{x^\nu}{\sigma^b}\right)+\derpar{}{x^\rho}\left(\sqrt{-\text{det}h}\, G_{\mu\nu}h^{ba}\derpar{x^\nu}{\sigma^b}\right)\derpar{x^\rho}{\sigma^a}
\\ 
& &+\sqrt{-\text{det}h}\left[ G_{\mu\nu}h^{ba} -G_{\mu\alpha}G_{\beta\nu}\derpar{x^\alpha}{\sigma^c}\derpar{x^\beta}{\sigma^i}\left(h^{ba}h^{ci}+h^{cb}h^{ai}-h^{ca}h^{bi}\right)\right]\frac{\partial^2x^\nu}{\partial\sigma^a\partial\sigma^b}=0 \ ,
\eeann
which are the Euler--Lagrange equations for this field theory.

\subsection{The String Hamiltonian Formulation}

The De Donder--Weyl Hamiltonian formalism is developed starting from the Legendre map $\mathscr{FL}:J^1E\rightarrow J^1E^*$ which satisfies
\beq
\label{eq:stringLrans}
\mathscr{FL}^*\sigma^a=\sigma^a \quad , \quad
\mathscr{FL}^*x^\mu=x^\mu \quad , \quad
\mathscr{FL}^*p^a_{\mu}=-T \sqrt{-\text{det} g}\ G_{\mu\nu}g^{ba}x_b^{\nu} \ .
\eeq
This Legendre map is invertible as expected from the regularity of the generalized Hessian (\ref{pbraneHessian}).
The De Donder--Weyl Hamiltonian is defined as 
\begin{equation*}
\mathscr{H}(\sigma^a,x^{\mu},p^a_{\mu})\equiv p^a_{\mu}(\mathscr{FL}^{-1})^* x_a^{\mu} - (\mathscr{FL}^{-1})^*\mathscr{L} \in C^\infty(J^1E^*)\ .
\end{equation*}
This Hamiltonian can be expressed in terms of the variables on $J^1E^*$ as \cite{ngbranes}
\begin{equation*}
\Pi^{ab}\equiv G^{\mu\nu}p^a_\mu p^b_\nu\ 
\Rightarrow \ \mathscr{FL}^*\Pi^{ab}=-T^2 \text{det}g\ g^{ba} \
\Longleftrightarrow \ \mathscr{FL}^*\text{det}\Pi=(-T^2\text{det}g)^2\text{det}(g^{-1})=T^4\text{det}g \ ,
\end{equation*}
from which it follows that
\begin{equation*}
(\mathscr{FL}^{-1})^*x_b^\nu=-\frac{1}{T}\sqrt{-\text{det}\Pi}\ G^{\mu\nu}\Pi_{ab} p^a_\mu \ \Longrightarrow \ 
\mathscr{H}(\sigma^a, x^\mu, p^a_\mu)= -\frac{1}{T}\sqrt{-\text{det}\Pi}\ , 
\end{equation*}
where
\begin{equation}
\label{invPi}
\Pi_{ab}=\frac{1}{\text{det}\Pi}\epsilon_{cb}\epsilon_{da}\Pi^{cd} \ .
\end{equation}
The Hamilton--Cartan $2$--form and the multisymplectic Hamilton--Cartan $3$-form on $J^1E^*$ are
\begin{subequations}
\beann
\Theta_{\mathscr{H}}&=&p^a_{\mu} \text{d}x^{\mu}\wedge \text{d}^1\sigma_a - \mathscr{H}\wedge\text{d}^2\sigma=p^a_\mu\text{d}x^\mu\wedge\text{d}^1\sigma_a+\frac{1}{T}\sqrt{-\text{det}\Pi}\ \text{d}^2\sigma \label{eq:stringOmegHa}\ , \\
\Omega_{\mathscr{H}}&=&
-\text{d}p^a_{\mu}\wedge \text{d}x^{\mu}\wedge \text{d}^1\sigma_a+\text{d}\mathscr{H}\wedge\text{d}^2\sigma \\
&=&-\text{d}p^a_{\mu}\wedge \text{d}x^{\mu}\wedge \text{d}^1\sigma_a-\frac{\sqrt{-\text{det}\Pi}}{T} \Pi_{ba}\left(\frac{1}{2}\partial_\mu G^{\rho\sigma}p^a_\rho p^b_\sigma \text{d}x^\mu + G^{\mu\nu}p^b_\nu\text{d}p^a_\mu\right)\wedge\text{d}^2\sigma \ .
\eeann
\end{subequations}

The field equations, $i(\textbf{X}_{\mathscr{H}})\Omega_{\mathscr{H}}=0$, can be computed using a representative of a class of $\bar{\pi}^1$-transverse
locally decomposable multivector fields 
$\textbf{X}_{\mathscr{H}}=X_0\wedge X_1\in \mathfrak{X}^2(J^1E^*)$, with components
given by the following local expression
\begin{equation*}
X_a=\frac{\partial}{\partial\sigma^a}+D_a^\mu\frac{\partial}{\partial x^\mu}+ H^c_{a\mu}\frac{\partial}{\partial p^c_\mu} \ .
\end{equation*}
Then,
\beann
0=i(\textbf{X}_\mathscr{H})\Omega_\mathscr{H}&=&
-\left(H^a_{a\mu}-\frac{\sqrt{-\text{det}\Pi}}{2T}\Pi_{ba} \derpar{G^{\rho\sigma}}{x^\mu} p^a_\rho p^b_\sigma\right)\text{d}x^\mu+\left(D_a^\mu+\frac{\sqrt{-\text{det}\Pi}}{T}\Pi_{ba}G^{\mu\nu}p^b_\nu\right)\text{d}p^a_\mu \\ & &
-\left[D_a^\mu H^a_{k\mu}-D_k^\mu H^a_{a\mu}+\frac{\sqrt{-\text{det}\Pi}}{T}\Pi_{ba}\left(\frac{1}{2}D_k^\mu\derpar{G^{\rho\sigma}}{x^\mu}p^a_\rho p^b_\sigma+H^a_{k\mu}G^{\mu\nu}p^b_\nu\right)\right]\text{d}\sigma^k \ .
\eeann
Setting differential forms separately equal to zero produces the following field equations:
\begin{subequations}
\label{stringHfieldeq}
\begin{align}
&H^a_{a\mu}-\frac{\sqrt{-\text{det}\Pi}}{2T}\Pi_{ba} \derpar{G^{\rho\sigma}}{x^\mu}p^a_\rho p^b_\sigma=0, \label{stringHfieldeqa} \\
&D_a^\mu+\frac{\sqrt{-\text{det}\Pi}}{T}\Pi_{ba}G^{\mu\nu}p^b_\nu=0, \label{stringHfieldeqb} \\
&D_a^\mu H^a_{k\mu}-D_k^\mu H^a_{a\mu}+\frac{\sqrt{-\text{det}\Pi}}{T}\Pi_{ba}\left(\frac{1}{2}D_k^\mu\derpar{G^{\rho\sigma}}{x^\mu}p^a_\rho p^b_\sigma+H^a_{k\mu}G^{\mu\nu}p^b_\nu\right)=0. \label{stringHfieldeqc}
\end{align}
\end{subequations}
Note that plugging (\ref{stringHfieldeqb}) into (\ref{stringHfieldeqc})
implies (\ref{stringHfieldeqa}), which means that (\ref{stringHfieldeqc}) is an identity as expected. 
Furthermore, the variational problem on $J^1E^*$ is solved by integral sections 
$\psi(\sigma)=(\sigma^a,x^\mu(\sigma), p^a_\mu(\sigma))$ of $\textbf{X}_\mathscr{H}$ for which $D_a^{\mu}=\derpar{x^\mu}{\sigma^a} , \ H_{a\mu}^{c}=\derpar{ p^c_\mu}{\sigma^a}$, 
and so the field equations become
\beann
&\displaystyle \derpar{p^a_{\mu}}{\sigma^a}-\frac{\sqrt{-\text{det}\Pi}}{2T}\Pi_{ba} \partial_\mu G^{\rho\sigma}p^a_\rho p^b_\sigma=0\ , 
\\
& \displaystyle \derpar{x^\mu}{\sigma^a}+\frac{\sqrt{-\text{det}\Pi}}{T}\Pi_{ba}G^{\mu\nu}p^b_\nu=0\ , 
\eeann
which are the corresponding Hamilton--De Donder--Weyl equations for the bosonic string; these equations can be plugged into another one to give the Euler--Lagrange equations as usual.

\subsection{Symmetries}

\subsubsection{Worldsheet Diffeomorphisms}

Consider worldsheet diffeomorphisms produced by
\begin{equation}
\label{eq:stringdiff}
\tilde{\sigma}^a =\sigma^a+\xi^a\ \Rightarrow\ \frac{\partial \tilde{\sigma}^a}{\partial \sigma^b}=\delta^a_b+\partial_b\xi^a \ .
\end{equation}
It follows that 
\begin{align*}
\begin{split}
\text{det}\tilde{g}=\left[ \text{det}\left(\frac{\partial \tilde{\sigma}^a}{\partial \sigma^b}\right)\right]^{-2}\text{det}g \quad ,\quad\text{d}^2\tilde{\sigma}=\text{det}\left(\frac{\partial \tilde{\sigma}^a}{\partial \sigma^b}\right)\text{d}^2\sigma \ ,
\end{split}
\end{align*}
and hence
\begin{equation*}
\textbf{L}=-T\sqrt{-\text{det}\tilde{g}}\ \text{d}^2\tilde{\sigma}=-T\sqrt{-\text{det}g}\ \text{d}^2\sigma\ .
\end{equation*}
The vector field $\xi\in\vf(\Sigma)$ which generates the worldsheet diffeomorphisms (\ref{eq:stringdiff}) is given by
\begin{equation*}
\xi=-\xi^a\derpar{}{\sigma^a}\in\vf(\Sigma)\ .
\end{equation*}
The lift of the vector field $\xi\in\mathfrak{X}(\Sigma)$ to $E=\Sigma\times M$,
obtained from \eqref{LieDerivSections1}, is 
\begin{equation*}
\xi_E=-\xi^a\frac{\partial}{\partial\sigma^a}\in\mathfrak{X}(E)\ ,
\end{equation*}
and the canonical lift of $\xi_E$ to $J^1E$ 
gives
\begin{equation*}
X_\xi=-\xi^a\frac{\partial}{\partial\sigma^a}+x_b^\mu\derpar{\xi^b}{\sigma^a}\frac{\partial}{\partial x_a^\mu} \ \in\mathfrak{X}(J^1E) \ .
\end{equation*}
The resulting field variation given by the Lie derivative (\ref{LieDerivSections1}) of the local sections $\phi:\Sigma\rightarrow E$ by the vector field $\xi$ is
\begin{equation*}
\delta X^\mu(\sigma)=-\xi^a\derpar{X^\mu}{\sigma^a}\ .
\end{equation*}
It follows that 
\begin{align*}
\mathpzc{L}_{X_\xi}\textbf{L}= 0 \ \Rightarrow\ \mathpzc{L}_{X_\xi}\Theta_\mathscr{L}=0 \ ,
\end{align*}
so the covariant momentum map is given as 
\begin{equation*}
J_\mathscr{L}(X_\xi)=-i(X_\xi)\Theta_\mathscr{L}=-T\sqrt{-\text{det}g}\left( \epsilon_{ac}\xi^cG_{\mu\nu}g^{ba}x_b^\nu\text{d}x^\mu+\xi^a\text{d}^1\sigma_a\right) \ \in \df^1(J^1E) \ .
\end{equation*}

In the Hamiltonian formalism, the vector field $Y_\xi=\mathscr{FL}_*X_\xi\in\vf(J^1E^*)$ which takes the form
\begin{equation*}
Y_\xi=-\xi^a\frac{\partial}{\partial \sigma^a}-\frac{1}{T}\sqrt{-\text{det}\Pi}\ G^{\mu\nu}\Pi_{bc}p^b_\nu\derpar{\xi^c}{\sigma_a}\frac{\partial}{\partial p^\mu_a}\ ,
\end{equation*}
generates the worldsheet diffeomorphisms on the multimomentum phase space $J^1E^*$ and is used to construct the covariant momentum map 
\begin{equation*}
J_\mathscr{H}(Y_\xi)=-i(Y_\xi)\Theta_\mathscr{H}=\epsilon_{ac}\xi^c p^a_\mu\text{d}x^\mu-\frac{1}{T}\sqrt{-\text{det}\Pi}\ \xi^a\text{d}^1\sigma_a\in\df^1(J^1E^*)\ .
\end{equation*}

\subsubsection{Spacetime Isometries}

Spacetime symmetries, which arise from performing transformations on the coordinates of $M$, are gauge symmetries in string theory since they are generated by vector fields on the configuration manifold $E$ which are vertical with respect to the projection onto the base space $\Sigma$.

Consider infinitesimal spacetime diffeomorphisms $x^\mu\rightarrow x^\mu+\zeta^\mu(x)$ generated by the vector field
\begin{equation*}
\zeta=-\zeta^\mu\frac{\partial}{\partial x^\mu}\in \mathfrak{X}(M) \ ,
\end{equation*}
which is a gauge transformation as the corresponding vector field $\zeta_E\in \mathfrak{X}(E)$ is $\pi$-vertical:
\begin{equation*}
\zeta_E=-\zeta^\mu\frac{\partial}{\partial x^\mu} \ .
\end{equation*}
The canonical lift to $J^1E$, given by the jet prolongation of $\zeta_E$, is
\begin{equation*}
X_\zeta=-\zeta^\mu\frac{\partial}{\partial x^\mu}-x^\nu_a\partial_\nu\zeta^\mu\frac{\partial}{\partial x^\mu_a}\in\mathfrak{X}(J^1E)\ .
\end{equation*}
The Lie derivative of $\textbf{L}=-T\sqrt{-\text{det}g}\ \text{d}^2\sigma$ with respect to $X_\zeta$ is
\begin{align*}
\mathpzc{L}_{X_\zeta}\textbf{L}=\frac{T}{2}\sqrt{-\text{det}g}\ g^{ba}x^\nu_bx^\mu_a\mathpzc{L}_\zeta G_{\mu\nu}\text{d}^2\sigma\ ,
\end{align*}
where $\mathpzc{L}_\zeta G_{\mu\nu}$ is the Lie derivative of $G_{\mu\nu}$ on $M$ with respect to 
$\displaystyle\zeta=-\zeta^\mu\derpar{}{x^\mu}\in\mathfrak{X}(M)$. Then, when $\zeta$ is a Killing vector field, 
\begin{align*}
\label{stringLagLie}
\mathpzc{L}_{X_\zeta}\textbf{L}=0 \ \Rightarrow\  \mathpzc{L}_{X_\zeta}\Theta_\mathscr{L}=0 \ ,
\end{align*}
and it therefore follows that the covariant momentum map is given by 
\begin{equation*}
J_\mathscr{L}(X_\zeta)=-i(X_\zeta)\Theta_\mathscr{L}=T\sqrt{-\text{det}g}\ G_{\mu\nu}G^{ba}x^\nu_b\xi^\mu\text{d}^1\sigma_a\in\df^1(J^1E)\ .
\end{equation*}

On the Hamiltonian side,
\begin{equation*}
Y_\zeta=\mathscr{FL}_* X_\zeta=-\xi^\mu\frac{\partial}{\partial x^\mu}+\frac{1}{T}\sqrt{-\text{det}\Pi}\, G^{\mu\nu}\Pi_{ba}\, p^b_\mu\,\derpar{\xi_\mu}{x^\nu}\frac{\partial}{\partial p^a_\mu}
\in\vf(J^1E^*)\ ,
\end{equation*}
so the covariant momentum map is given by 
\begin{equation*}
J_\mathscr{H}(Y_\zeta)=-i(Y_\zeta)\Theta_\mathscr{H}=-p^a_\mu\,\zeta^\mu\,\text{d}^1\sigma_a\in\df^1(J^1E^*)\ .
\end{equation*}

\subsection{Generalization to p-Branes (p \textgreater 1)}

The generalization to $p$-branes ($m=p+1$) is straightforward to perform by using the general Lagrangian density (\ref{eq:NG-L}) 
for which the corresponding Lagrangian energy $E_\mathscr{L}\in C^\infty(J^1E)$ is 
\begin{equation*}
E_{\mathscr{L}}\equiv \frac{\partial \mathscr{L}}{\partial x_a^\mu}x_a^\mu - \mathscr{L}=-T_p \sqrt{-\text{det} g}\ (g^{ba}g_{ab}-1)=-pT_p \sqrt{-\text{det} g}\ ,
\end{equation*}
where $g_{ab}$ is now an $m\times m$ matrix, so $g^{ba}g_{ab}=m=p+1$. Then
\beann
\Theta_\mathscr{L}&=&-T_p\sqrt{-\text{det}g}\ \left[G_{\mu\nu}g^{ba}x_b^\nu\text{d}x^\mu\wedge\text{d}^{m-1}\sigma_a-p\text{d}^m\sigma\right],\\
\Omega_\mathscr{L}&=&
T_p\Bigg\{\sqrt{-\text{det}g}\left[ G_{\mu\nu}g^{ba} -G_{\mu\alpha}G_{\beta\nu}x^\alpha_c x^\beta_i\left(g^{ba}g^{ci}+g^{cb}g^{ai}-g^{ca}g^{bi}\right)\right] \text{d}x^b_\nu\wedge\text{d}x^\mu\wedge\text{d}^{m-1}\sigma_a\\
& &+\derpar{}{x^\rho}\left(\sqrt{-\text{det}g}\, G_{\mu\nu}g^{ba}x^\nu_b\right)\text{d}x^\rho\wedge\text{d}x^\mu\wedge\text{d}^{m-1}\sigma_a\\
& &- \sqrt{-\text{det}g}\, \left[G_{\mu\nu}g^{ba} -G_{\mu\alpha}G_{\beta\nu}x^\alpha_c x^\beta_i\left(g^{ba}g^{ci}+g^{cb}g^{ai}-g^{ca}g^{bi}\right) \right]x^\mu_a\text{d}x^\nu_b\wedge\text{d}^m\sigma \\ 
& & -\left[
\derpar{}{\sigma^a}\left(\sqrt{-\text{det}g}\, G_{\mu\nu}g^{ba}x^\nu_b\right)
+\frac{1}{2}\sqrt{-\text{det}g}\,g^{ba} x^\alpha_ax^\beta_b\partial_{\mu} G_{\alpha\beta}
\right]\text{d}x^\mu\wedge\text{d}^m\sigma\Bigg\} \ .
\eeann
Proceeding similarly as above, take an $m$-multivector field 
$\textbf{X}_\mathscr{L}=X_0\wedge X_1\wedge\dotsb\wedge X_{m-1}$,
 with components 
\begin{equation*}
X_{a}=\frac{\partial}{\partial \sigma^a}+B_a^\mu\frac{\partial}{\partial x^{\mu}} + H_{ab}^\nu \frac{\partial}{\partial x_b^{\nu}}\ ,
\end{equation*}
to obtain the field equations, one of which is an identity, another gives the {\sc sopde} condition $B^\mu_a=x^\mu_a$, and the third equation upon plugging in the {\sc sopde} condition is
\beann
\derpar{}{x^\mu}\left(\sqrt{-\text{det}g}\right)-\derpar{}{\sigma^a}\left(\sqrt{-\text{det}g}\ G_{\mu\nu}g^{ba}x^\nu_b\right)-\derpar{}{x^\rho}\left(\sqrt{-\text{det}g}\ G_{\mu\nu}g^{ba}x^\nu_b\right)x^\rho_a
\\ -\sqrt{-\text{det}g}\left[ G_{\mu\nu}g^{ba} -G_{\mu\alpha}G_{\beta\nu}x^\alpha_c x^\beta_i\left(g^{ba}g^{ci}+g^{cb}g^{ai}-g^{ca}g^{bi}\right)\right]      H^\nu_{ab}=0 \ ,
\eeann
which becomes the Euler--Lagrange equations when working on integral sections of $\textbf{X}_\mathscr{L}$:
\beann
\derpar{}{x^\mu}\left(\sqrt{-\text{det}h}\right)-\derpar{}{\sigma^a}\left(\sqrt{-\text{det}h}\ G_{\mu\nu}h^{ba}\derpar{x^\nu}{\sigma^b}\right)-\derpar{}{x^\rho}\left(\sqrt{-\text{det}h}\ G_{\mu\nu}h^{ba}\derpar{x^\nu}{\sigma^b}\right)\derpar{x^\rho}{\sigma^a}
\\ -\sqrt{-\text{det}h}\left[ G_{\mu\nu}h^{ba} -G_{\mu\alpha}G_{\beta\nu}\derpar{x^\alpha}{\sigma^c} \derpar{x^\beta}{\sigma^i}\left(h^{ba}h^{ci}+h^{cb}h^{ai}-h^{ca}h^{bi}\right)\right] \derpar{^2x^\nu}{\sigma^a\partial\sigma^b} =0 \ .
\eeann

The Hamiltonian formulation is also carried out as in the string case with the Legendre map $\mathscr{FL}$ which, for $p$-branes, is
\begin{equation*}
\mathscr{FL}^*\sigma^a=\sigma^a \quad , \quad
\mathscr{FL}^*x^\mu=x^\mu \quad , \quad
\mathscr{FL}^*p^a_\mu=\frac{\partial\mathscr{L}}{\partial x^\mu_a}=-T_p\sqrt{-\text{det}g}\ G_{\mu\nu}g^{ba}x^\nu_b \ .
\end{equation*}
Next, define the matrix
\begin{equation*}
\Pi^{ab}=G^{\mu\nu}p^a_\mu p^b_\nu\ \Rightarrow\ \mathscr{FL}^*\Pi^{ab}=-T_p^2 \text{det}g\ g^{ba},
\end{equation*}
from which it follows that
\begin{equation*}
\left(\mathscr{FL}^{-1}\right)^*\text{det}g=-\frac{1}{T_p^2}\left(\frac{\text{det}\Pi}{-T_p^2}\right)^{1/p} \ ,
\end{equation*}
and hence
\begin{equation*}
\left(\mathscr{FL}^{-1}\right)^*x^\nu_b=-\sqrt{\left(\frac{\text{det}\Pi}{-T_p^2}\right)^{1/p}}G^{\mu\nu}\Pi_{ab}p^a_\mu.
\end{equation*}
The inverse matrix $\Pi_{ab}$ is given by the following general version of the identity (\ref{invPi}):
\begin{equation*}
\Pi_{ab}=-\frac{1}{p!\text{det}\Pi}\epsilon_{aa_2\dotsm a_m}\epsilon_{bb_2\dotsm b_m}\Pi^{b_2a_2}\dotsm\Pi^{b_ma_m} \ .
\end{equation*}
Then the De Donder--Weyl Hamiltonian for the $p$-brane theory is
\begin{equation*}
\mathscr{H}(\sigma^a,x^\mu,p^a_\mu)=-\sqrt{\left(\frac{\text{det}\Pi}{-T_p^2}\right)^{1/p}}\in C^\infty(J^1E^*) \ .
\end{equation*}
The De Donder--Weyl field equations are:
$$
\frac{\partial X^\mu}{\partial \sigma^a}=\left(\frac{\partial\mathscr{H} }{\partial  p^a_\mu}\right)\circ\psi\quad , \quad
\frac{\partial p^a_\mu}{\partial\sigma^a}=-\left(\frac{\partial\mathscr{H}}{\partial x^\mu}\right)\circ\psi\ ,
$$
however, the Hamiltonian is computationally more tedious to deal with than in the case of the string.

The multisymplectic symmetries are now worldvolume diffeomorphisms and spacetime isometires. The worldvolume diffeomorphsims are generated on $J^1E$ by the vector field
\begin{equation*}
X_\xi=-\xi^a\frac{\partial}{\partial\sigma^a}+x_b^\mu\derpar{\xi^b}{\sigma^a}\frac{\partial}{\partial x_a^\mu} \ \in\mathfrak{X}(J^1E) \ ,
\end{equation*}
and the corresponding momentum map $J_\mathscr{L}(X_\xi)=-i(X_\xi)\Theta_\mathscr{L}\in\df^{m-1}(J^1E)$ is 
\begin{equation*}
J_\mathscr{L}(X_\xi)=T_p\sqrt{-\text{det}g}\ \left[G_{\mu\nu}g^{ba}x_b^\nu\xi^c\text{d}x^\mu\wedge\text{d}^{m-2}\sigma_{ac}+p\xi^a\text{d}^{m-1}\sigma_a\right]\ .
\end{equation*}
Furthermore, as in the string theory case, the spacetime isometries which produce symmetries of the multisymplectic form $\Omega_\mathscr{L}$ are generated on $J^1E$ by the following vector field:
\begin{equation*}
X_\zeta=-\zeta^\mu\frac{\partial}{\partial x^\mu}-x^\nu_a\partial_\nu\zeta^\mu\frac{\partial}{\partial x^\mu_a}\in\mathfrak{X}(J^1E)\ .
\end{equation*}
The corresponding momentum map $J_\mathscr{L}=-i(X_\zeta)\Theta_\mathscr{L}\in\df^{m-1}(J^1E)$ is given as,
\begin{equation*}
J_\mathscr{L}(X_\zeta)=-T_p\sqrt{-\text{det}g}\ G_{\mu\nu}g^{ba}x_b^\nu\zeta^\mu\text{d}^{m-1}\sigma_a\ .
\end{equation*}

\section{(2+1)-dimensional Gravity}
\label{section:2+1Gravity}

Gravity with cosmological constant $\lambda$ in $2+1$ dimensions is developed in the tetrad formalism using the vielbein $e^a_\mu$
given by $g_{\mu\nu}=e^a_\mu e^b_\nu \eta_{ab}$
($\mu=0,1,2$, $a=0,1,2$),
and the Hodge dual spin connection $\omega^a_\mu=\frac{1}{2}\epsilon^{abc}\omega_{\mu bc}$,
which are treated as the variational fields of the theory as in \cite{WittenGravity}.

\subsection{Lagrangian Formulation}

The configuration bundle $E$ has coordinates $(x^\mu, e^a_\mu, \omega_\rho^c)$; the induced coordinates on $J^1E$ are $(x^\mu, e^a_\mu, \omega_\rho^c, e^a_{\sigma\mu}, \omega^c_{\sigma\rho})$ and $j^1\phi:M\rightarrow J^1E: x^\mu\mapsto \left(x^\mu, e^a_\mu(x), \omega_\rho^c(x), \partial_\sigma e^a_{\mu}(x), \partial_\sigma\omega^c_{\rho}(x)\right)$ are the first-order jet prolongations. 
The Lagrangian density is given by
\begin{equation*}
\textbf{L}=\mathscr{L}(x^\mu, e^a_\mu, \omega_\rho^c, e^a_{\sigma\mu}, \omega^c_{\sigma\rho})\text{d}^3x=\epsilon^{\mu\nu\rho}\left[2\eta_{ac}e^a_\mu\omega^c_{\nu\rho}+\epsilon_{abc}\left(e^a_\mu\omega^b_\nu \omega^c_\rho+\frac{1}{3}\lambda e^a_\mu e^b_\nu e^c_\rho\right)\right]\text{d}^3x \ .
\end{equation*}
It is well-known that this Lagrangian density can be written as 
a {\sl Chern--Simons theory} \cite{WittenGravity},
\begin{equation*}
\textbf{L}=\frac{1}{2}\Big\langle A\wedge\text{d}A+\frac{2}{3}A\wedge A\wedge A\Big\rangle \ ,
\end{equation*}
with gauge field $A=\left(e^a_\mu P_a + \omega^a_\mu J_a\right)\text{d}x^\mu$, where $J^a=\frac{1}{2}\epsilon^{abc}J_{bc}$ are the Lorentz generators in the dual form and $P^a$ are the translation generators which satisfy
\begin{equation}
\label{eq:CS-commutators}
\left[J_a,J_b\right]=\epsilon_{abc}J^c \quad , \quad \left[J_a,P_b\right]=\epsilon_{abc}P^c \quad , \quad  \left[P_a,P_b\right]=\lambda\epsilon_{abc}J^c \ .
\end{equation}
The invariant bilinear form on the Lie algebra of the gauge group is, in general, given as
\begin{equation*}
\left<J_a,P_b\right>=c_0\eta_{ab} \quad , \quad
\left<J_a,J_b\right>=c_1\eta_{ab} \quad ,\quad
\left<P_a,P_b\right>=\lambda c_1\eta_{ab} \ .
\end{equation*}
In this work, the choice $c_0=1,\ c_1=0$  is made and hence, 
\begin{equation*}
\left<J_a,P_b\right>=\eta_{ab} \quad , \quad \left<J_a,J_b\right>=\left<P_a,P_b\right>=0 \ .
\end{equation*}

This field theory is singular which can be seen directly from the following components of the Hessian matrix:
\begin{equation*}
\frac{\partial^2\mathscr{L}}{\partial e^a_{\mu\nu} \partial e^b_{\rho\sigma}}=0 \quad , \quad \frac{\partial^2\mathscr{L}}{\partial\omega^a_{\mu\nu}\partial\omega^b_{\rho\sigma}}=0 \quad , \quad \frac{\partial^2\mathscr{L}}{\partial e^a_{\mu\nu}\partial\omega^b_{\rho\sigma}}=0\ .
\end{equation*}
The null vectors of the Hessian matrix above are given as 
\begin{equation}
\label{CSnullvectors}
_{(1)}\gamma^a_{\mu\nu}\ =\ 
_{(2)}\gamma^a_{\mu\nu}\ =\
\begin{pmatrix}
\begin{pmatrix}
1&1&1\\
1&1&1\\
1&1&1
\end{pmatrix}
&
\begin{pmatrix}
1&1&1\\
1&1&1\\
1&1&1
\end{pmatrix}
&
\begin{pmatrix}
1&1&1\\
1&1&1\\
1&1&1
\end{pmatrix}
\end{pmatrix}^\textbf{T}\ ,
\end{equation}
respectively. The null vector above is expressed as a column vector in the latin (locally Minkowski tangent space) upper index while, at each row of the column vector, the $3\times 3$ matrices  are indexed by the roman lower indices. 

The corresponding Lagrangian energy $E_\mathscr{L}\in C^\infty(J^1E)$,
the Poincar\'e--Cartan $3$-form $\Theta_\mathscr{L}\in\df^3(J^1E)$, and the premultisymplectic 
Poincar\'e--Cartan $4$-form $\Omega_\mathscr{L}\in\df^4(J^1E)$ are given by
\beann
E_\mathscr{L}&=&\frac{\partial\mathscr{L}}{\partial e^a_{\sigma\mu}}e^a_{\sigma\mu}+\frac{\partial\mathscr{L}}{\partial \omega^c_{\sigma\rho}}\omega^c_{\sigma\rho}-\mathscr{L}(x^\mu, e^a_\mu, \omega_\rho^c, e^a_{\sigma\mu}, \omega^c_{\sigma\rho})=-\epsilon^{\mu\nu\rho}\epsilon_{abc}\Big(e^a_\mu\omega^b_\nu \omega^c_\rho+\frac{1}{3}\lambda e^a_\mu e^b_\nu e^c_\rho\Big) \ .
\\
\Theta_\mathscr{L}&=&\frac{\partial\mathscr{L}}{\partial e^a_{\sigma\mu}}\text{d}e^a_\mu\wedge\text{d}^2x_\sigma+\frac{\partial\mathscr{L}}{\partial\omega^c_{\sigma\rho}}\text{d}\omega^c_\rho\wedge\text{d}^2x_\sigma - E_\mathscr{L}\text{d}^3x=\widetilde{\Theta}_\mathscr{L}- E_\mathscr{L}\text{d}^3x \\
&=&2\epsilon^{\mu\nu\rho}\eta_{ac}e^a_\mu\text{d}\omega^c_\rho\wedge\text{d}^2x_\nu+\epsilon^{\mu\nu\rho}\epsilon_{abc}\Big(e^a_\mu \omega^b_\nu \omega^c_\rho + \frac{1}{3}\lambda e^a_\mu e^b_\nu e^c_\rho\Big)\text{d}^3x\ , \\
\Omega_\mathscr{L}&=&
-\text{d}\Big(\frac{\partial\mathscr{L}}{\partial e^a_{\sigma\mu}}\Big)\wedge\text{d}e^a_\mu\wedge\text{d}^2x_\sigma-\text{d}\Big(\frac{\partial\mathscr{L}}{\partial\omega^c_{\sigma\rho}}\Big)\wedge\text{d}\omega^c_\rho\wedge\text{d}^2x_\sigma + \text{d}E_\mathscr{L}\wedge\text{d}^3x \\
&=&
-2\epsilon^{\mu\nu\rho}\eta_{ac}\text{d}e^a_\mu\wedge\text{d}\omega^c_\rho\wedge\text{d}^2x_\nu-\epsilon^{\mu\nu\rho}\epsilon_{abc}\left(\omega^b_\nu\omega^c_\rho\text{d}e^a_\mu+2e^a_\mu\omega^b_\nu\text{d}\omega^c_\rho+\lambda e^b_\nu e^c_\rho \text{d}e^a_\mu\right)\wedge\text{d}^3x \\
&=&
\widetilde{\Omega}_\mathscr{L} + \text{d}E_\mathscr{L}\wedge\text{d}^3x \ .
\eeann
The field equations are obtained using multivector fields  written as $\textbf{X}_\mathscr{L}=X_0\wedge X_1\wedge X_2\in\mathfrak{X}^3(J^1E)$ with components given by
\beq
X_\nu=\frac{\partial}{\partial x^\nu}+B^a_{\nu\mu}\frac{\partial}{\partial e^a_\mu}+C^c_{\nu\rho}\frac{\partial}{\partial \omega^c_\rho}+D^a_{\nu\sigma\mu}\frac{\partial}{\partial e^a_{\sigma\mu}}+H^c_{\nu\sigma\rho}\frac{\partial}{\partial\omega^c_{\sigma\rho}} \ .
\label{dimgravmvf}
\eeq
The geometric field equation \eqref{equation:LEOM} gives,
\begin{align*}
\begin{split}
i(\textbf{X}_\mathscr{L})\Omega_\mathscr{L}&=
-2\epsilon^{\mu\nu\rho}\eta_{ac}\left(B^a_{\nu\mu}\text{d}\omega^c_\rho-C^c_{\nu\rho}\text{d}e^a_\mu\right)+\epsilon^{\mu\nu\rho}\epsilon_{abc}\left(\omega^b_\nu \omega^c_\rho\text{d}e^a_\mu+2e^a_\mu\omega^b_\nu\text{d}\omega^c_\rho+\lambda e^b_\nu e^c_\rho \text{d}e^a_\mu\right) \\
& \ \  -\epsilon^{\mu\nu\rho}\left[2\eta_{ac}\left(B^a_{\lambda\mu}C^c_{\nu\rho}-B^a_{\nu\mu}C^c_{\lambda\rho}\right)-\epsilon_{abc}\left(\omega^b_\nu\omega^c_\rho B^a_{\lambda\mu}+2e^a_\mu \omega^b_\nu C^c_{\lambda\rho}+\lambda e^b_\nu e^c_\rho B^a_{\lambda\mu}\right)\right]\text{d}x^\lambda =0\ .
\end{split}
\end{align*}
Setting differential forms separately equal to zero produces the following two independent field equations:
\begin{subequations}
\begin{align}
&\epsilon^{\mu\nu\rho}\left[\eta_{ac}C^c_{\nu\rho}+\frac{1}{2}\epsilon_{abc}\left(\omega^b_\nu\omega^c_\rho + \lambda e^b_\nu e^c_\rho\right)\right]=0 \ ,
\label{csfieldeq1} \\
&\epsilon^{\mu\nu\rho}\left[\eta_{ac}B^a_{\nu\mu}-\epsilon_{abc}e^a_\mu \omega^b_\nu\right]=0 \ .
\label{csfieldeq2}
\end{align}
\end{subequations}
These equations are compatible on all of $J^1E$ 
as no compatibility constraints arise.
Note that, as
\begin{equation*}
\text{ker}\ \widetilde{\Omega}_\mathscr{L}\cap\text{ker}\ \text{d}^3x= \left\langle \frac{\partial}{\partial e^a_{\rho\mu}},\frac{\partial}{\partial\omega^a_{\rho\mu}}\right\rangle \ ,
\end{equation*}
it follows from Proposition \ref{maintheor} that
$\displaystyle i\left(\frac{\partial}{\partial e^a_{\rho\mu}}\right)\text{d}E_\mathscr{L}=
i\left(\frac{\partial}{\partial\omega^a_{\rho\mu}}\right)\text{d}E_\mathscr{L}=0$ 
is satisfied automatically. 
Observe also that, as $\displaystyle \frac{\partial}{\partial e^a_{\rho\mu}},\frac{\partial}{\partial\omega^a_{\rho\mu}}\in\ker{\Omega}_\mathscr{L}$,
the coefficients $D^a_{\nu\sigma\mu},H^c_{\nu\sigma\rho}$
in \eqref{dimgravmvf} remain undetermined.

Nevertheless, solutions to the field equations
must be holonomic multivector fields $\textbf{X}_\mathscr{L}$ which thereby have components \eqref{dimgravmvf} for which $B^a_{\nu\mu}=e^a_{\nu\mu}$ and $C^c_{\nu\rho}=\omega^c_{\nu\rho}$; it follows that the field equations (\ref{csfieldeq1}) and (\ref{csfieldeq2}) become:
\begin{subequations}
\begin{align*}
&\epsilon^{\mu\nu\rho}\Big[\eta_{ac}\omega^c_{\nu\rho}+\frac{1}{2}\epsilon_{abc}\left(\omega^b_\nu\omega^c_\rho + \lambda e^b_\nu e^c_\rho\right)\Big]=0 \ , \\
&\epsilon^{\mu\nu\rho}\Big[\eta_{ac}e^a_{\nu\mu}-\epsilon_{abc}e^a_\mu \omega^b_\nu\Big]=0 \ .
\end{align*}
\end{subequations}
Both of these equations are {\sc sopde} constraints and define the submanifold $S_1\hookrightarrow J^1E$.
As usual, it is necessary to ensure the tangency of the multivector field $\textbf{X}_\mathscr{L}$ to the submanifold $S_{1}$. 
Recall that this is done by taking the Lie derivative of the constraints with respect to the multivector field components \eqref{dimgravmvf}, where $B^a_{\nu\mu}=e^a_{\nu\mu}$ and $C^c_{\nu\rho}=\omega^c_{\nu\rho}$, giving:
\begin{align*}
0=&\mathpzc{L}_{X_\sigma}\left[\epsilon^{\mu\nu\rho}\Big(\eta_{ac}\omega^c_{\nu\rho}+\frac{1}{2}\epsilon_{abc}(\omega^b_\nu\omega^c_\rho + \lambda e^b_\nu e^c_\rho)\Big)\right]=\epsilon^{\mu\nu\rho}\Big[\eta_{ac}H^c_{\sigma\nu\rho}+\epsilon_{abc}\left(\omega^b_\nu\omega^c_{\sigma\rho} + \lambda e^b_\nu e^c_{\sigma\rho}\right)\Big] \ , \\
0=&\mathpzc{L}_{X_\sigma}\Big[\epsilon^{\mu\nu\rho}(\eta_{ac}e^a_{\nu\mu}-\epsilon_{abc}e^a_\mu \omega^b_\nu)\Big]=\epsilon^{\mu\nu\rho}\Big[\eta_{ac}D^a_{\sigma\nu\mu}-\epsilon_{abc}e^a_{\sigma\mu} \omega^b_\nu-\epsilon_{abc}e^a_\mu \omega^b_{\sigma\nu}\Big] \ .
\end{align*}
These equations are relations for the coefficients $D^a_{\sigma\nu\mu}$ and $H^c_{\sigma\nu\rho}$ and are not constraints; it follows that there are no tangency constraints in this field theory.

Furthermore, the integral sections 
$(x^\mu,e^a_\mu(x),\omega_\rho^c(x),e^a_{\sigma\mu}(x),\omega^c_{\sigma\rho}(x))$
of these holonomic
multivector fields satisfy that
$B^a_{\nu\mu}=\partial_\nu e^a_\mu$ and $C^c_{\nu\rho}=\partial_\nu\omega^c_{\rho}$,
and therefore the equations \eqref{csfieldeq1} and \eqref{csfieldeq2} on $S_1$ become:
\beann
&\epsilon^{\mu\nu\rho}\left[\eta_{ac}\left(\derpar{\omega^c_{\rho}}{x^\nu}-\derpar{\omega^c_{\nu}}{x^\rho}\right)+\epsilon_{abc}\left(\omega^b_\nu\omega^c_\rho + \lambda e^b_\nu e^c_\rho\right)\right]=0 \ , 
\\
&\epsilon^{\mu\nu\rho}\left[\eta_{ac}\left(\derpar{e^a_\nu}{x^\mu}-\derpar{e^a_\mu}{x^\nu}\right) + \epsilon_{abc}\left(e^a_\mu \omega^b_\nu+\omega^a_\mu e^b_\nu\right)\right]=0 \ . 
\eeann
These are the well-known field equations for gravity in $2+1$ dimensions; the first equation is Einstein's equation with cosmological constant while the second guarantees that the spin connection is torsion free, thereby fully specifying the spin connection in terms of the dreibein $e^a_\mu$.

\subsection{Hamiltonian Formulation}

The Legendre map $\mathscr{FL}:J^1E\rightarrow J^1E^*$
associated to $\Lag$ gives the multimomenta
$$
\mathscr{FL}^*p_a^{\mu\nu}=\frac{\partial\mathscr{L}}{\partial e^a_{\mu\nu}}=0
\quad , \quad 
\mathscr{FL}^*\pi_a^{\mu\nu}=\frac{\partial\mathscr{L}}{\partial \omega^a_{\mu\nu}}=2\epsilon^{\mu\nu\rho}\eta_{ab}e^b_\rho \ .
$$
The relations above give the constraints
$p_a^{\mu\nu}=0$ and $\pi_c^{\mu\nu}-2\epsilon^{\mu\nu\rho}\eta_{ab}e^b_\rho=0$,
which define the primary constraint submanifold $\jmath_0:P_0\hookrightarrow J^1E^*$ on which the Hamiltonian formulation takes place. As expected from Proposition \ref{prop1}, the null vectors \eqref{CSnullvectors} are given by the partial derivatives of the primary constraints with respect to the multimomenta according to \eqref{gammarel}:
$$
_{(1)}\gamma^a_{\mu\nu}=\mathscr{FL}^*\derpar{}{p^{\mu\nu}_a}\left(p^{\mu\nu}_a\right)\quad ,\quad
_{(2)}\gamma^a_{\mu\nu}=\mathscr{FL}^*\derpar{}{\pi^{\mu\nu}_a}\left(\pi^{\mu\nu}_a-2\epsilon^{\mu\nu\rho}\eta_{ab}e^b_\rho\right)\ .
$$
The restricted Legendre map $\mathscr{FL}_0$ is given as $\mathscr{FL}=\jmath_0\circ\mathscr{FL}_0$ and the primary constraint submanifold $P_0$ is locally diffeomorphic to $E$
and the system is almost-regular.
Then, the local coordinates on $P_0$ are $(x^\mu, e^a_\mu, \omega_\rho^{c})$. 

The De Donder--Weyl Hamiltonian on $P_0$, which satisfies $E_\mathscr{L}=\mathscr{FL}_0^*\mathscr{H}_0$,
is given by 
\begin{equation*}
\mathscr{H}_0=-\epsilon^{\mu\nu\rho}\epsilon_{abc}\Big(e^a_\mu\omega^b_\nu \omega^c_\rho+\frac{1}{3}\lambda e^a_\mu e^b_\nu e^c_\rho\Big) \ .
\end{equation*}
The Hamilton--Cartan $3$-form $\Theta^0_\mathscr{H}\in\df^3(P_0)$ and the premultisymplectic 
Hamilton--Cartan $4$-form $\Omega^0_\mathscr{H}\in\df^4({P_0})$ are given by 
\beann
\Theta^0_\mathscr{H}&=&
\pi^c_{\sigma\rho}\text{d}\omega^c_\rho\wedge\text{d}^2x_\sigma-\mathscr{H}_0\text{d}^3x\\ &=&
2\epsilon^{\mu\nu\rho}\eta_{ac}e^a_\mu\text{d}\omega^c_\rho\wedge\text{d}^2x_\nu+\epsilon^{\mu\nu\rho}\epsilon_{abc}\Big(e^a_\mu \omega^b_\nu \omega^c_\rho + \frac{1}{3}\lambda e^a_\mu e^b_\nu e^c_\rho\Big)\text{d}^3x\ ,\\
\Omega^0_\mathscr{H}&=&
-\text{d}\pi^c_{\sigma\rho}\wedge\text{d}\omega^c_\rho\wedge\text{d}^2x_\sigma+\text{d}\mathscr{H}_0\wedge\text{d}^3x \\&=&
-2\epsilon^{\mu\nu\rho}\eta_{ac}\text{d}e^a_\mu\wedge\text{d}\omega^c_\rho\wedge\text{d}^2x_\nu-\epsilon^{\mu\nu\rho}\epsilon_{abc}\left(\omega^b_\nu\omega^c_\rho\text{d}e^a_\mu+2e^a_\mu\omega^b_\nu\text{d}\omega^c_\rho+\lambda e^b_\nu e^c_\rho \text{d}e^a_\mu\right)\wedge\text{d}^3x \ .
\eeann
As it is usual, the field equations are obtained using a multivector field $\textbf{X}_\mathscr{H}=X_0\wedge X_1\wedge X_2 \in \mathfrak{X}^3(P_0)$ with components given by
\begin{equation*}
X_\nu=\frac{\partial}{\partial x^\nu}+B^a_{\nu\mu}\frac{\partial}{\partial e^a_\mu}+C^c_{\nu\rho}\frac{\partial}{\partial \omega^c_\rho}\ .
\end{equation*}
Now the geometric field equation \eqref{dimgravmvfh} gives
\begin{align*}
i(\textbf{X}_\mathscr{H})\Omega_\mathscr{H}=
&2\epsilon^{\mu\nu\rho}\eta_{ac}\left(B^a_{\nu\mu}\text{d}\omega^c_\rho-C^c_{\nu\rho}\text{d}e^a_\mu\right)-\epsilon^{\mu\nu\rho}\epsilon_{abc}\Big(\omega^b_\nu \omega^c_\rho\text{d}e^a_\mu+2e^a_\mu\omega^b_\nu\text{d}\omega^c_\rho+\lambda e^b_\nu e^c_\rho \text{d}e^a_\mu\Big)+ \\
&\epsilon^{\mu\nu\rho}\left[2\eta_{ac}\left(B^a_{\lambda\mu}C^c_{\nu\rho}-B^a_{\nu\mu}C^c_{\lambda\rho}\right)+\epsilon_{abc}\Big(\omega^b_\nu\omega^c_\rho B^a_{\lambda\mu}+2e^a_\mu \omega^b_\nu C^c_{\lambda\rho}+\lambda e^b_\nu e^c_\rho B^a_{\lambda\mu}\Big)\right]\text{d}x^\lambda=0\ .
\end{align*}
Setting differential forms separately equal to zero yields
\beann
\epsilon^{\mu\nu\rho}\left[\eta_{ac}C^c_{\nu\rho}+\frac{1}{2}\epsilon_{abc}\left(\omega^b_\nu\omega^c_\rho + \lambda e^b_\nu e^c_\rho\right)\right]=0 \ , \\
\epsilon^{\mu\nu\rho}\left[\eta_{ac}B^a_{\nu\mu}-\epsilon_{abc}e^a_\mu \omega^b_\nu\right]=0 \ .
\eeann
Similarly to the Lagrangian formalism, no compatibility constraints arise.
Again, observe that
\begin{equation*}
\text{ker}\ \widetilde{\Omega}^0_\mathscr{H}\cap \text{ker}\ \text{d}^3x=\left\langle \frac{\partial}{\partial p_a^{\rho\mu}},\frac{\partial}{\partial\pi_a^{\rho\mu}}\right\rangle\ ,
\end{equation*}
from which it follows that 
$\displaystyle i\left(\frac{\partial}{\partial p_a^{\rho\mu}}\right)\text{d}\mathscr{H}_0=
i\left(\frac{\partial}{\partial\pi_a^{\rho\mu}}\right)\text{d}\mathscr{H}_0=0$ automatically. 
Then, by taking $(x^\mu, e^a_\mu(x), \omega_\rho^{bc}(x))$ to be integral sections of $\textbf{X}_\mathscr{H}$,
for which $B^a_{\nu\mu}=\partial_\nu e^a_\mu$ and $C^c_{\nu\rho}=\partial_\nu\omega^c_{\rho}$,
the field equations become again
\beann
\epsilon^{\mu\nu\rho}\left[\eta_{ac}\left(\derpar{\omega^c_{\rho}}{x^\nu}-\derpar{\omega^c_{\nu}}{x^\rho}\right)+\epsilon_{abc}\left(\omega^b_\nu\omega^c_\rho + \lambda e^b_\nu e^c_\rho\right)\right]=0\ , 
\\
\epsilon^{\mu\nu\rho}\left[\eta_{ac}\left(\derpar{e^a_\nu}{x^\mu}-\derpar{e^a_\mu}{x^\nu}\right) + \epsilon_{abc}\left(e^a_\mu \omega^b_\nu+\omega^a_\mu e^b_\nu\right)\right]=0 \ ,
\eeann
which are the same field equation obtained in the Lagrangian formulation.

\subsection{Symmetries}

\subsubsection{Spacetime Diffeomorphisms}

Spacetime diffeomorphisms are generated by the infinitesimal parameter $\xi^\mu$ via the following transformations:
\begin{equation*}
x^\mu\rightarrow x'^\mu= x^\mu + \xi^\mu(x) \ \Rightarrow \ e^a_\mu(x) \Longrightarrow e'^a_\mu(x')=\frac{\partial x^\mu}{\partial x'^\mu}e^a_\mu(x), \  \omega^a_\mu(x) \rightarrow \omega'^a_\mu(x')=\frac{\partial x^\mu}{\partial x'^\mu}\omega^a_\mu(x) \ ,
\end{equation*}
hence,
\beann
\delta e^a_\mu(x)=e'^a_\mu(x)-e^a_\mu(x) =-\xi^\nu\derpar{e^a_\mu}{x^\nu}-e^a_\nu\derpar{\xi^\nu}{x^\mu} \ , \\
\delta \omega^a_\mu(x)=\omega'^a_\mu(x)-\omega^a_\mu(x)=-\xi^\nu\derpar{\omega^a_\mu}{x^\nu}-\omega^a_\nu\derpar{\xi^\nu}{x^\mu} \ .
\eeann
Once again, these field variations are given by the Lie derivative of the local sections $\phi:M\rightarrow E$
generated by the vector field $\xi=-\xi^\mu \derpar{}{x^\mu}\in \mathfrak{X}(M)$ as defined in (\ref{LieDerivSections1}).
The lift of the vector field $\xi$ to the configuration bundle $E$
\cite{boundaries4,Krupka} is given by
\begin{equation*}
\xi_E= -\xi^\mu  \derpar{}{x^\mu} + e^a_\nu \partial_\mu\xi^\nu \frac{\partial }{\partial e^a_\mu} + \omega^a_\nu\partial_\mu \xi^\nu\frac{\partial}{\partial \omega^a_\mu} \in \mathfrak{X}(E) \ .
\end{equation*}
The canonical lift $X_\xi\in \mathfrak{X}(J^1E)$ of $\xi_E$ to $J^1E$ is given as
\begin{equation*}
X_\xi= -\xi^\mu \derpar{}{x^\mu}+ e^a_\nu \partial_\mu\xi^\nu \frac{\partial }{\partial e^a_\mu}+ \omega^a_\nu\partial_\mu \xi^\nu\frac{\partial}{\partial \omega^a_\mu} +\left(e^a_{\sigma\rho}\derpar{\xi^\rho}{x^\mu}-e^a_{\rho\mu}\derpar{\xi^\rho}{x^\sigma}\right)\frac{\partial}{\partial e^a_{\sigma\mu}}
+\left(\omega^a_{\sigma\rho}\derpar{\xi^\rho}{x^\mu}-\omega^a_{\rho\mu}\derpar{\xi^\rho}{x^\sigma}\right)\frac{\partial}{\partial \omega^a_{\sigma\mu}}\ .
\end{equation*}
It follows that $\mathpzc{L}_{X_{\xi}}\Theta_\mathscr{L}=0$ so the momentum map $J_\mathscr{L}(X_\xi)\in \df^2(J^1E)$ is given by
\begin{align*}
\begin{split}
J_\mathscr{L}(X_\xi)&=-i(X_\xi)\Theta_\mathscr{L} \\
&=-2\epsilon^{\mu\nu\rho}\eta_{ac}e^a_\mu\left( \xi^\sigma \text{d}\omega^c_\rho \wedge\text{d}^1x_{\nu\sigma}+\omega^c_\sigma\partial_\rho\xi^\sigma\text{d}^2x_\nu\right)+\epsilon^{\mu\nu\rho}\epsilon_{abc}\left(e^a_\mu\omega^b_\nu\omega^c_\rho+\frac{1}{3}\lambda e^a_\mu e^b_\nu e^c_\rho\right)\xi^\sigma \text{d}^2x_\sigma \ .
\end{split}
\end{align*}
Furthermore, since $P_0$ is locally diffeomorphic to $E$, the vector field $Y_\xi=\mathscr{FL}_{0*}X_\xi$ which generates the spacetime diffeomorphisms on $P_0$ is written as
\begin{equation*}
Y_{\xi}= -\xi^\mu \derpar{}{x^\mu}+ e^a_\nu \partial_\mu\xi^\nu \frac{\partial }{\partial e^a_\mu}+ \omega^a_\nu\derpar{\xi^\nu}{x^\mu}\frac{\partial}{\partial \omega^a_\mu} \in \mathfrak{X}(P_0) \ ,
\end{equation*}
and since $\mathpzc{L}_{Y_{\xi}}\Theta^0_\mathscr{H}=0$, the momentum map $J_\mathscr{H}(Y_\xi)\in \df^2(P_0)$ is given by
\begin{align*}
\begin{split}
J_\mathscr{H}(\xi)&=-i(Y_\xi)\Theta^0_\mathscr{H} \\
&=-2\epsilon^{\mu\nu\rho}\eta_{ac}e^a_\mu\left( \xi^\sigma \text{d}\omega^c_\rho \wedge\text{d}^1x_{\nu\sigma}+\omega^c_\sigma\partial_\rho\xi^\sigma\text{d}^2x_\nu\right)+\epsilon^{\mu\nu\rho}\epsilon_{abc}\left(e^a_\mu\omega^b_\nu\omega^c_\rho+\frac{1}{3}\lambda e^a_\mu e^b_\nu e^c_\rho\right)\xi^\sigma \text{d}^2x_\sigma \ .
\end{split}
\end{align*}

\subsubsection{Local Poincar\'e Transformations}

Local Poincar\'e transformations generated by the infinitesimal parameter $\Lambda(x)= P_a\rho^a(x)+ J_a\tau^a(x)$ induce transformations of the gauge field given as
\begin{equation*}
\delta A_\mu(x)=-D_\mu\Lambda(x)=-\partial_\mu\Lambda-[A_\mu,\Lambda],
\end{equation*}
where  $\tau^a(x)$ are the infinitesimal generators of local Lorentz transformations and $\rho^a(x)$ are the infinitesimal generators of local translations. It follows from the Lie algebra (\ref{eq:CS-commutators}) that
\begin{equation*}
\delta e^a_\mu(x)=-\derpar{\rho^a}{x^\mu}-\epsilon_{abc}e^b_\mu\tau^c-\epsilon_{abc}\omega^b_\mu\rho^c \quad , \quad
\delta\omega^a_\mu(x)=-\derpar{\tau^a}{x^\mu}-\epsilon_{abc}\omega^b_\mu\tau^c-\lambda\epsilon_{abc}e^b_\mu\rho^c \ .
\end{equation*}
The vector field on the configuration bundle $E$ which generates the local Poincar\'e transformations is 
\begin{equation*}
\zeta_E=\left(\derpar{\rho^a}{x^\mu}+\epsilon^a_{\ bc}e^b_\mu\tau^c+\epsilon^a_{\ bc}\omega^b_\mu\rho^c\right)\frac{\partial}{\partial e^a_\mu}+\left(\derpar{\tau^a}{x^\mu}+\epsilon^a_{\ bc}\omega^b_\mu\tau^c+\lambda\epsilon^a_{\ bc} e^b_\mu\rho^c\right)\frac{\partial}{\partial\omega^a_\mu} \ .
\end{equation*}
The canonical lift of $\zeta_E$ to $J^1E$ is
\begin{align*}
\begin{split}
X_\zeta&=\left(\derpar{\rho^a}{x^\mu}+\epsilon^a_{\ bc}e^b_\mu\tau^c+\epsilon^a_{\ bc}\omega^b_\mu\rho^c\right)\frac{\partial}{\partial e^a_\mu}+\left(\derpar{\tau^a}{x^\mu}+\epsilon^a_{\ bc}\omega^b_\mu\tau^c+\lambda\epsilon^a_{\ bc} e^b_\mu\rho^c\right)\frac{\partial}{\partial\omega^a_\mu}\\ &+\epsilon^a_{\ bc}\left(e^b_\mu\derpar{\tau^c}{x^\sigma}+\omega^b_\mu\derpar{\rho^c}{x^\sigma}+e^b_{\sigma\mu}\tau^c\right)\frac{\partial}{\partial e^a_{\sigma\mu}}+\epsilon^a_{\ bc}\left(\omega^b_\mu\derpar{\tau^c}{x^\sigma}+\lambda e^b_\mu\derpar{\rho^c}{x^\sigma}+\omega^b_{\sigma\mu}\tau^c\right)\frac{\partial}{\partial \omega^a_{\sigma\mu}}\ ,
\end{split}
\end{align*}
while the projection $\mathscr{FL}_{0*}X_\zeta\equiv Y_\zeta$ onto $P_0$ is
\begin{equation*}
Y_\zeta=\left(\derpar{\rho^a}{x^\mu}+\epsilon^a_{\ bc}e^b_\mu\tau^c+\epsilon^a_{\ bc}\omega^b_\mu\rho^c\right)\frac{\partial}{\partial e^a_\mu}+\left(\derpar{\tau^a}{x^\mu}+\epsilon^a_{\ bc}\omega^b_\mu\tau^c+\lambda\epsilon^a_{\ bc} e^b_\mu\rho^c\right)\frac{\partial}{\partial\omega^a_\mu}\ .
\end{equation*}
The momentum map $J_\mathscr{L}(X_\zeta)\in\df^2(J^1E)$ on the multivelocity phase space is given by
$$
J_\mathscr{L}(X_\zeta)=-i(Y_\zeta)\Theta_\mathscr{L}
=-2\epsilon^{\mu\nu\rho}\eta_{ac}e^a_\mu\left( \derpar{\tau^c}{x^\mu}+\epsilon^c_{\ db}\omega^d_\mu\tau^b+\lambda\epsilon^c_{\ db} e^d_\mu\rho^b\right)\text{d}^2x_\nu \ .
$$
and the momentum map $J_\mathscr{H}(Y_\zeta)\in \df^2(P_0)$ on the multimomentum phase space is given by the same expression
$$
J_\mathscr{H}(Y_\zeta)=-i(Y_\zeta)\Theta^0_\mathscr{H}
=-2\epsilon^{\mu\nu\rho}\eta_{ac}e^a_\mu\left( \derpar{\tau^c}{x^\mu}+\epsilon^c_{\ db}\omega^d_\mu\tau^b+\lambda\epsilon^c_{\ db} e^d_\mu\rho^b\right)\text{d}^2x_\nu \ .
$$


\section{Conclusions and Outlook}

In this paper, a geometric approach has been used for performing a full constraint analysis of field theories on both the multivelocity and multimomentum phase spaces where the Lagrangian and De Donder--Weyl Hamiltonian formalisms take place respectively.
The multivelocity and multimomentum phase spaces are,  respectively, the jet and dual jet bundles of the field configuration manifold $E$ which is regarded as a fiber bundle over spacetime (or over the string worldsheet in the case of string theory). These phase spaces are equipped with
multisymplectic or premultisymplectic forms which are then used to produce the field equations of the theory under investigation, reveal the symmetries of the field theory, and carry out a full constraint analysis of the singular field theories. 

The approach for carrying out the premultisymplectic constraint analysis of field theories involves the use of multivector fields 
to represent geometrically the solutions of field equations.
So, an easy procedural geometric technique for finding the constraints locally has been described,
and some new properties of the constraints are also exposed
(Propositions \ref{maintheor} and \ref{prop1}).
The converse of Proposition \ref{maintheor}, which states that the compatibility constraint submanifold $C_1$ is completely determined by \eqref{ncond}, is left for further research
along with other results and features regarding the geometric constraint algorithm.

Furthermore, it is shown how further constraint submanifolds 
(defined by the so-called {\sc sopde} constraints) may be found in the Lagrangian formalism when imposing the holonomic condition which guarantees that the Lagrangian field equations are second-order partial differential equations. 
The field equations, i.e. the well-known Euler--Lagrange equations in the Lagrangian case and the Hamilton--De Donder--Weyl equations in the Hamiltonian case, are obtained from the (pre)multisymplectic variational principle in which multivector fields of a particular type serve as solutions to the field equations. 
By example, in the field theories analyzed in Sections \ref{canonicalKG} and \ref{CarrollMagnetic},
it is shown that when {\sc sopde} constraints exist, the vector fields which generate symmetries of the field theory under investigation may be projectable via the Legendre map only on the {\sc sopde} constraint submanifold. 
The geometric constraint analysis finalizes by imposing stability of the solutions to the field equations; this is done by demanding tangency of the solution multivector fields to all constraint submanifolds present in the system until no new constraint submanifolds are produced. It is also worth noting that this work presents a new proven proposition, Proposition \ref{prop1}, which states that the multi-Hessian (\ref{multiHessian}) (which characterizes the Legendre map to the multimomentum phase space) has null vectors which are given by the partial derivative of the primary constraints with respect to the multimomenta. 

The technique used in this paper for performing the geometric constraint analysis described above is illustrated in various field theory examples, each of which highlight different aspects of the geometric constraint structure. 
The scalar field theories presented in this work include a new approach to the study of the canonical Klein--Gordon Lagrangian as well as the first premultisymplectic analysis of Carrollian scalar field theories.
The multisymplectic treatment of bosonic string theory from the Nambu--Goto action is given 
along with the generalization to the theory of bosonic $p$-branes. 
The regularity of the Nambu--Goto action with respect to the De Donder--Weyl Legendre map is shown geometrically by inspection of the multisymplectic forms that characterize the theory's multivelocity and multimomentum phase spaces.
The connection between the regularity of bosonic string theory in the De Donder--Weyl formalism and its singular behavior in the canonical formulation remains unexplored as, in general, no direct relationship has yet been established between the premultisymplectic constraint analysis presented in this work and the canonical constraint structure arising from the canonical Legendre map.
Understanding the connection between the constraint structures in these two formalisms is left for future research.
Finally, the premultisymplectic construction of Chern--Simons gravity in $2+1$ dimensions is given 
and it is shown that in the Lagrangian formalism there are only {\sc sopde} constraints present,
while in the De Donder--Weyl Hamiltonian formalism there are only primary constraints which are produced by the singular Legendre map.

It is interesting to note that General Relativity is singular (in the De Donder--Weyl sense) in any spacetime dimension.
This was previously shown in $3+1$ dimensions in \cite{art:Capriotti2,first,GR1,GR2} and here it was shown in $2+1$ dimensions.
The analysis of General Relativity given here and in \cite{art:Capriotti2,GR2} involves treating the connection as an independent variational field;
in this setting, the Einstein--Hilbert Lagrangian is singular as it is linear in the multivelocities, thereby leading to a singular multi-Hessian \eqref{multiHessian}. 
When only the metric is taken to be the fundamental field of the theory (as the connection is assumed to be Levi--Civita),
the Einstein--Hilbert Lagrangian
is singular for similar reasons as shown in the premultisymplectic setting in \cite{first,GR1}.
No complications arise in generalizations of higher dimension as the singular structure of the relevant Hessians is independent of spacetime dimension.
Similarly, 
theories of massive gravity, multi-gravity, and all theories of the Chern--Simons type (including higher-spin gravity in $(2+1)$ dimensions) are De Donder--Weyl singular due to linear dependence of the corresponding Lagrangians on the multivelocities. 

Given that the constraint structure of field theories in the De Donder--Weyl formalism is only understood geometrically at this point in time, 
it would be interesting to develop an algebraic understanding of the constraints {\sl a la} Dirac. Such a treatment of constraints requires the construction of Poisson brackets on some phase space. 
In the canonical formalism for field theories, Poisson brackets are well understood in the context of symmetries and constraints on the covariant phase space by Lee and Wald \cite{leewaldcov} for example; for further insight into such covariant phase spaces (and related ones) see \cite{BHS-1991}.
In the De Donder--Weyl formalism however, the construction of Poisson brackets it still a topic under investigation with many papers published on the matter (see, for example, \cite{multisympcov,MMMT-86,K-1998,FPR-2005,G-2021,GMM-2022} and references therein). Covariant Poisson brackets on jet bundles (where the Lagrangian formulation of field theories takes place) has also been discussed extensively in the literature (see for example \cite{ACDI-2017} and references therein) along with the development of the BRST-BV formalism \cite{M-1994}. However, the understanding of the BRST-BV formalism in terms of the (pre)multisympletic structures on jet bundles is not yet fully understood and is left for further research; we also leave gauge fixing in the De Donder--Weyl formulation of the field theories studied in this paper for future work. 


\section*{Acknowledgments}

We acknowledge conversations on non-Lorentzian topics with Eric Bergshoeff, Roberto Casalbuoni,
Jos\'e Figueroa-O'Farrill,  Axel Kleinschmidt, and Alfredo P\'erez.
We are indebted to Prof. Victor Tapia for having drawn our attention to an error in the first calculation of the formula \eqref{pbraneHessian}.
We acknowledge the financial support of the 
{\sl Ministerio de Ciencia, Innovaci\'on y Universidades} (Spain), projects PGC2018-098265-B-C33 and PID2021-125515NB-C21,
and the financial support for research groups AGRUPS-2022 of the Universitat Polit\`ecnica de Catalunya (UPC).
The work of JG has been supported in part by MINECO FPA2016-76005-C2-1-P and PID2019- 105614GB-C21 and from the State Agency for Research of the Spanish Ministry of Science and Innovation through the Unit of Excellence Maria de Maeztu 2020-203 award to the Institute of Cosmos Sciences (CEX2019-000918-M).


\end{document}